\documentclass[reprint,superscriptaddress,amsmath,amssymb,aps,prb]{revtex4-2}
\usepackage{graphicx}
\usepackage[caption=false]{subfig}
\usepackage{dcolumn}
\usepackage{bm}
\usepackage{hyperref}

\usepackage{xcolor}

\newcommand{\be}{\begin{equation}}
\newcommand{\ee}{\end{equation}}
\def\a{\alpha}
\def\b{\beta}

\def\t{\tau}

\def\o{\omega}

\def\O{\Omega}

\def\ra{\rightarrow}

\def\bk{{\bf k}}

\def\bQ{{\bf Q}}
\def\bA{{\bf A}}

\DeclareMathOperator{\Tr}{Tr}

\begin{document}

\preprint{2D}

\title{Two-dimensional THz spectroscopy in electronic systems: \\a many-body diagrammatic approach}

\author{Jacopo Fiore}
	\email{jacopo.fiore@uniroma1.it}
	\affiliation{Department of Physics, ``Sapienza'' University of Rome, P.le
		A. Moro 5, 00185 Rome, Italy}
	\author{Niccolò Sellati}
	\affiliation{Department of Physics, ``Sapienza'' University of Rome, P.le
		A. Moro 5, 00185 Rome, Italy}
	\author{Mattia Udina}
    \affiliation{Institut de Physique et Chimie des Mat\'{e}riaux de Strasbourg (UMR 7504),
Universit\'{e} de Strasbourg and CNRS, Strasbourg, 67200, France}
    \affiliation{Laboratoire Mat\'{e}riaux et Ph\'{e}nom\`{e}nes Quantiques,
Universit\'{e} Paris Cit\'{e}, CNRS, 75205 Paris, France}
	\affiliation{Department of Physics, ``Sapienza'' University of Rome, P.le
		A. Moro 5, 00185 Rome, Italy}
	\author{Lara Benfatto}
	\email{lara.benfatto@roma1.infn.it}
	\affiliation{Department of Physics, ``Sapienza'' University of Rome, P.le
		A. Moro 5, 00185 Rome, Italy}

\date{\today}

\begin{abstract}

The term two-dimensional coherent spectroscopy (2DCS) usually refers to experimental setups where a coherently generated electric field in a sample is recorded over many runs as a function of two time variables: the delay $\tau$ between two consequent excitation pulses and the time $t$ over which the signal is emitted. Even if its implementation in the femtosecond time domain for investigation of vibrational molecular states was developed more than two decades ago, its experimental development in the THz domain with application to interacting electronic systems is still in its infancy. The present work aims at providing a theoretical framework for the description and interpretation of 2DCS by using the same many-body language based on a perturbative diagrammatic expansion that has been largely developed and applied in the literature to linear spectroscopy. By focusing on the case of centrosymmetric systems, we show how the problem of the theoretical interpretation of the 2D maps can be recast in the solution of two different but complementary problems. The first one is the evaluation of a third-order response function to the gauge field, whose derivation in the velocity gauge leads to a semi-analytical expression for the computation of the 2D maps {in the case of simple band dispersions and Gaussian-like pulse envelope}. This has the two-fold advantage of reducing considerably the computational complexity and to guide the assignment of the spectral features to microscopic processes, as we demonstrate explicitly for {a toy model of electrons in a semiconducting-like band structure}. The second one is a careful description of multi-wave propagation effects inside the material, that for bulk systems can completely screen the intrinsic properties of the nonlinear response, as we show for the case of soft superconducting Josephson plasmons. Our results provide a theoretical foundation prone to further extension to several interacting systems and offer a flexible method for modeling realistically the nonlinear responses across arbitrary spectral widths of the driving fields.

\end{abstract}

\maketitle

\section{\label{sec:intro}Introduction}

Recent experimental advances in the generation and control of phase-stable light pulses in the THz regime opened the way to the exploration of a wide range of phenomena connected to coherent driving of {excitations} in this energy window \cite{hirori_Appl.Phys.Lett.11,leitenstorfer_J.Phys.D:Appl.Phys.23}. On the one hand, THz pulses can be used to selectively drive infrared active phonon modes, with the aim of inducing time-dependent modifications of the lattice structure that are not possible at equilibrium \cite{forst_NaturePhys11,kampfrath_NaturePhoton13}. On the other hand, in the THz regime one can directly access the continuum of low-energy electronic excitations of metals, without inducing optical transitions to higher energy bands, as happens instead with light pulses in the visible \cite{giannetti_Adv.Phys.16}. Such a perspective is particularly promising when the metallic state undergoes a phase transition to a lower symmetry state, as happens across a superconducting (SC) or charge-density wave (CDW) transition. In this case, not only the fermionic excitation spectrum is modified by the opening of a gap, but simultaneously new collective modes connected to the order-parameter fluctuations emerge \cite{anderson_Phys.Rev.58,volkov_Sov.J.Exp.Theor.Phys.74,littlewood_Phys.Rev.B82,sun_Phys.Rev.Res.20}. THz spectroscopy, and in particular its nonlinear counterpart, then arises as a preferential tool to coherently drive these collective excitations by avoiding the chain of relaxation processes occurring when the exciting pulse lies in the visible range \cite{matsunaga_Phys.Rev.Lett.13, matsunaga_Science14}.

From the theoretical side, the investigation of systems with a continuum of low-energy excitations, which eventually develop a gap in the THz range below a phase transition, has been so far based on effective models where only states near the Fermi level are considered. The nonlinear current generated by the intense THz field is computed in the vector-potential gauge $\mathbf{A}(t)$, coupled to the Hamiltonian via the minimal coupling shift $\mathbf{p}-e/c\mathbf{A}$ of the momentum operator. In centrosymmetric systems, the leading nonlinear current can then be expressed as:
\begin{align}
\label{eq:jnl}
J^{(3)}(\omega)=\int d\omega_i\,&K^{(3)}(\omega_1,\omega_2,\omega_3)A(\omega_1)A(\omega_2)A(\omega_3)\nonumber\\
\times&\delta(\omega-\omega_1-\omega_2-\omega_3).
\end{align}
where space indices are omitted for simplicity, $A(\omega)$ is the Fourier transform of the perturbing THz field and $K^{(3)}$ stands for the nonlinear optical kernel, obtained as a third-order response function to the perturbation $A(t)$. 
Despite the simplification of considering only low-energy electronic states, the computation of $K^{(3)}$ requires a careful {account for} processes of {\em diamagnetic} or {\em paramagnetic} nature, as the extensive literature on superconductors has shown \cite{cea_Phys.Rev.B16,murotani_Phys.Rev.B19,silaev_Phys.Rev.B19,tsuji_Phys.Rev.Research20,haenel_Phys.Rev.B21,seibold_Phys.Rev.B21,udina_FaradayDiscuss.22}. The terminology relies on the fundamental light-electron coupling obtained via the minimal-coupling scheme, i.e.\ 
\be
H_{\text{em}}=-\mathbf{j}\cdot \mathbf{A}+ \rho \mathbf{A}^2=H_{\text{para}}+H_{\text{dia}}
\ee
where $\mathbf{j}$ is the electronic current and $\rho$ the electronic density. In the paramagnetic process one photon excites an electron-hole pair, while in the diamagnetic one this occurs via two photons coming at the same time. The main consequence for what concerns the nonlinear response $K^{(3)}$ is that diamagnetic interactions give rise to two-point (Kubo-like) correlation functions, which share some analogies with those measured in the linear response, while paramagnetic interactions require one to compute up to four-point correlation functions, whose frequency properties are not accessible by linear spectroscopy. 

As the benchmark case of superconductors has shown, the two classes of processes (diamagnetic vs.\ paramagnetic) provide different and complementary information on the interactions at play in the systems. Theoretical predictions have been tested, for example, against experiments of high harmonic generation, where a single light pulse impinges on the sample and the nonlinear electric field generated by the nonlinear current at different harmonics is measured \cite{matsunaga_Science14,chu_NatCommun20,wang_Phys.Rev.B22,reinhoffer_Phys.Rev.B22}. 

However, a complete experimental characterization of the different microscopic processes would require knowledge of the dependence of $K^{(3)}$ on {\em three} different frequencies, which is not possible simply reading $J^{(3)}$ on a single $\omega$, as in Eq.\ \ref{eq:jnl}.  Recent experimental advances on the so-called two-dimensional THz spectroscopy (2DTS) made a significant step forward in this direction \cite{lu_MultidimensionalTime-ResolvedSpectroscopy19,liu_npjQuantumMater.25,huang_NatRevPhys26}. The experimental setup is based on the detection of the nonlinear signal induced by two strong light pulses delayed by an interval $\tau$, by recording the response as a function of the acquisition time $t$ and the delay $\tau$ \cite{mukamel_95,hamm_11,cundiff_PhysicsToday13,smallwood_LaserPhotonicsRev.18}. When 2D temporal maps are transformed in the frequency domain, it is possible to read out different frequency combinations separately in the nonlinear kernel $K^{(3)}$. 
 
In preliminary applications of 2DTS to electronic systems, glimpses of novel effects have been seen, including the possibility to experimentally disentangle signatures of superconducting amplitude modes \cite{luo_Nat.Phys.23,kim_Sci.Adv.24,katsumi_Phys.Rev.Lett.24,yuan_Sci.Adv.24,katsumi_Phys.Rev.Lett.25,huang_Sci.Adv.25,cheng_Phys.Rev.B25}, to address separately the role of disorder and inhomogeneity \cite{wan_Phys.Rev.Lett.19,mahmood_Nat.Phys.21,liu_Nat.Phys.24}, to distinguish energy relaxation from momentum relaxation pathways \cite{barbalas_Phys.Rev.Lett.25,chaudhuri_25}, to probe coupling between distinct modes in magnetic materials \cite{lu_Phys.Rev.Lett.17,mashkovich_Science21,zhang_Nat.Phys.24}. However, a general theoretical framework capable of providing a link between these complex experimental protocols and the properties of the higher-order correlation functions involved in the measurements is still under construction \cite{mootz_CommunPhys22,hart_Phys.Rev.B23,puviani_Phys.Rev.B23,mootz_Phys.Rev.B24,salvador_25,chen_npjComputMater25,ono_Phys.Rev.Lett.25,chen_25,tsuji_25}. This is in part due to the fact that 2DTS has traditionally a long history of applications to molecular systems \cite{mukamel_95,hamm_11,kuehn_J.Phys.Chem.B11,cundiff_PhysicsToday13}, where the fundamental light-matter process encompasses a dipolar-like excitation of vibrational states, which is usually described with the density-matrix evolution in the time domain for a nondispersive two-level system.

The present manuscript aims to provide a completely different perspective, motivated by the preliminary results of the emerging field of 2DTS in electronic systems. Our goal is to provide a theoretical description of 2DTS protocols based on the computation of third-order correlation functions in the frequency domain via a diagrammatic approach, able to account for the features related to the existence of a continuum of excitations and their diamagnetic or paramagnetic coupling to light. This goal will be achieved by addressing two complementary aspects. The first one concerns the interplay between the spectral content of the pump field and the resonances of the nonlinear optical kernel. Such an interplay will be discussed within a paradigmatic {toy model for a semiconducting-like band structure.} We will show that diamagnetic and paramagnetic processes, connected to two-photon (diamagnetic) or single-photon (paramagnetic) excitation mechanisms, lead to markedly different signatures in the 2DTS maps, which can be used experimentally to test their relative importance. 

The second aspect concerns the connection between the nonlinear response induced within the system and the nonlinear signal detected outside the sample, taking into account propagation effects \cite{huber_TheJournalofChemicalPhysics21,frenzel_Sci.Adv.23,sellati_npjQuantumMater.25}. For thin samples and materials with a featureless refractive index in the spectral range of the pump, such propagation effects play a minor role; conversely, for thick samples and in the presence of dipolar-active modes, they can even mask the intrinsic response of the system. This problem can be efficiently addressed by perturbatively solving Maxwell's equations in the presence of a nonlinear source current coming from Eq.\ \ref{eq:jnl}, along the lines suggested in some recent publications \cite{zhang_NationalScienceReview23, fiore_Phys.Rev.B24, gomezsalvador_Phys.Rev.B24}. In this manuscript, we will provide a general solution applicable to different experimental geometries once the frequency-dependent refractive index $n(\omega)$ of the material is known, and we will show the extreme sensitivity of 2DTS maps to such propagation effects.  As a benchmark example of their relevance, we will discuss the case of soft Josephson plasmons, which represent the fundamental collective excitation of the phase degrees of freedom in layered superconductors, like cuprate superconductors. Josephson plasmons provide a natural source for nonlinear optical response, since their dynamics in the presence of a THz pump polarized along the stacking direction is described by an intrinsically anharmonic Hamiltonian:
\be
\label{eq:josephson}
H=J \cos \left(\phi-\frac{2e}{\hbar c}A\right).
\ee
Here $\phi$ denotes the interlayer phase difference between stacked planes coupled by a Josephson interaction $J$. In this context as well, the problem has been mostly studied in the past by solving in the time domain the nonlinear sine-Gordon equation for the current obtained from Eq.\ \ref{eq:josephson} \cite{bulaevskii_Phys.Rev.B94,machida_PhysicaC:Superconductivity00,savelev_NaturePhys06,savelev_Rep.Prog.Phys.10,laplace_Adv.Phys.X16}. However, recent measurements through a 2DTS setup \cite{liu_Nat.Phys.24,taherian_npjQuantumMater.25} turned the attention to the interpretation of the data in the 2D frequency maps. According to the same general principles outlined above, one should address the problem by taking into account that the Josephson model in Eq.\ \ref{eq:josephson} admits a diamagnetic-like nonlinear coupling $\sim \phi^2 A^2$ of the plasma modes to the light pulses \cite{gabriele_NatCommun21}. As we show below, this simple reasoning completely fails in providing an explanation for the experiments that crucially requires to account for the modification of the driving pulse when it propagates inside the material. More specifically, we will show how simulations of the 2DTS maps generated by the nonlinear driving of Josephson plasmons in cuprates can be extremely sensitive to fine details of the modeling of the propagation effects. This finding suggests that some extra care should be used when the refractive index of the system is strongly frequency dependent in the spectral range covered by the pump, as it occurs for example in the case of infrared-active vibrational modes \cite{huber_TheJournalofChemicalPhysics21,frenzel_Sci.Adv.23,sellati_npjQuantumMater.25}.

\subsection{{Synopsis of the paper}}

\begin{figure}
\includegraphics[width=\columnwidth]{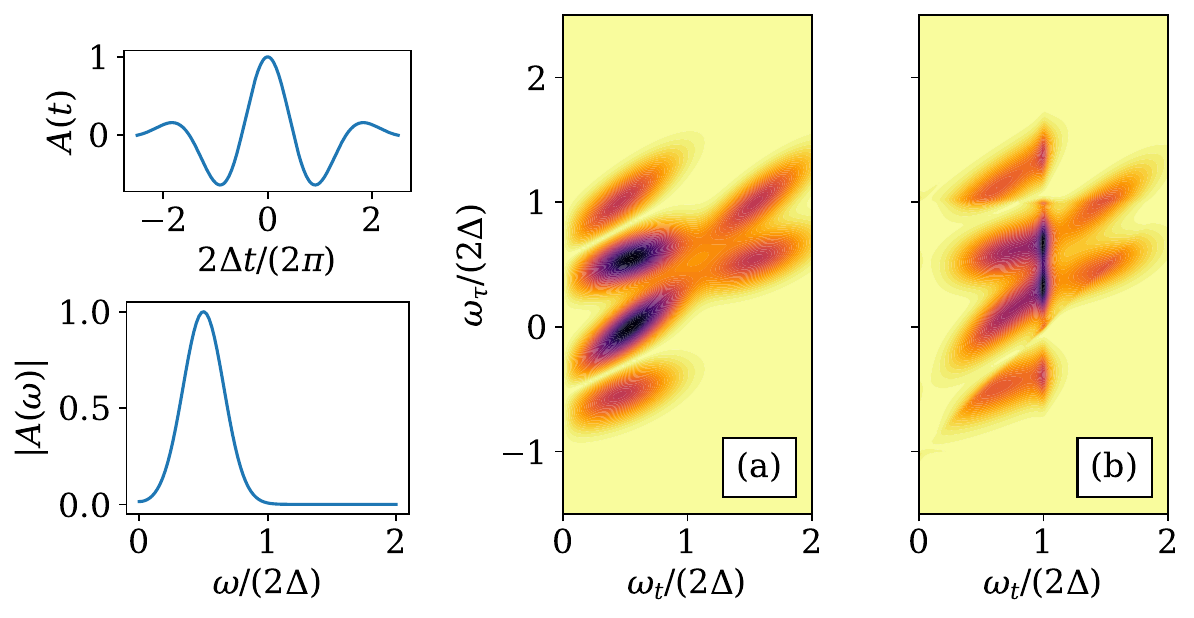}
\caption{\label{fig:syn} Time-domain envelope of $A(t)$ representing the typical waveform for realistic few-cycles pulses used in the experiments (top). The Fourier transform (bottom) has a finite spectral width around the central frequency $\Omega$, set in this example at $\Omega=\Delta$. Panel (a): 2D map obtained from Eq.\ \ref{eq:tot} for a frequency-independent $K_{\mathcal{S}}^{(3)}$. In  2D experiments with pulses having a finite duration in time the convolution of the $A_{0,\tau}(\omega)$ spectra in Eq.\ \ref{eq:tot} sets the region in the 2D map where one can access the system excitations, encoded in the nonlinear kernel $K_{\mathcal{S}}^{(3)}$. The 2D features represent a broadened version of the spot-like response expected for a monochromatic pulse, and shown in Fig.\ \ref{fig:mono} below. Panel (b): when the full structure of $K_{\mathcal{S}}^{(3)}$ is included in Eq.\ \pageref{eq:tot} the spectral weight gets redistributed in the 2D plane. In the present case we show the results for the semiconducting-like model derived in Sec.\ \ref{sec:mod}  where the main resonances follow the spectral gap at $2\Delta$. The parameters are the same used in Fig.\ \ref{fig:semitot}(b) below.}
\end{figure}

The present summary aims at providing the pipeline of the main results of the manuscript to guide the reader through the various sections. The starting point will be the simple case of transmission through a non-absorbing thin film, discussed in Secs.\ \ref{sec:gen} and \ref{sec:mod}, where the signal $E_{2D}(t,\tau)$ measured in 2D spectroscopy can be mapped into the nonlinear third-order current generated inside the material $J^{(3)}_{2D}(t,\tau)$
\begin{equation}
\label{eq:appr}
E_{2D}(t,\tau)\sim J^{(3)}_{2D}(t,\tau).    
\end{equation}
As it is customary in the context of electronic systems, the current will be computed in response to an applied gauge potential $A(t)$. The Fourier transform $J^{(3)}_{2D}(\omega_t,\omega_\tau)$ will then be shown to depend on the convolution between the spectral content of the pump and probe pulses, denoted with $A_{\tau}(\omega)$ and $A_{0}(\omega)$, respectively, and the nonlinear (symmetric) kernel $K_\mathcal{S}^{(3)}$, which encodes all the information on the resonant excitations of the system. The quantity $J^{(3)}_{2D}$ depends generically on the product of three electromagnetic fields: one can distinguish processes scaling as $A_\tau A_\tau A_0$ from processes scaling as $A_\tau A_0 A_0$. By making the dependence on the detection frequency $\omega_t$ and on the delay frequency $\omega_\tau$ explicit, we show in Eqs.\  \ref{eq:j00tau} and \ref{eq:j0tautau} that one can express these contributions as:
\begin{widetext}
\begin{align}
\label{eq:tot}
E_{2D}(\omega_t,\omega_\tau)\sim 
{A_0(\omega_t-\omega_\tau)}
\int d\omega\,{K^{(3)}_{\mathcal{S}}}(\omega_t-\omega_\tau,\omega,\omega_\tau-\omega){A_\tau(\omega)A_\tau(\omega_\tau-\omega)}\nonumber\\
+{A_\tau(\omega_\tau)}\int d\omega\,{K^{(3)}_{\mathcal{S}}}(\omega_\tau,\omega,\omega_t-\omega_\tau-\omega)
{A_0(\omega)A_0(\omega_t-\omega_\tau-\omega)}.
\end{align}    
\end{widetext}
We notice that the first term of Eq.\ \ref{eq:tot} would be also present in usual pump-probe experiments where the probe field is weaker than the pump one, while the second only arises when the two delayed pulses have the same strength. 

Eq.\ \ref{eq:tot} will be used to discuss the outcome of 2D maps as one varies the bandwidth of the pulses, encoded in the $A_{\tau,0}(\omega)$ functions, and the nature of the excitations of the system, described by the kernel ${K^{(3)}_{\mathcal{S}}}$. In Sec.\ \ref{sec:gen} we discuss standard results for a monochromatic pulse where $A_{\tau,0}(\omega)$ only contains $\delta(\omega\pm \Omega)$. In this case the 2D maps present simple spots at multiples of the pulse frequencies $\pm\Omega$, see Fig.\ \ref{fig:mono}. In the complementary case of an impulsive electric field, that can be  approximated with a delta function in time, the 2DTS is dominated by the typical excitations processes described by the nonlinear kernel ${K^{(3)}_{\mathcal{S}}}$. As a working example we discuss in Sec.\ \ref{sec:imp} the non-dispersive two-level system, showing how one can reproduce within a diagrammatic language the standard results derived in the literature by means of a density-matrix formalism in real time, see Fig.\ \ref{fig:imp}. At this stage, it should be emphasized that in realistic experiments the impulsive limit is rarely achieved. The first consequence of Eq.\ \ref{eq:tot} is that one should preliminarily study the convolution spectra of the pump and probe pulses. For realistic single-cycle THz pulses the 2D map obtained   by setting ${K^{(3)}_{\mathcal{S}}}\sim const$ in Eq.\ \ref{eq:tot} contains elongated features in correspondence of multiples of the central pulse frequency, as shown in Fig.\ \ref{fig:syn}(a). As a consequence, any feature of the kernel which resides outside this convolution will not be accessible in the experimental situation of interest. When the resonances of the kernel overlap with the 2D map of the pulses the spectral weight is redistributed, and additional features highlighting the system excitations appear, as shown in Fig.\ \ref{fig:syn}(b).

In Sec.\ \ref{sec:mod} we extend the formalism to dispersive electronic systems and consider as an example a toy model for a semiconducting-like band structure. We discuss the main steps needed to obtain a semi-analytical form of the 2D signal, that are encoded in the set of Eqs.\ \ref{eq:scheme}-\ref{eq:jdiatot2}. The main goal of this Section is to highlight the different signatures in the 2D maps that can be ascribed to diamagnetic-like vs.\ paramagnetic-like processes. As we shall discuss in the following, these subsets of processes are interesting \textit{per se} since they can be relevant in contexts beyond the toy model considered here.

A final aspect addressed in the manuscript is the generalization of Eq.\ \ref{eq:appr} to absorbing and bulk systems. In this case, one cannot simply approximate the 2D detected signal $E_{2D}(t,\tau)$ by the nonlinear current generated inside the material, since the latter depends on the local value of the gauge potential within the sample. We will show that one can extend Eq.\ \ref{eq:tot} to include two ingredients: (i) the boundary conditions at the sample/vacuum interfaces, that connect the external field $A_{\text{ext}}$ to the internal one via the refractive index $n(\omega)$ of the material; (ii) the propagation effects inside the sample, that modify the four-wave mixing process behind the generation of a nonlinear current. Eq.\ \ref{eq:tot} can then be generalized to:
\begin{widetext}
\begin{align}
E_{2D}(\omega_t,\omega_\tau)\sim 
\int d\omega\,{K^{(3)}_{\mathcal{S}}}(\omega_t-\omega_\tau,\omega,\omega_\tau-\omega)M(\omega_t-\omega_\tau,\omega,\omega_\tau-\omega){A_{0,\rm ext}(\omega_t-\omega_\tau)}{A_{\tau,\rm ext}(\omega)A_{\tau,\rm ext}(\omega_\tau-\omega)}+\nonumber\\
\int d\omega\,{K^{(3)}_{\mathcal{S}}}(\omega_\tau,\omega,\omega_t-\omega_\tau-\omega)M(\omega_\tau,\omega,\omega_t-\omega_\tau-\omega){A_{\tau,\rm ext}(\omega_\tau)}
{A_{0,\rm ext}(\omega)A_{0,\rm ext}(\omega_t-\omega_\tau-\omega)}.\label{eq:genfre}
\end{align}    
\end{widetext}
where $M(\omega_1,\omega_2,\omega_3)$ is a function of $n(\omega)$ that depends on the detection scheme (reflection or transmission), which can be thought of as a nonlinear generalization of the usual Fresnel coefficients. The explicit form of Eq.\ \ref{eq:genfre} to the case of reflection from a bulk sample is presented in Eq.\ \ref{eq:genout}, and its analogous in transmission is given in the Supplementary Material (SM) \cite{fiore_}, where we also provide detailed steps for its derivation.
As a case study of the relevance of this generalization we discuss in Secs.\ \ref{sec:plasmon1}-\ref{sec:plasmon3} the nonlinear response of superconducting Josephson plasmons in cuprate superconductors. In Sec.\ \ref{sec:con} we provide the final comments and an outlook on possible extensions of the present work. Since the manuscript aims at a pedagogical discussion of several relevant aspects, we provide further technical details, along with useful formulas for the computation of the 2D maps, in the SM \cite{fiore_}.

\section{\label{sec:gen}General formalism of 2DTS}

\begin{figure}
\includegraphics[width=\columnwidth]{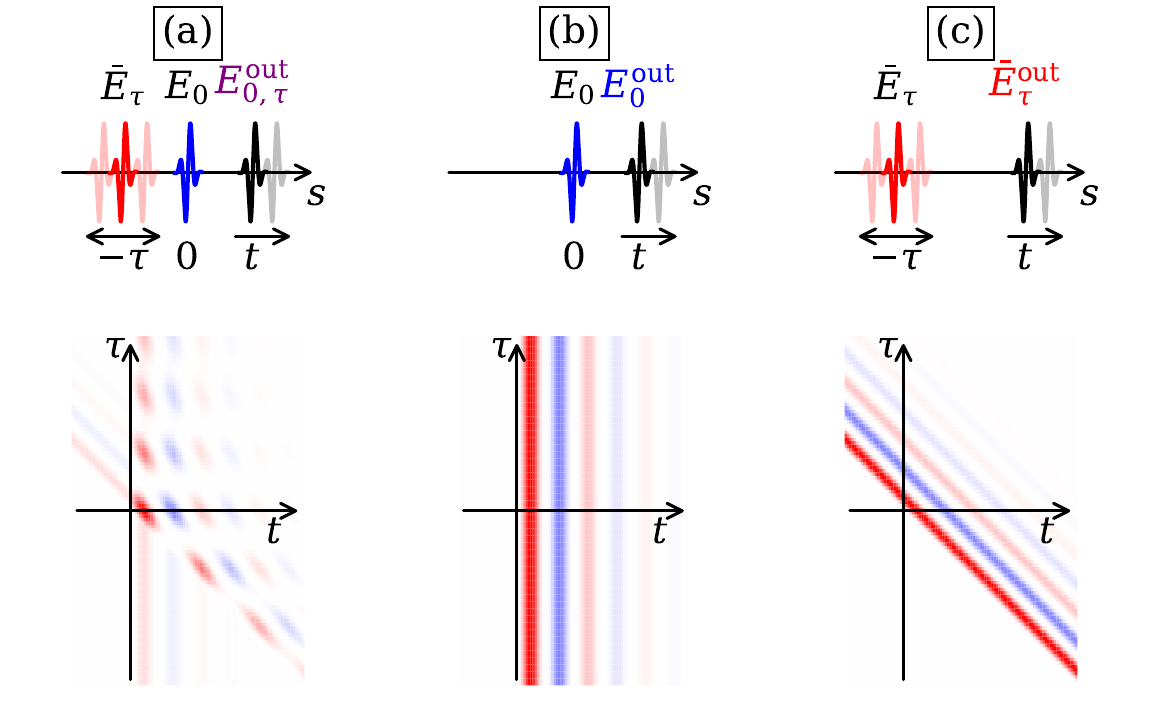}
\caption{\label{fig:exp2d} Schematic representation of the three signals that generate the experimental 2D map according to Eq.\ \ref{eq:2dexp}. Top row: the field $E_0$ (blue) is kept fixed at $s=0$, the field $\bar{E}_\tau$ (red) is centered at $s=-\tau$ and the outgoing signal is collected at time $s=t$. Both $t$ and $\tau$ are scanned (light shading), with the latter being either positive ($E_\tau$ coming first) or negative ($E_0$ coming first). The signal is collected when both fields are present, $E_{0,\tau}^{\text{out}}$ in panel (a), or when only one of the two is present, $E_{0}^{\text{out}}$ in panel (b) and $E_{\tau}^{\text{out}}$ in panel (c). Bottom row: resulting signal for the three situations assuming a nonlinear polarization as generated by the two-level system discussed in Section \ref{sec:imp}.}
\end{figure}

As a starting point, we need to connect the outcome of a 2D measurement to a third-order response function of the system. Let us consider the general 2D protocol sketched in Fig.\ \ref{fig:exp2d}. Here we denote with $E_0(s)$ the ``fixed'' field, whose envelope is centered at $s=0$, and $\bar{E}_{\tau}(s)$ the ``moving'' one, which is centered at $s=-\tau$. Notice that we do not restrict here the time ordering of the two pulses, which can be varied according to the sign of $\tau$: we thus have that either $\bar{E}_\tau$ or $E_0$ come first for $\tau>0$ or $\tau<0$, respectively \cite{liu_J.Chem.Phys.25}. The experimental procedure consists in collecting the generated response at a time $s=t$ when both fields are present, $E^{\text{out}}_{0,\tau}(t,\tau)$ in Fig.\ \ref{fig:exp2d}(a), or when only one of them illuminates the system, $E^{\text{out}}_{0}(t)$ and $E^{\text{out}}_{\tau}(t,\tau)$ in Fig.\ \ref{fig:exp2d}(b)-(c), respectively. The generated radiation is collected as a function of $(t,\tau)$ and the differential signal is recorded as:
\be
\label{eq:2dexp}
E_{2D}(t,\tau)\equiv E^{\text{out}}_{0,\tau}(t,\tau)-E^{\text{out}}_{0}(t)-E^{\text{out}}_{\tau}(t,\tau).
\ee
{In practice, the three signals entering Eq.\ \ref{eq:2dexp} are extracted using a differential chopping scheme combined with phase-resolved electro-optic sampling of the emitted THz field \cite{mahmood_Nat.Phys.21,katsumi_Phys.Rev.Lett.24}. Here, the two excitation pulses are selectively modulated, and one can isolate the nonlinear contribution corresponding to the combined action of both fields, which is mathematically equivalent to the subtraction in Eq.\ \ref{eq:2dexp}. This equivalence holds provided that the system response is stationary across laser shots and that relaxation to equilibrium occurs on a timescale shorter than the pulse repetition rate, conditions typically satisfied in THz 2D spectroscopy experiments employing low to moderate fluences.}

In the simple case of non-absorbing thin films the $E_{2D}(t,\tau)$ can be approximated with the (nonlinear)  {current $J_{2D}(t,\tau)$} induced in the system, $E_{2D}(t,\tau)\sim {J_{2D}(t,\tau)}$, neglecting thus any effect of the propagation of the electric fields inside and outside the sample{, as detailed in the SM \cite{fiore_}}. The analysis of Sec.\ \ref{sec:prop} will clarify the limits of such approximation. Consequently,  one can \emph{theoretically} compute the quantity:
\be
\label{eq:2d}
{J_{2D}(t,\tau)=J[A_{0,\tau}](t,\tau)-J[A_{0}](t)-J[\bar{A}_{\tau}](t,\tau).}
\ee
where each term represents the induced {current $J$} in the presence of two {($A_{0,\tau}$)} or one {($A_{0}$, $\bar{A}_{\tau}$)} driving fields, {here represented by their associated gauge potential, related by $E(t)=-\partial_tA$}. The signal collected in time is usually Fourier transformed to the conjugated frequencies domain $(t,\tau)\rightarrow(\omega_t,\omega_\tau)$ and represented as a 2D map. {Additional experimental data processing, such as phase untwisting \cite{mahmood_Nat.Phys.21,hart_Phys.Rev.B23} and time or frequency windowing \cite{liu_Nat.Phys.24,liu_J.Chem.Phys.25}, can in principle be incorporated within the present framework depending on the specific experimental realization.} 

Let us first perform the Fourier transform with respect to $t$ of the 2D signal in Eq.\ \ref{eq:2d} by explicitly writing the electric fields entering each term. For the first term we have the sum {$A(\omega,\tau)=A_0(\omega)+\bar A_\tau(\omega)$}, while {$A(\omega,\tau)=A_{0}(\omega)$} or {$A(\omega,\tau)=\bar A_{\tau}(\omega)$} for the remaining two terms. Since the Fourier transform of {$\bar{A}_{\tau}(t)$} is simply {$\bar{A}_\tau(\omega)=e^{-i\omega\tau}A_{\tau}(\omega)$}, where {$A_{\tau}(\omega)$} is the spectrum of the delayed field for $\tau=0$, we see that the $\tau$ dependence of the {$J_{2D}(\omega_t,\tau)$} only enters via the phase factor $e^{-i\omega \tau}$ of the delayed pulse. Moreover, one can easily see that the linear response cancels out in Eq.\ \ref{eq:2d}, since contributions to the {current} at first order in {$A$}, {$J^{(1)}(\omega)=K^{(1)}(\omega)A(\omega)$}, are simply additive. As a consequence, the lowest-order nonlinear process in a centrosymmetric material generating the 2D signal comes from a third-order {current}. Let us write this response in full generality as follows:
\begin{align}
\label{eq:j3ft}
{J^{(3)}}(\omega_t,\tau)=\int d\omega_i\,&\delta(\omega_t-\omega_1-\omega_2-\omega_3){K^{(3)}}(\omega_1,\omega_2,\omega_3)\nonumber\\
\times&{A(\omega_1,\tau)A(\omega_2,\tau)A(\omega_3,\tau)}.
\end{align}
{Here ${K^{(3)}}$ is again the nonlinear response kernel to the gauge potential $A$ of Eq.\ \ref{eq:jnl}}, that depends in the most general case on the three frequencies $\omega_i$, $i=1,2,3$. We note that in principle in Eq.\ \ref{eq:j3ft} there is a further summation over the cartesian components of the {kernel} and {gauge} fields, that we keep aside for the moment to make the notation lighter. By replacing in Eq.\ \ref{eq:j3ft} the {$A(\omega,\tau)$} obtained for each term of  Eq.\ \ref{eq:2d} we  note that the third-order terms proportional to {$A_0^3$} and {$A_\tau^3$} drop out, leaving two contributions:
\be
{J^{(3)}_{2D}=J^{(3)}_{00\tau}+J^{(3)}_{0\tau\tau}.}
\ee
The first corresponds to two interactions with field $E_0$ and one with field {$A_{\tau}$}, which we will denote as {$J^{(3)}_{00\tau}$}. The other corresponds to two interactions with field {$A_{\tau}$} and one with field {$A_0$}, which we will denote accordingly as {$J^{(3)}_{0\tau\tau}$}. They read, as a function of $\omega_t$ and $\tau$,
\begin{widetext}
\begin{align}
\label{eq:j2dwt1}
{J^{(3)}_{00\tau}}(\omega_t,\tau)&=\int d\omega_i\,{K^{(3)}}(\omega_1,\omega_2,\omega_3)\delta(\omega_t-\omega_1-\omega_2-\omega_3)\left[e^{-i\omega_1\tau}{A_\tau(\omega_1)A_0(\omega_2)A_0(\omega_3)}+\text{perm.}\right],\\
\label{eq:j2dwt2}
{J^{(3)}_{0\tau\tau}}(\omega_t,\tau)&=\int d\omega_i\,{K^{(3)}}(\omega_1,\omega_2,\omega_3)\delta(\omega_t-\omega_1-\omega_2-\omega_3)\left[e^{-i(\omega_1+\omega_2)\tau}{A_\tau(\omega_1)A_\tau(\omega_2)A_0(\omega_3)}+\text{perm.}\right],
\end{align}
\end{widetext}
where by ``perm.'' we mean analogous terms where we permute the frequency indices $123\to231,321$ in square brackets. 
	
Let us focus on the first contribution, {$J^{(3)}_{00\tau}(\omega_t,\omega_\tau)$}. The Fourier transform $\tau\to\omega_\tau$ in Eq.\ \ref{eq:j2dwt1} will produce a $\delta(\omega_i-\omega_\tau)$, which can be used to eliminate one integral over frequency and set one of the $\omega_i$ frequencies to $\omega_\tau$. Since there is still the energy-conservation delta to exploit in Eq.\ \ref{eq:j2dwt1}, we can use it to set the sum of the remaining two frequencies, both coming from {$A_0$}, to $\omega_t-\omega_\tau$, remaining with only one integral over frequency. We notice that since this process is symmetric with respect to the frequency choice $\omega_i$, since all of them appear in Eq.\ \ref{eq:j2dwt1}, we can introduce an {explicitly} symmetrized form of {the nonlinear response kernel}, 
\begin{align}
\label{eq:chi3sy}
{K^{(3)}_{\mathcal{S}}}(\omega_\tau,\omega,\omega_t-\omega_\tau-\omega)&={K^{(3)}}(\omega_\tau,\omega,\omega_t-\omega_\tau-\omega)\nonumber\\
&+{K^{(3)}}(\omega,\omega_\tau,\omega_t-\omega_\tau-\omega)\nonumber\\
&+{K^{(3)}}(\omega,\omega_t-\omega_\tau-\omega,\omega_\tau).
\end{align}
Since in the present framework the {$K^{(3)}$ kernel} is {\textit{already}} symmetric  with respect to \textit{any} permutation of the incoming fields \cite{mukamel_95,rostami_AnnalsofPhysics21}, {as detailed in the SM}, the previous expression reduces to ${K^{(3)}_{\mathcal{S}}}(\omega_\tau,\omega,\omega_t-\omega_\tau-\omega)=3{K^{(3)}}(\omega_\tau,\omega,\omega_t-\omega_\tau-\omega)$. We can then write out
\begin{align}
\label{eq:j00tau}
{J^{(3)}_{00\tau}}(\omega_t,\omega_\tau)={A_\tau(\omega_\tau)}\int d\omega\,&{K^{(3)}_{\mathcal{S}}}(\omega_\tau,\omega,\omega_t-\omega_\tau-\omega)\nonumber\\
\times&{A_0(\omega)A_0(\omega_t-\omega_\tau-\omega)}.
\end{align}
A similar analysis can be carried out for Eq.\ \ref{eq:j2dwt2}, {$J^{(3)}_{0\tau\tau}$}. In this case, one gets from the $\tau$ Fourier transform a $\delta(\omega_\tau-\omega_i-\omega_j)$, with $i\neq j$. But then the energy-conservation delta will require the only photon from {$A_0$} to be at frequency $\omega_t-\omega_\tau$. As before we can then write out
\begin{align}
\label{eq:j0tautau}
{J^{(3)}_{0\tau\tau}}(\omega_t,\omega_\tau)&={A_0(\omega_t-\omega_\tau)}\nonumber\\
\times\int d\omega\,&{K^{(3)}_{\mathcal{S}}}(\omega_t-\omega_\tau,\omega,\omega_\tau-\omega){A_\tau(\omega)A_\tau(\omega_\tau-\omega)}.
\end{align}
We immediately notice that the two contributions can be transformed into each other by means of the exchange $\omega_\tau\leftrightarrow\omega_t-\omega_\tau$ and {$A_0(\omega)\leftrightarrow A_\tau(\omega)$}. In the following, we specifically analyze two limiting cases relevant to experimental situations: two time-delayed almost monochromatic pulses and two time-delayed almost impulsive fields.

\subsection{Monochromatic limit}
	
In the case of multicycle THz pulses one can approximately represent the field as monochromatic in order to get an intuition on the structure of the 2D map. We can then put  {$A_0(t)=\bar{A}_0\cos(\Omega t)$} and {$\bar{A}_\tau(t)=\bar{A}_\tau\cos\left(\Omega (t+\tau)\right)$}, using two different prefactors to distinguish the two classes of process. From the previous Eq.\ \ref{eq:j00tau} and Eq.\ \ref{eq:j0tautau} we obtain the analytical expressions
\begin{widetext}
\begin{align}
\label{eq:j2dmono}
{J^{(3)}_{2D}}(\omega_t>0,\omega_\tau)&=\delta(\omega_t-\Omega)\left[2\delta(\omega_\tau-\Omega)+\delta(\omega_\tau+\Omega)\right]{K^{(3)}_{\mathcal{S}}}(\Omega,\Omega,-\Omega){\bar{A}_0^2\bar{A}_\tau}+\delta(\omega_t-3\Omega)\delta(\omega_\tau-\Omega){K^{(3)}_{\mathcal{S}}}(\Omega,\Omega,\Omega){\bar{A}_0^2\bar{A}_\tau}\nonumber\\
&+\delta(\omega_t-\Omega)\left[2\delta(\omega_\tau)+\delta(\omega_\tau-2\Omega)\right]{K^{(3)}_{\mathcal{S}}}(\Omega,\Omega,-\Omega){\bar{A}_0\bar{A}_\tau^2}+\delta(\omega_t-3\Omega)\delta(\omega_\tau-2\Omega){K^{(3)}_{\mathcal{S}}}(\Omega,\Omega,\Omega){\bar{A}_0\bar{A}_\tau^2}.
\end{align}
\end{widetext}

\begin{figure}
\includegraphics[width=\columnwidth]{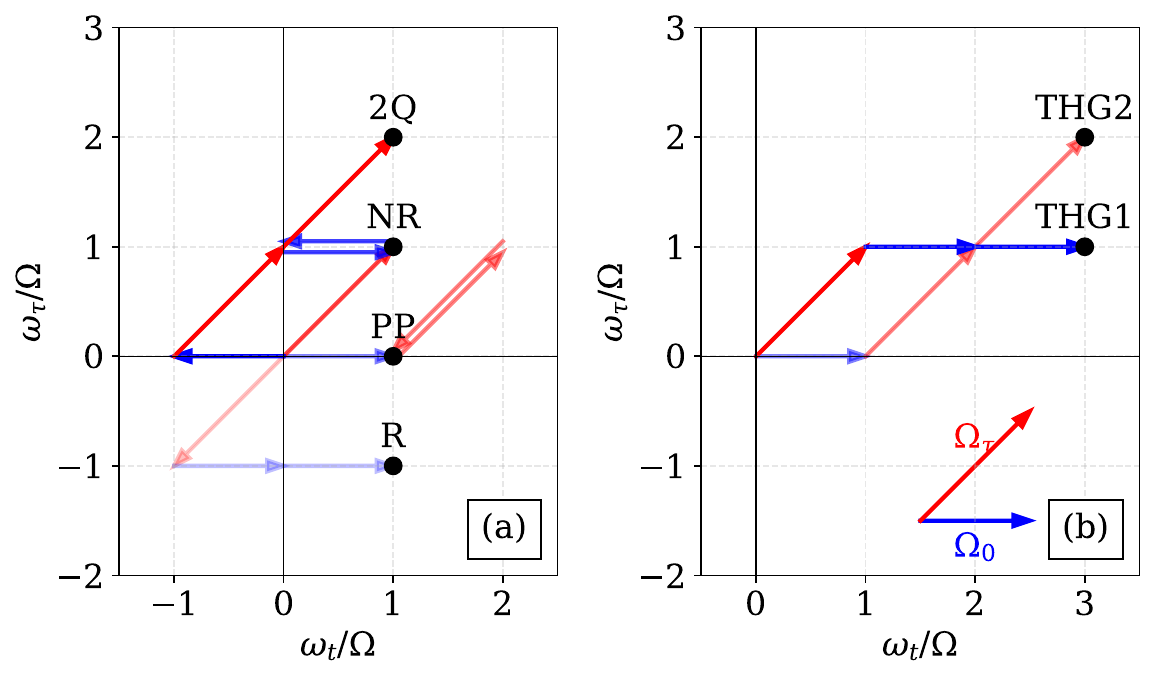}
\caption{\label{fig:mono} Schematic outcome of a 2D experiment performed via monochromatic driving pulses $E(t)\sim\cos(\Omega t)$, obtained by using the frequency-vector representation introduced in the text  for $E_0$ ($\overrightarrow{\Omega}_0$, blue arrow) and $\bar{E}_\tau$ ($\overrightarrow{\Omega}_\tau$, red arrow). Panel (a): representation of the processes providing an oscillation at $\omega_t=\Omega$, i.e.\ the nonlinear first harmonic. Panel (b): same representation for the third harmonic signal at $\omega_t=3\Omega$. Different shades of the arrows correspond to different ``paths'' to reach the six spots, which are named according to Table \ref{tab:names}.}
\end{figure}

\begin{table}
\caption{\label{tab:names}Summary of the spots obtained with monochromatic pulses. The labeling of the first-harmonic spots follows the usual convention appropriate in the impulsive limit for $\tau>0$ for systems with discrete energy levels, with the correspondence $\Omega \leftrightarrow 2\Delta$, $2\Delta$ being the level spacing \cite{kuehn_J.Phys.Chem.B11,mahmood_Nat.Phys.21,liu_npjQuantumMater.25}.}
\begin{ruledtabular}
\begin{tabular}{ccc}
\textrm{Position}&
\textrm{Name}&
\textrm{Abbreviation}\\
\colrule
$(\omega_t,\omega_\tau)=(\Omega,2\Omega)$ & \textrm{Two-quantum} & \textrm{2Q} \\
$(\omega_t,\omega_\tau)=(\Omega,\Omega)$ & \textrm{Non-rephasing} & \textrm{NR} \\
$(\omega_t,\omega_\tau)=(\Omega,0)$ & \textrm{Pump-probe} & \textrm{PP} \\
$(\omega_t,\omega_\tau)=(\Omega,-\Omega)$ & \textrm{Rephasing} & \textrm{R} \\
$(\omega_t,\omega_\tau)=(3\Omega,\Omega)$ & \textrm{Third harmonic} & \textrm{THG1} \\
$(\omega_t,\omega_\tau)=(3\Omega,2\Omega)$ & \textrm{Third harmonic} & \textrm{THG2} \\
\end{tabular}
\end{ruledtabular}
\end{table}

The final 2D map in the half-plane $\omega_t>0$ contains then six spots, four of them aligned along the vertical direction of the first harmonic $\omega_t=\Omega$ ($\omega_\tau=-\Omega,0,\Omega,2\Omega)$, see panel (a) in Fig.\ \ref{fig:mono}, and two of them aligned along the line $\omega_t=3\Omega$ ($\omega_\tau=\Omega,2\Omega$), as in panel (b) of the same figure. We note that various naming conventions are used in the literature to denote the spots in the two-dimensional plane, most of which originate from the time-domain evolution of systems driven by impulsive fields \cite{kuehn_J.Phys.Chem.B11,mahmood_Nat.Phys.21,liu_npjQuantumMater.25}. The terms ``rephasing'', ``non-rephasing'', ``pump–probe'', and ``two-quantum'' acquire specific physical meanings only when the temporal ordering of the pulses can be uniquely determined. {In the strictly monochromatic limit, where the pulses have infinite duration, the order of arrival cannot be inferred from the relative position of their envelopes. Nevertheless, since experimental narrowband pulses possess finite envelopes, we adopt in Fig.\ \ref{fig:mono} the same convention\cite{kuehn_J.Phys.Chem.B11,mahmood_Nat.Phys.21,liu_npjQuantumMater.25}  established in the impulsive limit for $\tau>0$ according to Table \ref{tab:names}. }

In panels (a) and (b) of Fig.\ \ref{fig:mono} we also introduce a compact way of representing the nonlinear process that produces the various spots on the 2D map by means of a frequency vector scheme \cite{woerner_NewJ.Phys.13,mahmood_Nat.Phys.21}. Since an oscillation in time at frequency $\Omega$ can be represented as a vector in the conjugated frequency domain oriented in the direction perpendicular to the wavefronts, we introduce two vectors $\overrightarrow{\Omega}_0=(\Omega,0)$ and $\overrightarrow{\Omega}_\tau=(\Omega,\Omega)$ in the $(\omega_t,\omega_\tau)$ plane to represent (positive frequency) photons coming from {$A_0(t)$} or {$\bar{A}_{\tau}(t)$}, respectively. {More specifically, for a monochromatic field $A_0(\tau)=\bar A_0 \cos(\Omega t)$ the Fourier components at $\pm \Omega$ are represented by the vectors $
\pm \overrightarrow{\Omega}_0$, and analogously for the $A_\tau (t)$}.  The outcome of the nonlinear process can then be pictorially computed by the vector sum of two $\pm\overrightarrow{\Omega}_0$ and one $\pm\overrightarrow{\Omega}_\tau$ for the {$A_0^2A_\tau$} contribution, or one $\pm\overrightarrow{\Omega}_0$ and two $\pm\overrightarrow{\Omega}_\tau$ for the piece {$A_0A^2_\tau$}.

In the perfectly monochromatic limit the spectral content of the 2D signal is uniquely set by the spectrum of the incoming field. In particular, at fixed $\omega_t=\Omega$ or $\omega_t=3\Omega$, the relative weight of the spots along the $\omega_\tau$ direction only depends on the combinatorial prefactors and field amplitudes determined by Eq.\ \ref{eq:j2dmono}. Here, the specific form of the third-order kernel only controls the relative weight between the first and third harmonic through {$K^{(3)}(\Omega,\Omega,-\Omega)$} and {$K^{(3)}(\Omega,\Omega,\Omega)$}, respectively. It is worth noting that the third-harmonic {kernel $K^{(3)}(\Omega,\Omega,\Omega)$} can also be extracted by single-pulse experiments, where one only measures the electric field transmitted or reflected from a sample \cite{matsunaga_Science14,chu_NatCommun20,wang_Phys.Rev.B22,reinhoffer_Phys.Rev.B22}, since higher harmonics of the incoming pulse can only arise from nonlinear processes. This configuration can be ultimately obtained as the $\tau=0$ limit of the two-pulse protocol presented above, which in frequency domain amounts to integrate in $\omega_\tau$ the previous Eq.\ \ref{eq:j2dmono}. On the other hand, in one-dimensional protocols the \emph{nonlinear first-harmonic}, weighted by {$K^{(3)}(\Omega,\Omega,-\Omega)$}, is superimposed to the linear response, that is usually much larger. As a consequence, even in the monochromatic limit the 2D protocol contains additional information as compared to single-pulse protocols \cite{katsumi_Phys.Rev.Lett.24,katsumi_Phys.Rev.Lett.25}.

\subsection{\label{sec:imp}Impulsive limit}

\begin{figure}
\includegraphics[width=\columnwidth]{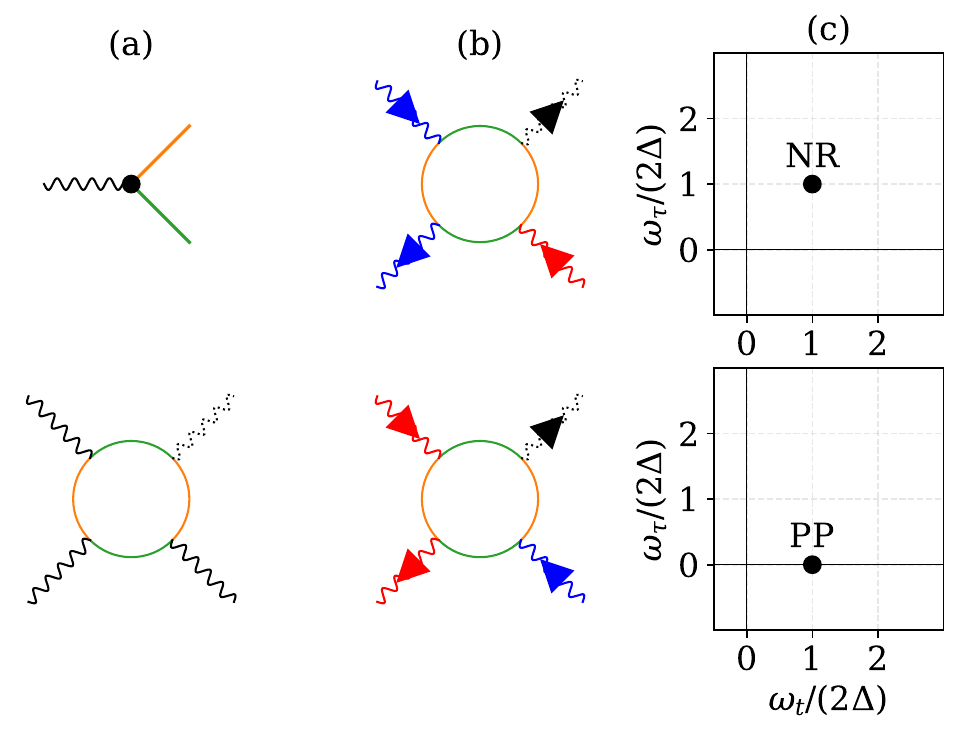}
\caption{\label{fig:flow} {Flowchart of the construction of the 2D map within the diagrammatic approach. (a) The starting building block (top) is the light-matter interaction vertex: for the two-level system of Eq.\ \ref{eq:ham} it is the dipolar interaction of Eq.\ \ref{eq:intv} where the electric field (wavy line) generates transitions between the lower (solid orange line) and upper (solid green line) level. The third order nonlinear response is then built from diagrams with four light-matter insertions (bottom). The dotted wavy line denotes the detection frequency $\omega_t=\sum_i\omega_i$, while each electromagnetic field with frequency $\omega_i$ appears as a solid wavy line. (b) To get the 2D map the latter black lines are colored as $E_0$ (blue) or $E_\tau$ (red), to form $P^{(3)}_{00\tau}$ in Eq.\ \ref{eq:i00t} (top) or $P^{(3)}_{0\tau\tau}$ in Eq.\ \ref{eq:i0tt} (bottom), here exemplified for a specific field combination. In the impulsive limit the Fourier transforms of the fields are constant in frequency and the values on the external legs are set by the transition energy between $-\Delta$ and $+\Delta$. (c) The detection at $\omega_t>0$ imposes $\omega_t=2\Delta$, consequently, the upper diagram carries $\omega_\tau=2\Delta$ (red line) leading to the NR peak at $(\omega_t,\omega_\tau)=(2\Delta,2\Delta)$; the lower diagram shows $\omega_t-\omega_\tau=2\Delta$ (blue line), i.e.\ $\omega_\tau=0$, generating the PP spot at $(\omega_t,\omega_\tau)=(2\Delta,0)$.}}
\end{figure}

A second typical approximation, used for experiments with single-cycle THz pulses \cite{mahmood_Nat.Phys.21,tsuji_25}, is the impulsive limit, modeled by delta-like fields in the time domain $E(t)\sim\delta(t)$, which are Fourier transformed to constants in frequency space. In this case, by direct inspection of Eqs.\ \ref{eq:j00tau} and \ref{eq:j0tautau} it is understood that the behavior of the 2D map will {mostly} depend on the frequency structure of the {nonlinear conductivity $\sigma^{(3)}(\omega_1,\omega_2,\omega_3)\sim K^{(3)}/(\omega_1\omega_2\omega_3)$, due to the relation $E\sim i\omega A$}. The analysis of the 2D maps obtained for a {representative} dispersive electronic systems will be the subject of the next Section. However, to set the stage for this analysis we will first discuss a simpler and paradigmatic case, i.e.\ a nondispersive two-level system, which has been discussed in the literature within the time-dependent density-matrix formalism \cite{mukamel_95,hamm_11,mahmood_Nat.Phys.21}. We will see that one can easily map these results into a collection of Feynman diagrams decorated with different pulse sequences, representing the various excitation pathways. As a starting model we then consider a collection of $N$ non-interacting two-level systems ($\lvert0\rangle,\lvert1\rangle$) hosting $N$ spinless electrons in the ground state at $T=0$. Setting the zero of the energy at the chemical potential, the Hamiltonian can be written as
\be
\label{eq:ham}
H=\sum_n
\begin{pmatrix}
c_{n0}^{\dagger}\,{,}& c_{n1}^{\dagger}
\end{pmatrix}
\begin{pmatrix}
-\Delta&0\\
0&\Delta
\end{pmatrix}
\begin{pmatrix}
c_{n0}\\
c_{n1}
\end{pmatrix},
\ee
where $2\Delta$ is the level spacing and $n=1\dots N$ labels the electronic creation and annihilation operators {$c^{\dagger}_{n0,1}$ and $c_{n0,1}$}. The system can be coupled to the electromagnetic field through the dipolar-like interaction term
\be
\label{eq:intv}
H_{\text{int}}(t)=\mu E(t)
\begin{pmatrix}
0&1\\
1&0
\end{pmatrix},
\ee
where $\mu$ is the dipole moment of the transition. The diagrammatic representation of the light-electron coupling vertex is shown in {Fig.\ \ref{fig:flow}}.

Performing a conventional perturbative expansion in powers of the electric field, one can obtain the third-order nonlinear susceptibility in real frequencies which reads:
\begin{align}
\label{eq:chi32l}
\chi^{(3)}(\omega_1,\omega_2,\omega_3)&\sim\frac{1}{\omega_{123}-2\Delta}\frac{1}{\omega_2+2\Delta}\nonumber\\
&\times\left(\frac{1}{\omega_1-2\Delta}+\frac{1}{\omega_3-2\Delta}\right)-(\Delta\to-\Delta)\nonumber\\
&+\text{perm.},
\end{align}
where $\omega_{123}=\sum_i\omega_i$ and all real frequencies $\omega_i$ are meant to include a vanishingly small imaginary part $i\eta$ after the analytical continuation. {Before proceeding further, we notice that from this nonlinear susceptibility we could immediately obtain the nonlinear kernel $K^{({3})}(\omega_1,\omega_2,\omega_3)\sim\omega_{123}\omega_1\omega_2\omega_3\chi^{(3)}(\omega_1,\omega_2,\omega_3)$, that we used in the previous derivations to connect the applied \textit{gauge field} $A$ to the experimentally-relevant induced \textit{current} $J_{2D}$. However, just for the rest of this section, we switch the focus on the induced \textit{polarization} $P_{2D}$ in response to the \textit{electric field} to make a more explicit connection to the results of Ref.\ \cite{mahmood_Nat.Phys.21} obtained for the two-level system. As one can check, the conversion factor between the two in frequency space, $J_{2D}(\omega_t,\omega_\tau)\sim\omega_tP_{2D}(\omega_t,\omega_\tau)$, will not influence the position of the relevant resonances in the 2D plane. In time domain, $J_{2D}(t,\tau)\sim\partial_tP_{2D}(t,\tau)$, thus the time derivative will just change the phase of the oscillations.} 

{To get an intuition for how this nonlinear response can be built up in the diagrammatic representation, we summarize in Fig.\ \ref{fig:flow} the flowchart of its construction. Starting from the light-matter vertex, that in the case of the model in Eq.\ \ref{eq:intv} is a dipolar-like interaction between the electric field $E$ (wavy lines) and the two-level system (solid lines), one builds up Feynman diagrams with four external legs representing the perturbing fields at $\omega_i$ and the generated one at $\omega_t=\sum_i \omega_i$ (dashed wavy line). The loop accounts for the propagation of fermionic states at $\pm \Delta$ between two electric field pulses, whose explicit expression is the susceptibility of Eq.\ \ref{eq:chi32l}. Finally, we decorate the external legs with the detection frequency $\omega_t$, that always occurs on the dashed wavy line, and with the combination of frequencies assigned to the $E_0$ (blue) and $E_\tau$ (red) pulses, in the spirit of the definitions \ref{eq:j00tau} and \ref{eq:j0tautau} above.}

\begin{figure}
\includegraphics[width=\columnwidth]{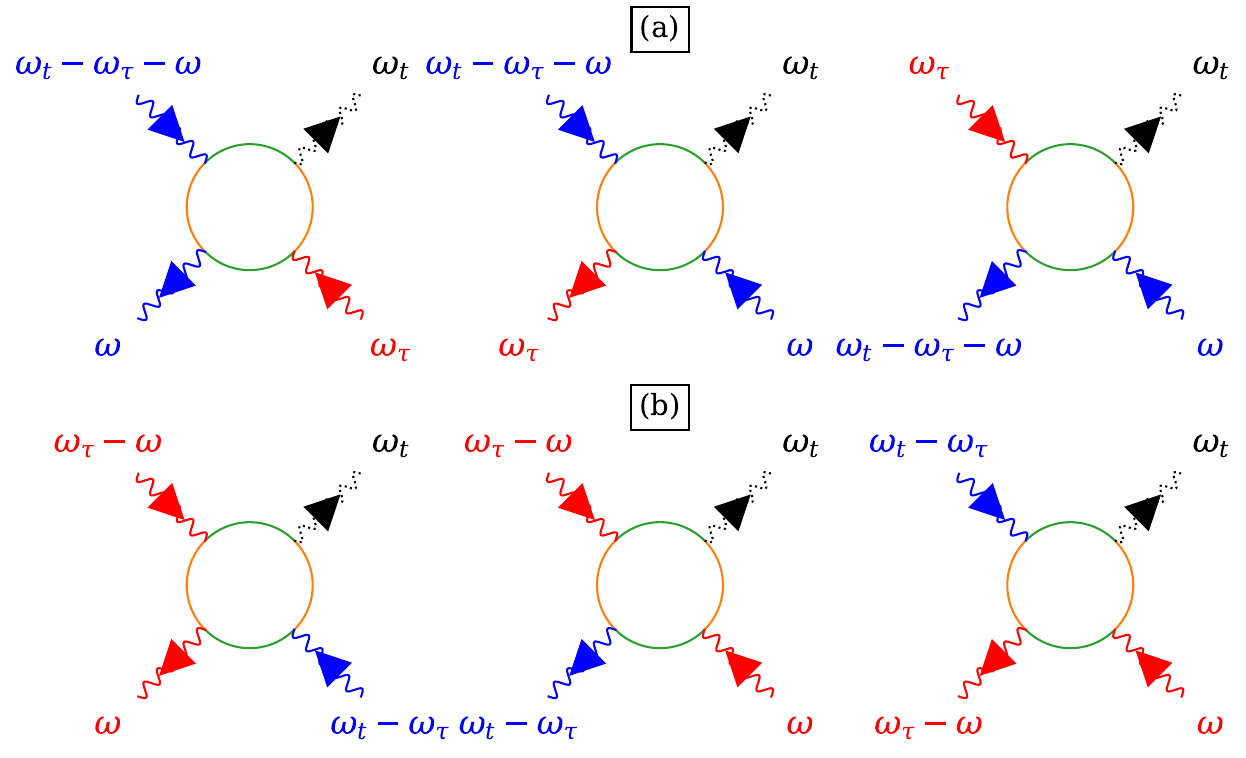}
\caption{\label{fig:assf} {Representation of all the possible ways of decorating the first term of the susceptibility in Eq.\ \ref{eq:chi32l} with combinations of $E_0$ (blue) and $E_\tau$ (red). Panel (a): diagrams relevant for $P^{(3)}_{00\tau}$ in Eq.\ \ref{eq:i00t}. Panel (b): those relevant for $P^{(3)}_{0\tau\tau}$ in Eq.\ \ref{eq:i0tt}.}}
\end{figure}

{A more complete representation of this process is provided in Fig.\ \ref{fig:assf}, where we accounted for all possible permutations of the external light legs. Notice that each of them leads to a different process, since excitations between the levels read out different frequency values. This representation helps us to highlight the absorption/emission of photons at each interaction since we use different colors, orange or green, to distinguish propagators of the eigenstates at energy $-\Delta$ or $+\Delta$, respectively.} Their alternation is a consequence of the off-diagonal form of $H_{\text{int}}$ in Eq.\ \ref{eq:intv}, while their ordering is forced by energy conservation if we set $\omega_t>0$ for detection. We stress that the direction of the arrow in the blue and red photon lines (incoming or outgoing) denotes positive/negative frequencies, whose value in this impulsive limit is fixed by the difference ($+2\Delta$ or $-2\Delta$) between the energies of the propagators connecting each $H_\text{int}$ insertion. Moreover, since in the impulsive limit $E_{0,\tau}(\omega)=\bar{E}_{0,\tau}$  are constant, the $\omega-$integral in Eqs.\ \ref{eq:j00tau} and \ref{eq:j0tautau} with the form of Eq.\ \ref{eq:chi32l} can be performed by residues. We report, for instance, the contribution stemming from diagram $\text{(i)}$, obtained by the assignment $\omega_{123}\ra \omega_t$,  $\omega_1\ra\omega_\tau$, $\omega_2\ra \omega$, $\omega_3\ra \omega_t-\omega_\tau-\omega $:
\begin{align}
\label{eq:diag1}
{\text{(i)}}&\sim\int d\omega\,\frac{1}{\omega_t+i\eta-2\Delta}\frac{1}{\omega+i\eta+2\Delta}\nonumber\\
&\times\left(\frac{1}{\omega_\tau+i\eta-2\Delta}+\frac{1}{\omega_t-\omega_\tau-\omega+i\eta-2\Delta}\right)\nonumber\\
&\sim\frac{1}{\omega_t+i\eta-2\Delta}\left(\frac{2}{\omega_t-\omega_\tau+i\eta}+\frac{1}{\omega_\tau+i\eta-2\Delta}\right).
\end{align}
Summing all the pieces together, the resulting 2D map will be given by 
\begin{widetext}
\begin{align}
P^{(3)}_{00\tau}(\omega_t,\omega_\tau)&\sim\left(\frac{1}{\omega_t+i\eta-2\Delta}-\frac{1}{\omega_t+i\eta+2\Delta}\right)\left[\frac{2}{\omega_t-\omega_\tau+i\eta}+\frac{1}{\omega_\tau+i\eta-2\Delta}+\frac{1}{\omega_\tau+i\eta+2\Delta}\right]\bar{E}_0^2\bar{E}_{\tau},\label{eq:i00t}\\
P^{(3)}_{0\tau\tau}(\omega_t,\omega_\tau)&\sim\left(\frac{1}{\omega_t+i\eta-2\Delta}-\frac{1}{\omega_t+i\eta+2\Delta}\right)\left[\frac{2}{\omega_\tau+i\eta}+\frac{1}{\omega_t-\omega_\tau+i\eta-2\Delta}+\frac{1}{\omega_t-\omega_\tau+i\eta+2\Delta}\right]\bar{E}_0\bar{E}_{\tau}^2.\label{eq:i0tt}
\end{align}
\end{widetext}

\begin{figure*}
\includegraphics[width=\textwidth]{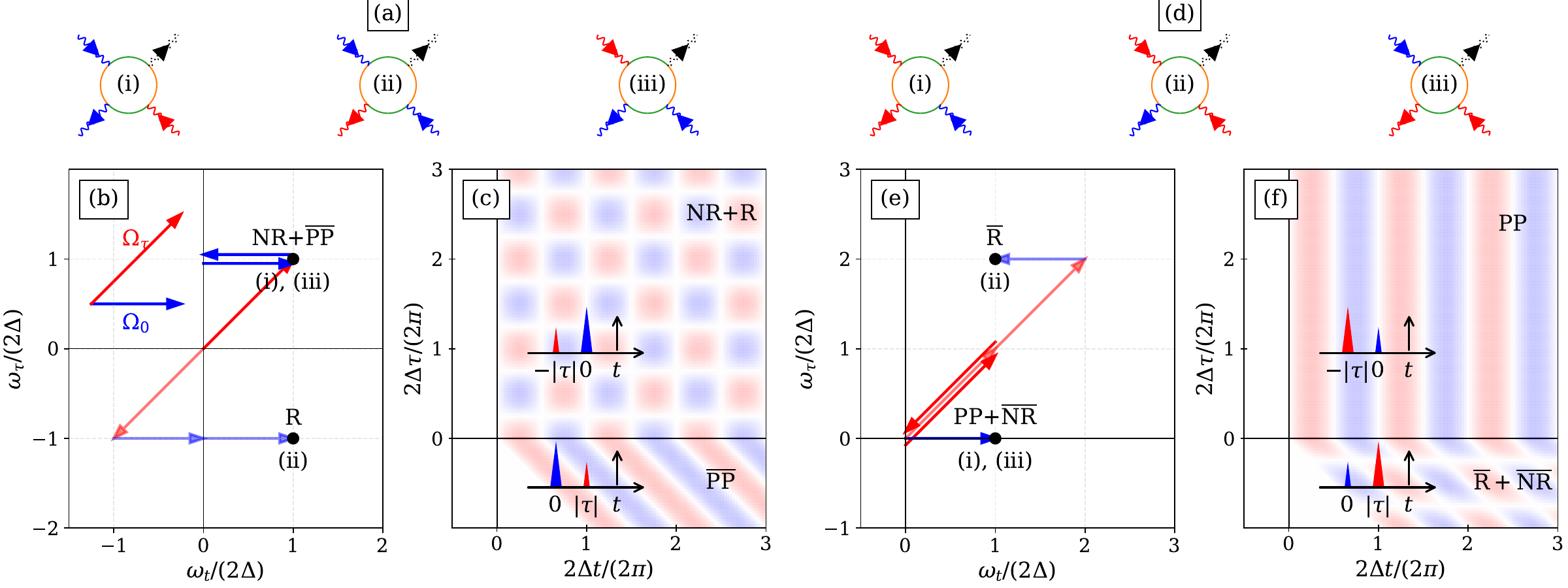}
\caption{\label{fig:imp} Schematic outcome for the 2D experiment performed on a two-level system by using impulsive fields $E(t)\sim\delta(t)$, where in panels (a)-(c) we analyzed the contributions of $P_{00\tau}$ in Eq.\ \ref{eq:i00t} and in panels (d)-(f) those of $P_{0\tau\tau}$, see Eq.\ \ref{eq:i0tt}. Panels (a) and (d): diagrammatic representation of the nonlinear processes obtained by distributing photons coming from $E_0$ (blue) and $E_\tau$ (red) in the arguments of $\chi^{(3)}$, which alternates propagators at energy $-\Delta$ (orange) and $+\Delta$ (green), following Fig.\ \ref{fig:assf}(a)-(b); each process is labelled as (i), (ii) and (iii). Panels (b) and (e): corresponding frequency-vector scheme, where we indicated how diagrams on the upper row map in frequency space. The names of the peaks are given according to the order of the pulses; for $\tau>0$ ($E_\tau$ first) we have $\text{R}$, $\text{PP}$ and $\text{NR}$, also used in Table \ref{tab:names}, for $\tau<0$ ($E_0$ first) we have $\overline{\text{R}}$, $\overline{\text{PP}}$ and $\overline{\text{NR}}$. Panel (c) and (f): time-domain simulation of the 2D map, which shows the causality conditions embedded in the various processes, with indications of the corresponding contributions coming from the peaks in frequency space.}
\end{figure*}

When $\eta\to 0$ the various fractions produce four peaks which for $\omega_t>0$ are located at the intersections between the line $\omega_t=2\Delta$, which accounts for the prefactor in round brackets, and the lines maximizing each addend in square brackets. For $P^{(3)}_{00\tau}$ in Eq.\ \ref{eq:i00t} this occurs at $\omega_\tau=\pm2\Delta$ and $\omega_t-\omega_\tau=0$, while for $P^{(3)}_{0\tau\tau}$ in Eq.\ \ref{eq:i0tt} this leads to $\omega_\tau=0$ and $\omega_t-\omega_\tau=\pm2\Delta$. The four resulting spots are located at coordinates $(2\Delta,-2\Delta)$, $(2\Delta,0)$, $(2\Delta,2\Delta)$ and $(2\Delta,4\Delta)$. These resonances can be attributed to the various combinations of fixed pulses $E_0$ (blue) and delayed pulses $E_\tau$ (red), as explicitly shown in Fig.\ \ref{fig:imp}(b) and (e) in the frequency vector scheme. Indeed, in analogy with Fig.\ \ref{fig:mono}(a), the 2D map can still be generated by a combination of frequency vectors, but, in the impulsive limit applied to the two-level system, the \emph{length} of the arrow is dictated  by the $2\Delta$ energy, setting the scale of allowed interband transitions. In other words, in this case $\overrightarrow{\Omega}_0=(2\Delta,0)$ and $\overrightarrow{\Omega}_\tau=(2\Delta,2\Delta)$: by locating the spots on the 2D map, one immediately reads out the resonance frequency of the system under investigation. As a further remark, we stress that although each diagram in Fig.\ \ref{fig:imp}(a) and (d) produces a different spot in Fig.\ \ref{fig:imp}(b) and (e), the inverse is not true, as evident for $\text{(i)}$ and $\text{(iii)}$ that both appear in $(2\Delta,2\Delta)$ and $(2\Delta,0)$. This is due to the fact that the integration over $\omega$ exemplified in Eq.\ \ref{eq:diag1} leads to the same frequency content for $\text{(i)}$ and $\text{(iii)}$. 

In this regard, it is also instructive to compute the nonlinear signal in time-domain by going back from $(\omega_t,\omega_\tau)$ to $(t,\tau)$. Due to the analytic continuation in Eq.\ \ref{eq:i00t} and Eq.\ \ref{eq:i0tt}, the position of the poles in the upper or lower half-plane for complex $\omega_t,\omega_\tau$ is extremely important to preserve the causality structure reflecting the ordering of the electric fields in time. Close inspection shows that the Fourier transform of terms containing $\omega_t-\omega_\tau$ leads to $\theta(t+\tau)\theta(-\tau)$, which produces signal in the region where $E_0$ comes before $E_\tau$, while from terms containing only $\omega_\tau$ we get $\theta(t)\theta(\tau)$, which is nonzero when $E_\tau$ comes before $E_0$. We stress that this information, although already contained in the frequency structure of the diagrams in Fig.\ \ref{fig:imp}(a) and (d), \textit{cannot} be obtained at once by looking at the order of the field lines in the drawing. As an example, looking again at Eq.\ \ref{eq:diag1}, we see that in this process both denominators $\omega_t-\omega_\tau$ and $\omega_\tau$ are present, so that the process contributes to both time ordering $\tau>0$ and $\tau<0$. For this reason, to label the spots we follow the notation of Ref.\ \cite{mahmood_Nat.Phys.21}, where both cases are covered. 

Let us first compute the signal in time by Fourier transforms of Eqs.\ \ref{eq:i00t} and \ref{eq:i0tt}, which are shown in Fig. \ref{fig:imp}(c) and (d), respectively. For two interactions with $E_0$ and one with $E_\tau$ we have:
\begin{align}
\label{eq:00tau}
P^{(3)}_{00\tau}(t,\tau)&\sim\bar{E}_0^2\bar{E}_{\tau}\left[\theta(t)\theta(\tau)\sin(2\Delta t)\cos(2\Delta \tau)\right.\nonumber\\
&\left.+\theta(t+\tau)\theta(-\tau)\sin(2\Delta(t+\tau))\right],
\end{align}
reported in Fig.\ \ref{fig:imp}(c). Two interactions with $E_\tau$ and one with $E_0$ produce instead
\begin{align}
\label{eq:0tautau}
P^{(3)}_{0\tau\tau}(t,\tau)&\sim\bar{E}_0\bar{E}_{\tau}^2\left[\theta(t)\theta(\tau)\sin(2\Delta t)\right.\nonumber\\
&\left.+\theta(t+\tau)\theta(-\tau)\sin(2\Delta(t+\tau))\cos(2\Delta \tau)\right],
\end{align}
presented in Fig.\ \ref{fig:imp}(f). As stated before, the notation of Table \ref{tab:names} refers to the processes obtained for $\tau>0$, i.e. the first lines of Eq.\ \ref{eq:00tau} and \ref{eq:0tautau}, and lead to the NR, R and PP spots of Figs.\ \ref{fig:imp}(b) and (e). When $\tau<0$ we are exchanging the role of the ``pump'' and ``probe'' fields with respect to the usual sense of pump-probe experiments. Then one sees that when the field amplitudes are exchanged, $\bar{E}_0\leftrightarrow\bar{E}_{\tau}$ the 2D map in the time domain for $\tau<0$ is connected to the one in $\tau>0$ by the relation $P_{2D}^{(3)}(t,-\lvert\tau\rvert)=P_{2D}^{(3)}(t-\lvert\tau\rvert,\lvert\tau\rvert)$. This {is equivalent to mapping} the upper (lower) half-plane of Fig.\ \ref{fig:imp}(c) into the lower (upper) half-plane of Fig.\ \ref{fig:imp}(f), which is the time-domain equivalent of the transformation $\omega_t-\omega_\tau\leftrightarrow\omega_\tau$ in frequency plane mentioned after Eqs.\ \ref{eq:j00tau} and \ref{eq:j0tautau}. Such a symmetry justifies the classification of the frequency spots originating from $\tau<0$ as $\overline{\text{PP}}$, $\overline{\text{NR}}$ and $\overline{\text{R}}$ in Figs.\ \ref{fig:imp}(b) and (e). As a last remark, we notice that in the two-level system the two-quantum process at $\tau>0$ never occurs, since we have no terms with structure $\sim\theta(t)\theta(\tau) \cos(4\Delta \tau)$ in time domain, corresponding to a frequency structure $\sim(\omega_\tau-4\Delta+i\eta)^{-1}(\omega_t-2\Delta+i\eta)^{-1}$. To get such a resonance an excitation at energy $4\Delta$ is needed, which is absent in the two-level system.
{We finally observe that Eqs.\ \ref{eq:00tau} and \ref{eq:0tautau} are identical to the ones obtained in Ref.\ \cite{mahmood_Nat.Phys.21}, by computing the nonlinear polarization via an explicit solution of the density-matrix evolution in time domain. This equivalence is expected, since for a simple non-interacting and non-dispersive system as the one of Eq.\ \ref{eq:ham} an analytical solution can be easily derived both in time and frequency space. As a consequence, the two theoretical approaches have the same predictive power as compared to experiments. The real issue is how to extend the calculations to more complex electronic systems, that is the purpose of the rest of the paper.}

\section{\label{sec:mod}2DTS of dispersive electronic systems: paramagnetic vs.\ diamagnetic processes}

As we have shown in the previous Section, in the impulsive limit the 2D map is dominated by the frequency content of the nonlinear optical kernel, that in turn is dominated for the two-level system by the energy difference $2\Delta$ between the two nondispersive eigenstates. 
We want now to generalize this notion to the case of a many-body electronic system. {In particular, we will consider a toy model for a semiconducting-like band structure hosting an optically-active gap}, in order to establish analogies and differences with the standard two-level model \cite{mukamel_95,hamm_11,kuehn_J.Phys.Chem.B11,cundiff_PhysicsToday13}. As we shall see, two main differences must be considered: (i) in the usual representation of the gauge potential, the light-matter interaction admits multi-photon vertices in addition to the single-photon one, dipole-like interaction of Eq.\ \ref{eq:intv}, the first one being a diamagnetic-like two-photon vertex present already in the continuum limit; (ii) the energy dispersion of valence and conduction states leads to a modification of the 2D features found for the two-level case even for the single-photon processes. 

To make a concrete example we will compute the $K^{(3)}$ third-order response to the gauge field $\bA$ by focusing on {a two-dimensional system displaying a band gap at $2\Delta$. The fermionic spectrum is equivalent to the one obtained at mean-field level for a perfectly-nested charge-density-wave CDW phase with nesting vector $\mathbf{Q}$ on a two-dimensional square lattice \cite{cea_Phys.Rev.B14}.} By introducing the spinor
\be
\Psi_{\mathbf{k}}^{\dagger}=
\begin{pmatrix}
c_{\mathbf{k}}^{\dagger}\,{,}&c_{\mathbf{k}+\mathbf{Q}}^{\dagger}\\
\end{pmatrix}
\ee
the system is described by the spinless Hamiltonian
\begin{align}
\label{eq:hamcdw}
H&=\sum_{\mathbf{k}\in\text{RBZ}}
\Psi_{\mathbf{k}}^{\dagger}
\begin{pmatrix}
\epsilon_{\mathbf{k}}&\Delta\\
\Delta&\epsilon_{\mathbf{k}+\mathbf{Q}}\\
\end{pmatrix}		
\Psi_{\mathbf{k}},
\end{align}
with a nearest-neighbors hopping $\bar{t}$ on a square lattice, such that the band dispersion is 
\be
\label{eps}
\epsilon_\mathbf{k}=-2\bar{t}\cos(k_xa)-2\bar{t}\cos(k_ya)
\ee
where $a$ is the distance between atomic sites, from now on set to unity. Since for $\mathbf{Q}=(\pi,\pi)$ the band structure is perfectly nested, $\epsilon_{\mathbf{k}+\mathbf{Q}}=-\epsilon_{\mathbf{k}}$, the $\mathbf{k}$ summation in Eq.\ \ref{eq:hamcdw} is restricted to the  reduced Brillouin zone (RBZ). 
{As mentioned above, in the CDW case $\Delta$ can be thought as a mean-field order parameter connected to the macroscopic average $\langle c^\dagger_\bk c_{\bk+\bQ}\rangle$. The model in Eq.\ \ref{eq:hamcdw} is however used here to provide us with a pedagogical example, useful to highlight the role of dispersive excitations and the interplay between the frequency structure of nonlinear kernel and that of the pulse. To this aim we will also introduce a phenomenological and constant damping of the electronic excitations, and we will neglect the additional contribution due to collective modes, present in a realistic CDW case \cite{cea_Phys.Rev.B14}. As we shall see, even within these approximations the computation of the 2D maps for arbitrary field spectrum can be cumbersome, and we will discuss possible semi-analytical approximations for it. In addition, we will vary the ratio $\bar t/\Delta$ from low ($\bar t/\Delta=0.1$ to large ($\bar t/\Delta=2$) values in order to highlight the role of dispersive electronic excitations as compared to the two-level systems.}

The fermionic Green's function in the Matsubara formalism for the Nambu spinor $\Psi_\mathbf{k}$ shows poles at $\pm E_{\mathbf{k}}=\pm\sqrt{\epsilon_\mathbf{k}^2+\Delta^2}$ and reads 
\begin{align}
\label{eq:g0}
G_0(i\omega_n,\mathbf{k})&\equiv
\begin{pmatrix}
i\omega_n-\epsilon_\mathbf{k}&-\Delta\\
-\Delta&i\omega_n+\epsilon_\mathbf{k}
\end{pmatrix}^{-1} \nonumber \\
&\equiv
\mathcal{U}_{\mathbf{k}}^{\dagger}\tilde{G}_0(i\omega_n,\mathbf{k})\mathcal{U}_{\mathbf{k}},
\end{align}
where $i\omega_n$ are fermionic Matsubara frequencies. Here we introduced the usual Bogoliubov rotation matrices $\mathcal{U}_{\mathbf{k}}=u_{\mathbf{k}}\sigma_0+i\sigma_2 v_{\mathbf{k}}$, with $u_{\mathbf{k}}^2-v_{\mathbf{k}}^2=\epsilon_{\mathbf{k}}/E_{\mathbf{k}}$ and $2u_{\mathbf{k}}v_{\mathbf{k}}=\Delta/E_{\mathbf{k}}$, which bring the Green's function $G_0$ in its diagonal form $\tilde{G}_0$:
\be
\tilde G_0(i\omega_n,\mathbf{k})=
\begin{pmatrix}
\frac{1}{i\omega_n-E_\bk} & 0\\
0&\frac{1}{i\omega_n+E_\bk}
\end{pmatrix}.
\ee
To compute the nonlinear response of the system we will follow the standard procedure of coupling the electrons to the gauge potential via the Peierls substitution, namely $\epsilon_{\mathbf{k}}\to\epsilon_{\mathbf{k}-e/c\mathbf{A}(t)}$. Such an approach, which extends to the lattice case the usual minimal coupling, leads to a set of coupling terms between the gauge field and electrons obtained by expansion in a power series of $\mathbf{A}$. 
It should be noted that the choice of using the gauge potential $\mathbf{A}(t)$ instead of the electric field $\mathbf{E}(t)$ corresponds to using the velocity instead of the length (or dipole) gauge in the standard nomenclature \cite{aversa_Phys.Rev.B95, ventura_Phys.Rev.B17}. For dispersive electronic systems, this choice from one side avoids the computation of the dipole matrix elements, involving the position operator $\mathbf{r}$, which is cumbersome for electronic systems \cite{sipe_Phys.Rev.B93, mikhailov_Phys.Rev.B16}. On the other hand, it leads to a simpler frequency structure of the response function, which is rewritten in terms of simple poles. This is particularly relevant for numerical simulations with realistic pulses, since we can obtain a semi-analytical expression by approximating the gauge field with a gaussian spectrum, $A(t)\propto e^{-t^2/(2\tau_p^2)}\cos{(\Omega t)}$. The price to pay--as already happens in the linear response--is an increase in the number of diagrams needed to describe the optical properties of the material \cite{passos_Phys.Rev.B18, parker_Phys.Rev.B19}, as we shall discuss here. 

\begin{figure}
\centering
\includegraphics[width=\columnwidth]{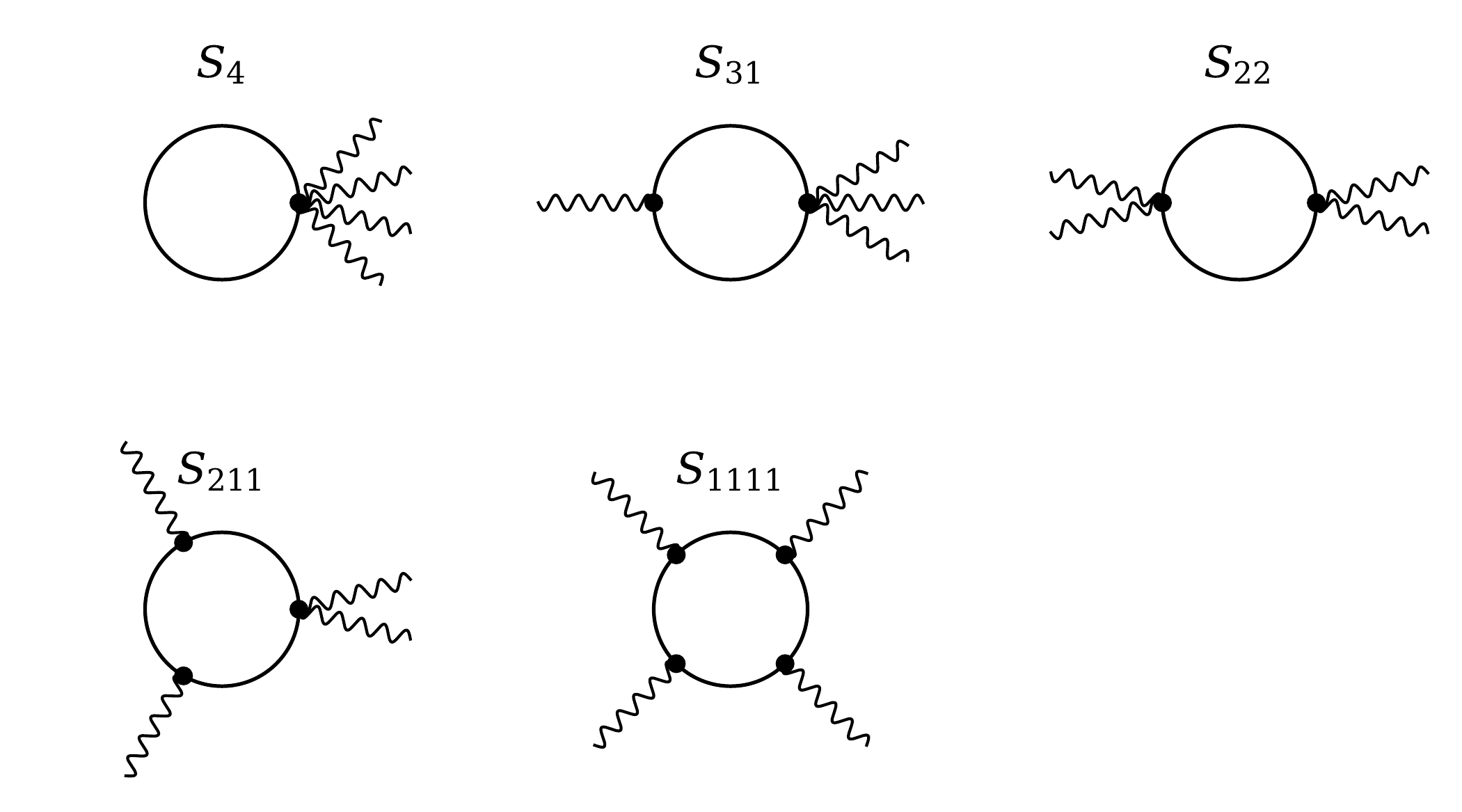}
\caption{Diagrams contributing to $S[A^4]$ in Eq.\ \ref{eq:s4}, obtained by combinations of the multi-photon vertices in Eq.\ \ref{eq:sesemi}. The wavy lines denote the gauge field, the bullets denote the $\sigma_3$ vertex of Eq.\ \ref{eq:sasigma}, and black lines denote the electronic Green's function in Nambu space, see Eq.\ \ref{eq:g0}.}
\label{fig:diasemi}
\end{figure}

Since the effective electronic model in Eq.\ \ref{eq:hamcdw} is quadratic, once the gauge field has been introduced by Peierls substitution the fermionic degrees of freedom can be integrated out exactly. This procedure can be efficiently carried out in a path-integral approach \cite{nagaosa_99, cea_Phys.Rev.B16}, where the quantum action  $S[\mathbf{A}]$ for electromagnetic field can be expressed by means of a perturbative series: 
\be
S[\mathbf{A}]=\sum_n\frac{1}{n}\Tr{\left(G_0\Sigma_A\right)^n},
\ee
where the trace here means summing over Nambu, energy and momenta degrees of freedom. Here $\Sigma_A$ acts as a fermionic self-energy due to the coupling with the gauge field, that upon expansion in power series of $\mathbf{A}$ provides all the possible coupling terms between $\bf A$ and the electrons:
\be
\label{eq:sasigma}
\Sigma_A(i\Omega,\mathbf{k})=\sum_{n>0}\Sigma_{A}^{(n)}(i\Omega,\mathbf{k})\sigma_3, 
\ee
which for the {semiconducting-like model} under consideration has a $\sigma_3$ structure in Nambu space. A whole set of multi-photon vertices appears
\be
\label{eq:sesemi}
\Sigma^{(n)}_A(i\Omega,\mathbf{k})=\frac{(-e)^n}{n!c^n}\sum_{\alpha_1,\dots,\alpha_n}\frac{\partial^n\epsilon_{\mathbf{k}}}{\partial k_{\alpha_1}\dots\partial k_{\alpha_n}}A^{n}_{\alpha_1\dots\alpha_n}(i\Omega),
\ee
where $A^n_{\alpha_1\dots\alpha_n}(i\Omega)$, with $i\Omega$ bosonic Matsubara frequency, denotes the Fourier transform of $A_{\alpha_1}(t)\cdots A_{\alpha_n}(t)$ in imaginary time, with $\alpha_i=x,y$ cartesian components of the gauge field.
The first two terms $n=1$ and $n=2$ represent the lattice equivalent of the paramagnetic and diamagnetic terms already present in continuum models \cite{cea_Phys.Rev.B16}, so that they describe single-photon and two-photon {vertices} in the diagrammatic representation, respectively. 
Since the current is obtained by a functional derivative with respect to the gauge field:
\be
J_\alpha(t)=\frac{\delta S[A]}{\delta A_\alpha(t)},
\ee
we need to consider nonlinear processes up to the fourth order in the effective action for $A$ to compute the generated current. {The derivation of the nonlinear response in a Feynman diagram representation follows then the same logic outlined above while discussing the two-level system. However, in addition to the single-photon vertex already considered in Fig.\ \ref{fig:flow}, one must consider now multi-photon vertices, corresponding to the case where multiple light pulse at the same time drive a particle-hole excitation. As a consequence, fourth-order terms in the field $A$ do not necessarily correspond to four-point susceptibilities, as it was the case for the two-level system shown in Fig.\ \ref{fig:flow}.} Since in a lattice model vertices with arbitrary number of $A$ lines are allowed, see Eq.\ \ref{eq:sesemi}, we  have overall five contribution to the fourth-order effective action, represented by the Feynman diagrams of Fig.\ \ref{fig:diasemi}:
\be
\label{eq:s4}
S[A^4]=S_4+S_{22}+S_{13}+S_{112}+S_{1111}.
\ee
{By close inspection of Fig.\ \ref{fig:diasemi} one then sees that $S_4$ is a constant, independent on the incoming frequencies of the $A(\omega_i)$ fields. $S_{22}$, and $S_{31}$ are two-point correlation functions, which only depend on the sum of the frequencies of the external field lines, as it is the case for standard linear-response functions. Finally, $S_{211}$ and $S_{1111}$ are three-point and four-point correlation functions, respectively, with a separate dependence on the frequencies of the external legs. These different structures will be responsible for specific signatures in the 2DCS maps \cite{fiore_}.}

The overall derivation of the nonlinear current can be carried out in a similar way also for other types of ordered phases treated at mean-field level, as e.g.\ the magnetic phases of Rice-Mele-Hubbard models \cite{ono_Phys.Rev.Lett.25}, or excitonic insulators \cite{tanabe_Phys.Rev.B21}, as well as superconductors \cite{cea_Phys.Rev.B16,murotani_Phys.Rev.B19,tsuji_Phys.Rev.Research20,haenel_Phys.Rev.B21}. They are formally characterized by the same effective electronic gapped spectrum $\pm E_{\mathbf{k}}=\pm\sqrt{\epsilon_\mathbf{k}^2+\Delta^2}$, but they generally produce a different structure in Nambu space for the gauge-field coupling of Eq.\ \ref{eq:sasigma}. Once these differences are considered, the strategy for computing the $K^{(3)}$ kernel remains the same.

To find the explicit expression of the diagrams in Fig.\ \ref{fig:diasemi} we will need to compute $n$-point correlation functions of the form
\begin{widetext}
\begin{align}
X^{(n)}(i\Omega_1,\dots,i\Omega_n,\mathbf{k})&=\sum_{i\omega_m}\Tr{\left[G_0(i\omega_m,\mathbf{k})\sigma_3G_0(i\omega_m+i\Omega_1,\mathbf{k})\sigma_3\dots G_0(i\omega_m+i\Omega_{1\dots n},\mathbf{k})\sigma_3\right]} \nonumber \\
&\equiv\sum_{i\omega_m}\Tr{\left[\tilde{G}_0(i\omega_m,\mathbf{k})\Lambda_{\mathbf{k}\mathbf{k}}^{(3)}\dots \tilde{G}_0(i\omega_m+i\Omega_{1\dots n},\mathbf{k})\Lambda_{\mathbf{k}\mathbf{k}}^{(3)}\right]},
\end{align}
\end{widetext}
where we denote $i\Omega_{1\dots n}=i\Omega_1+\dots+i\Omega_n$ and we also introduce the matrix $\Lambda_{\mathbf{k}\mathbf{k}}^{(3)}=\mathcal{U}_{\mathbf{k}}\sigma_3\mathcal{U}_{\mathbf{k}}^{\dagger}$ representing the light-matter coupling in the band basis. Due to the structure in Nambu space, we have $\Lambda_{\mathbf{k}\mathbf{k}}^{(3)}=\epsilon_{\mathbf{k}}/E_{\mathbf{k}}\sigma_3-\Delta/E_{\mathbf{k}}\sigma_1$. As we shall see in the following, the
$\sigma_1$ term is crucial to allow for optically active one-photon transitions between the occupied $-E_{\mathbf{k}}$ and the empty $E_{\mathbf{k}}$ band. 
To see this explicitly, let us examine the form of $X^{(2)}(i\Omega_{ij},\mathbf{k})$:
	\begin{widetext}
	\be
	\label{eq:kbm}
		X^{(2)}(i\Omega_{ij},\mathbf{k})=\sum_{i\omega_n}\Tr\left[\begin{pmatrix}
			\frac{1}{i\omega_n-E_\mathbf{k}} & 0 \\
			0 &\frac{1}{i\omega_n+E_\mathbf{k}} \\
		\end{pmatrix}\Lambda_{\mathbf{k}\mathbf{k}}^{(3)}\begin{pmatrix}
			\frac{1}{i\omega_n+i\Omega_{ij}-E_\mathbf{k}} & 0 \\
			0 &\frac{1}{i\omega_n+i\Omega_{ij}+E_\mathbf{k}} \\
		\end{pmatrix}\Lambda_{\mathbf{k}\mathbf{k}}^{(3)}\right]
	\ee
	\end{widetext}
As stated above, in Nambu space the matrix $\Lambda_{\mathbf{k}\mathbf{k}}^{(3)}$ is a linear combination of $\sigma_3$ (diagonal) or $\sigma_1$ (off-diagonal). Since the matrix $\tilde{G}_0$ is diagonal, the only contributions that survive in the trace are obtained from the product of the two terms containing $\sigma_3$ or the two terms containing $\sigma_1$. The former contribution describes an {\em intraband} process involving scattering of electrons and holes belonging to the same bands,  while the latter describes an {\em interband} process involving a scattering between the two copies of the {semiconducting-like} bands. By evaluating  the Matsubara sum in Eq.\ \ref{eq:kbm}, one sees that only the interband part survives, as expected for transitions at long wavelength and finite frequency. 

In the following {we will analyze first the contributions to the nonlinear kernel stemming from the two terms in the effective action $S_{22}$, (fully-diamagnetic), and $S_{1111}$, (fully-paramagnetic), see Figs.\ \ref{fig:diasemi} and \ref{fig:diasemi}, respectively, and we will then show results obtained by including all the contributions in Fig.\ \ref{fig:semitot}. The interest in discussing the different signatures in the 2D map of these two contributions is motivated by the fact that in other models the only relevant nonlinearities can be ascribed to a fully-diamagnetic or a fully-paramagnetic response. For example, in superconductors only diamagnetic-like contributions are present in the clean limit \cite{cea_Phys.Rev.B16}, while the fully paramagnetic response dominates the nonlinear current at strong disorder. In addition, a diamagnetic-like response appears} whenever the interaction between light and matter can be mapped in an effective process where two photons excite simultaneously two collective modes. This is typically the case relevant for the nonlinear driving of infrared-active optical phonons \cite{udina_Phys.Rev.B19,lopez_24,basini_Phys.Rev.B24}, but also for the soft Josephson plasmons, discussed in Sec.\ \ref{sec:prop} below. {For what concerns the fully-paramagnetic term, it is interesting to compare its outcome with the response of the two-level Hamiltonian, }where only single-photon dipole-like transitions are allowed. It is thus possible to make a direct comparison between the many-body situation where a continuum, dispersive set of excited states is present and the discrete, nondispersive case of the two-level system. {In addition, previous work by several groups has shown that in disordered superconductors the nonlinear response is dominated by paramagnetic-like terms \cite{cea_Phys.Rev.B16,murotani_Phys.Rev.B19,silaev_Phys.Rev.B19,tsuji_Phys.Rev.Research20,haenel_Phys.Rev.B21,seibold_Phys.Rev.B21,udina_FaradayDiscuss.22}. Even though in this case a realistic computation should account for the effect of disorder on the energy eigenstates, the present results provide one with a hint on the expected 2D maps as a function of the pulse spectral width. Finally, we stress that, for what concerns the toy model in Eq.\ \ref{eq:hamcdw},} in order to formally ensure the invariance between calculations in the two gauges, length $E(t)$ and velocity $A(t)$, one has to \emph{sum} all the diagrams stemming from Eq.\ \ref{eq:s4}. For this reason, we provide in the SM \cite{fiore_} the calculation of all these terms, and we verify that gauge-invariance relations are satisfied \cite{paul_J.Phys.:Condens.Matter24}.  {These results are used to compute each term separately, also shown in the SM at \cite{fiore_}, and to obtain the full 2D map, whose differences with respect to the subset of fully-paramagnetic and fully-diamagnetic term is discussed below.}

\subsection{Fully-paramagnetic contribution}
	
	Let us start by analyzing the fully-paramagnetic term, which is derived from $S_{1111}$ in Fig.\ \ref{fig:diasemi} as
	\begin{align}
	\label{eq:jpara}
		J_{\alpha}^{(3),\text{para}}(i\Omega)&=\frac{\delta S_{1111}[A^4]}{\delta A_{\alpha}(-i\Omega)} \nonumber \\
		&\equiv\sum_{i\Omega_i}K_{\alpha\beta\gamma\delta}^{(3),\text{para}}(i\Omega_1,i\Omega_2,i\Omega_3) \nonumber \\
		&\times A_\beta(i\Omega_1)A_\gamma(i\Omega_2)A_\delta(i\Omega_3)\delta(i\Omega-i\Omega_{123}),
	\end{align}
	where
	\begin{align}
	K_{\alpha\beta\gamma\delta}^{(3),\text{para}}(i\Omega_1,i\Omega_2,i\Omega_3)&=\sum_{\mathbf{k},i\omega_n}\frac{\partial\epsilon_\mathbf{k}}{\partial k_{\alpha}}\frac{\partial\epsilon_\mathbf{k}}{\partial k_{\beta}}\frac{\partial\epsilon_\mathbf{k}}{\partial k_{\gamma}}\frac{\partial\epsilon_\mathbf{k}}{\partial k_{\delta}}\nonumber\\
	&\times X^{(4)}(i\Omega_1,i\Omega_2,i\Omega_3,\mathbf{k}).
	\end{align}
	In this case, one realizes that the trace will give nonzero results whenever one selects in $\Lambda_{\mathbf{k}\mathbf{k}}^{(3)}$ four $\sigma_3$, two $\sigma_3$ and two $\sigma_1$, or four $\sigma_1$ contributions. The first comes from fully intraband scattering processes and it vanishes after the Matsubara summation, as happened for $X^{(2)}$. The second one admits a mixture of interband and intraband scattering processes, with the various terms differing in the ordering of the scattering events. The third is equivalent to purely interband transitions between the CDW bands, mediated by four $\sigma_1$ vertices. After lengthy algebra, one can recollect \cite{fiore_} the various terms in a fully symmetric fashion under the exchange of $i\Omega_i$,
	\begin{align}
	X^{(4)}(i\Omega_1,i\Omega_2,i\Omega_3,\mathbf{k})&=\tanh{(\beta E_{\mathbf{k}}/2)}\nonumber \\
	&\times\left[\frac{\Delta^4}{E_{\mathbf{k}}^4}K^{\text{inter}}(E_\mathbf{k},i\Omega_1,i\Omega_2,i\Omega_3)\right. \nonumber \\
	&\left.+\frac{2\Delta^2\epsilon_{\mathbf{k}}^2}{E_{\mathbf{k}}^4}K^{\text{mi}}(E_\mathbf{k},i\Omega_1,i\Omega_2,i\Omega_3)\right].
	\end{align}
	Here we define the interband contribution
	\begin{align}
	\label{eq:k3inter}
		\bar{K}^{\text{inter}}(E_\mathbf{k},i\Omega_i)&=\left(\frac{1}{i\Omega_{123}-2E_{\mathbf{k}}}-\frac{1}{i\Omega_{123}+2E_{\mathbf{k}}}\right)\nonumber\\
        &\times\sum_{i\neq j}\left(\frac{1}{i\Omega_i-2E_{\mathbf{k}}}\frac{1}{i\Omega_j+2E_{\mathbf{k}}}\right.\nonumber\\
        &+\left.\frac{1}{i\Omega_i+2E_{\mathbf{k}}}\frac{1}{i\Omega_j-2E_{\mathbf{k}}}\right),
	\end{align}
	which has the same {frequency} structure of Eq.\ \ref{eq:chi32l}. We also have the mixed one  
	\begin{align}
	\label{eq:k3mix}
		\bar{K}^{\text{mi}}(E_\mathbf{k},i\Omega_i)&=\left(\frac{1}{i\Omega_{123}-2E_{\mathbf{k}}}-\frac{1}{i\Omega_{123}}\right)\sum_{i\neq j}\frac{1}{i\Omega_{ij}-2E_{\mathbf{k}}}\nonumber\\
		&\times\left(\frac{1}{i\Omega_i-2E_{\mathbf{k}}}+\frac{1}{i\Omega_{j}-2E_{\mathbf{k}}}\right)\nonumber\\
		&-(E_{\mathbf{k}}\to-E_{\mathbf{k}}).
	\end{align}
	
	The effects of the band dispersion and of the light polarization can be encoded in a tensorial density of states defined as:
	\be
    \label{eq:tenspara}
	\tilde{\rho}_{\alpha\beta\gamma\delta}^{\text{para}}(\epsilon)\equiv\sum_{\mathbf{k}}\delta(\epsilon-\epsilon_{\mathbf{k}})\frac{\partial\epsilon_\mathbf{k}}{\partial k_{\alpha}}\frac{\partial\epsilon_\mathbf{k}}{\partial k_{\beta}}\frac{\partial\epsilon_\mathbf{k}}{\partial k_{\gamma}}\frac{\partial\epsilon_\mathbf{k}}{\partial k_{\delta}}.
	\ee
	In this way, it is possible to rewrite the full paramagnetic term as an integral over the energy $E=\sqrt{\epsilon^2+\Delta^2}$, reading
    \begin{align}
    \label{eq:k3semi}
	K_{\alpha\beta\gamma\delta}^{(3),\text{para}}(i\Omega_i)\equiv\int dE\,\left[\right.&\rho_{\alpha\beta\gamma\delta}^{\text{inter}}(E)\bar{K}^{\text{inter}}(E,i\Omega_i)\nonumber\\+&\rho_{\alpha\beta\gamma\delta}^{\text{mi}}(E)\bar{K}^{\text{mi}}(E,i\Omega_i)\left.\right],
	\end{align}
    where we defined for later convenience 
    \begin{align}
    \rho_{\alpha\beta\gamma\delta}^{\text{inter}}(E)&=\frac{E}{\sqrt{E^2-\Delta^2}}\tanh{\left(\beta E/2\right)}\frac{\Delta^4}{E^4}\nonumber\\
    &\times\tilde{\rho}_{\alpha\beta\gamma\delta}^{\text{para}}\left(\sqrt{E^2-\Delta^2}\right),\label{rhointer}\\
    \rho_{\alpha\beta\gamma\delta}^{\text{mi}}(E)&=\frac{E}{\sqrt{E^2-\Delta^2}}\tanh{\left(\beta E/2\right)}\frac{2(E^2-\Delta^2)\Delta^2}{E^4}\nonumber\\
    &\times\tilde{\rho}_{\alpha\beta\gamma\delta}^{\text{para}}\left(\sqrt{E^2-\Delta^2}\right).\label{rhomix}
    \end{align}
    
	In the case of a square lattice and with the nearest-neighbor band dispersion in Eq.\ \ref{eps} one can obtain analytical expressions for the tensorial density of states Eq.\ \ref{eq:tenspara}, as shown in the SM \cite{fiore_}.

\subsection{Fully-diamagnetic contribution}
	The fully-diamagnetic contribution $S_{22}$ leads to the nonlinear current
	\begin{align}
	\label{eq:j3dia}
	J_{\alpha}^{(3),\text{dia}}(i\Omega)&=\frac{\delta S_{22}[A^4]}{\delta A_{\alpha}(-i\Omega)} \nonumber \\
	&\equiv\sum_{i\Omega^{\prime}}K^{(3),\text{dia}}_{\alpha\beta\gamma\delta}(i\Omega^{\prime})A_{\beta}(i\Omega-i\Omega^{\prime})A^2_{\gamma\delta}(i\Omega^{\prime}), 
	\end{align}
	where we defined as $A_{\gamma\delta}^2(i\Omega)$ the Fourier transform of $A_{\gamma}(\tau)A_{\delta}(\tau)$ in Matsubara frequencies, namely
	\be
	A_{\gamma\delta}^2(i\Omega)=\sum_{i\Omega^{\prime}}A_{\gamma}(i\Omega^{\prime})A_{\delta}(i\Omega-i\Omega^{\prime}).
	\ee
	By means of a change of variable, the  expression in Eq.\ \ref{eq:j3dia} can also be rewritten in the completely symmetric form
	\begin{align}
	J_{\alpha}^{(3),\text{dia}}(i\Omega)&=\sum_{i\Omega_i}K^{(3),\text{dia}}_{\alpha\beta\gamma\delta}(i\Omega_{ij})A_{\beta}(i\Omega_1)A_{\gamma}(i\Omega_2)A_{\delta}(i\Omega_3) \nonumber \\
	&\times\delta(i\Omega-i\Omega_{123}),
	\label{eq:jdia}
	\end{align}
	where $i\Omega_{ij}$ is the sum of any two incoming frequencies. Considering the structure of the $S_{22}$ diagram in Fig.\ \ref{fig:diasemi} one obtains
	\be
	K_{\alpha\beta\gamma\delta}^{(3),\text{dia}}(i\Omega_{ij})=\sum_{\mathbf{k}}\frac{\partial^2\epsilon_{\mathbf{k}}}{\partial k_{\alpha}\partial k_{\beta}}\frac{\partial^2\epsilon_{\mathbf{k}}}{\partial k_{\gamma}\partial k_{\delta}}X^{(2)}(i\Omega_{ij},\mathbf{k}),
	\ee
	where $X^{(2)}$ has the typical structure of a two-point correlation function:
	\begin{align}
	K_{\alpha\beta\gamma\delta}^{(3),\text{dia}}(i\Omega_{ij})&=\sum_{\mathbf{k}}\frac{\partial^2\epsilon_{\mathbf{k}}}{\partial k_{\alpha}\partial k_{\beta}}\frac{\partial^2\epsilon_{\mathbf{k}}}{\partial k_{\gamma}\partial k_{\delta}}\tanh{(\beta E_{\mathbf{k}}/2)}\nonumber\\
	&\times\frac{\Delta^2}{E_{\mathbf{k}}^2}\left(\frac{1}{i\Omega_{ij}-2E_{\mathbf{k}}}-\frac{1}{i\Omega_{ij}+2E_{\mathbf{k}}}\right).
	\end{align}
As before, the frequency-dependent part of the response in encoded in the structure:
	\be
    \label{eq:kdia}
	K^{\text{dia}}(E_\bk,i\Omega_{ij})=\frac{1}{i\Omega_{ij}-2E_\bk}-\frac{1}{i\Omega_{ij}+2E_\bk},
	\ee
while the effect of the electronic dispersion is accounted by a tensorial density of states, defined for the diamagnetic vertex as
	\be
    \label{eq:tensdia}
	\tilde{\rho}_{\alpha\beta\gamma\delta}^{\text{dia}}(\epsilon)\equiv\sum_{\mathbf{k}}\delta(\epsilon-\epsilon_{\mathbf{k}})\frac{\partial^2\epsilon_{\mathbf{k}}}{\partial k_{\alpha}\partial k_{\beta}}\frac{\partial^2\epsilon_{\mathbf{k}}}{\partial k_{\gamma}\partial k_{\delta}},
	\ee
	In full analogy with Eq.\ \ref{eq:k3semi}, one can put:
	\begin{align}
	\label{eq:k3diasemi}
	K_{\alpha\beta\gamma\delta}^{(3),\text{dia}}(i\Omega_{ij})=\int dE\,{\rho}_{\alpha\beta\gamma\delta}^{\text{dia}}(E) \bar{K}^{\text{dia}}\left(E,i\Omega_{ij}\right), 
	\end{align}
    where we also introduced the quantity
    \begin{align}
    \label{rhodia}
    {\rho}_{\alpha\beta\gamma\delta}^{\text{dia}}(E)&=\frac{E}{\sqrt{E^2-\Delta^2}} \tanh{\left(\beta E/2\right)}\frac{\Delta^2}{E^2}\nonumber\\
    &\times\tilde{\rho}_{\alpha\beta\gamma\delta}^{\text{dia}}\left(\sqrt{E^2-\Delta^2}\right).
    \end{align}

    Also in this case, for a square lattice and with the nearest-neighbor band dispersion in Eq.\ \ref{eps}, one can obtain analytical expressions for the tensorial density of states Eq.\ \ref{eq:tensdia}, which is also shown in the SM \cite{fiore_}.
	
\subsection{Analytical continuation and 2D map}

Once computed the third order kernel $K^{(3)}$ in Matsubara frequencies, we need  to perform the analytical continuation to real frequencies, $i\Omega_i\to\omega_i+i\eta$, with $\eta>0$ \cite{evans_Proc.Phys.Soc.66,rostami_AnnalsofPhysics21}, to obtain the 2D map. The nonlinear current can then be expressed as a function of $(\omega_t,\omega_\tau)$ by using the {prescription in Eqs.\ \ref{eq:j00tau} and \ref{eq:j0tautau}}. Assuming for simplicity $A_\tau(\omega)=A_0(\omega)=A(\omega)$, we can write a contribution to the 2D map for each diagram in the effective action in Eq.\ \ref{eq:s4}:
	\begin{align}
	\label{eq:j2drep}
	J_{00\tau}^{(3)}(\omega_t,\omega_\tau)=A(\omega_\tau)\int d\omega\,&K^{(3)}_{\mathcal{S}}(\omega,\omega_\tau,\omega_t-\omega_\tau-\omega)\nonumber\\
	\times &A(\omega)A(\omega_t-\omega_\tau-\omega),
	\end{align}
with an analogous expression for $J^{(3)}_{0\tau\tau}$, along the lines of Eq.\ \ref{eq:j0tautau}. 

In analogy to what we found before in the case of the two-level system, we expect that in the impulsive limit the 2D map will be determined by the frequency resonances of the kernel. In the non-interacting electronic model that we are considering, the frequency dependence of the kernel is described by a combination of simple poles at $\pm 2E$, and the possible $2E$ values are weighted by a tensorial density of states $\rho(E)$, see Eq.\ \ref{eq:k3semi} for the paramagnetic contribution and Eq.\ \ref{eq:k3diasemi} for the diamagnetic one. To get an analytical intuition on the structure of the 2D map, it is then convenient to recast the nonlinear signal by factorizing the energy integration from the signatures in the $(\omega_t,\omega_\tau)$ planes obtained for each value of $E$. In other words, the integral one has to compute in Eq.\ \ref{eq:j2drep} has the form 
	\be
	\label{eq:scheme}
	J^{(3)}_{00\tau}(\omega_t,\omega_\tau)\equiv\int dE\, \rho(E)\bar{J}_{00\tau}(E,\omega_t,\omega_\tau)
	\ee
where for each value of $E$ the quantity $\bar{J}_{00\tau}(E,\omega_t,\omega_\tau)$ is obtained as
	\begin{align}
    \label{eq:schemebis}
	\bar{J}_{00\tau}(E,\omega_t,\omega_\tau)=A(\omega_\tau)\int d\omega&\bar{K}_{\mathcal{S}}(E,\omega,\omega_\tau,\omega_t-\omega_\tau-\omega)\nonumber\\
	\times &A(\omega)A(\omega_t-\omega_\tau-\omega).
	\end{align}
 Because $\bar{K}_{\mathcal{S}}(E, \omega,\omega_\tau,\omega_t-\omega_\tau-\omega)$ is a product of simple poles, see e.g.\ Eqs.\ \ref{eq:k3inter} and \ref{eq:k3mix} for the paramagnetic term, when $A(\omega)$ has a gaussian envelope the $\omega-$ integration in Eq.\ \ref{eq:schemebis} can be expressed in terms of Faddeeva functions. As we shall see below, the obtained {\em analytical} expression for each value of $E$ remains to be numerically integrated over the tensorial density of states to provide the final result.
 
 Let us first analyze the paramagnetic contribution. By following the procedure outlined above \cite{fiore_} one can write, for the fully-interband contribution at a fixed $E$ value, the expression
	\begin{widetext}
	\begin{align}
	\label{eq:j00taupara}
		\bar{J}^{\text{inter}}_{00\tau}(E,\omega_t,\omega_\tau)&=A(\omega_\tau)\left(\frac{1}{\omega_t+i\eta-2E}-\frac{1}{\omega_t+i\eta+2E}\right)\nonumber\\
		&\times\left[\frac{F^{\text{inter}}_{00\tau}(2E)}{\omega_\tau+i\eta-2E}+\frac{F^{\text{inter}}_{00\tau}(-2E)}{\omega_\tau+i\eta+2E}+\frac{F^{\text{inter}}_{00\tau}(2E)+F^{\text{inter}}_{00\tau}(-2E)}{\omega_t-\omega_\tau+i\eta}\right],\\
	\label{eq:j0tautaupara}
		\bar{J}^{\text{inter}}_{0\tau\tau}(E,\omega_t,\omega_\tau)&=A(\omega_t-\omega_\tau)\left(\frac{1}{\omega_t+i\eta-2E}-\frac{1}{\omega_t+i\eta+2E}\right)\nonumber\\
		&\times\left[\frac{F^{\text{inter}}_{0\tau\tau}(2E)}{\omega_t-\omega_\tau+i\eta-2E}+\frac{F^{\text{inter}}_{0\tau\tau}(-2E)}{\omega_t-\omega_\tau+i\eta+2E}+\frac{F^{\text{inter}}_{0\tau\tau}(2E)+F^{\text{inter}}_{0\tau\tau}(-2E)}{\omega_\tau+i\eta}\right],
	\end{align}
	\end{widetext}
	where the function $F^{\text{inter}}_{00\tau}$ reads 
	\be
	\label{eq:fadd}
	F_{00\tau}^{\text{inter}}(2E)=\int d\omega\,\frac{A(\omega)A(\omega_t-\omega_\tau-\omega)}{\omega+i\eta+2E},
	\ee
	while the function $F^{\text{inter}}_{0\tau\tau}$ is obtained from $F^{\text{inter}}_{00\tau}$ by exchanging $\omega_t-\omega_\tau\leftrightarrow\omega_\tau$.
    
    {Let us now assume that the envelope of the incoming gauge potential can be well approximated by a gaussian, i.e.\ $A(t)=\bar{A}e^{-t^2/(2\tau_p^2)}\cos(\Omega t)$. For sufficiently short-in-time pulses, where one can approximate $A(\omega)=\bar{A}$ as a frequency independent constant over the energy scales of the system}, the $F^{\text{inter}}$ functions can be evaluated through residues and they simply reduce to a constant. This has the remarkable consequence that Eq.\ \ref{eq:j00taupara} and Eq.\ \ref{eq:j0tautaupara} {produce the same results} obtained in the two-level system, described by Eq.\ \ref{eq:i00t} and Eq.\ \ref{eq:i0tt}, respectively. In other words, we recover the same frequency structure of the two-level case, provided that $\Delta$ is replaced by $ E$. We then expect, in analogy with Fig.\ \ref{fig:imp}(b) and (e), four peaks in the half-plane $\omega_t>0$ at coordinates $(2E,-2E)$, $(2E,0)$, $(2E,2E)$ and $(2E,4E)$. The calculation of the contribution coming from $\bar{K}^{\text{mi}}$, reported in the SM \cite{fiore_}, can instead be shown to lead to enhanced signals in the 2D map at the intercepts of the lines $\omega_t=0,2E$ with $\omega_\tau=2E$ or $\omega_t-\omega_\tau=2E$. 

    \begin{figure}
	\centering
	\includegraphics[width=\columnwidth]{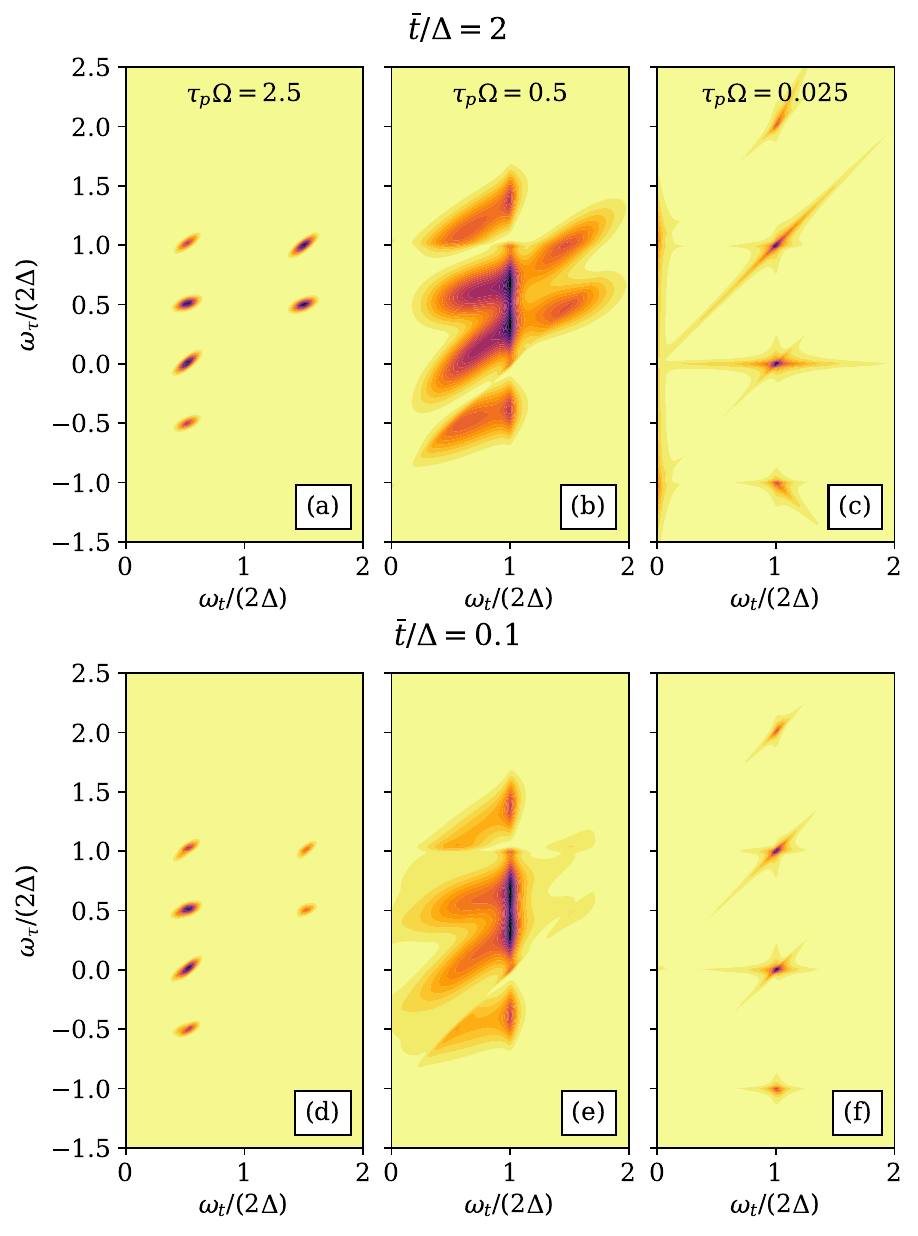}	
	\caption{{Panels (a)-(c): 2D map of $|J_{2D}^{\text{para}}(\omega_t,\omega_{\tau})|$ obtained for the fully-paramagnetic process setting $\bar{t}/\Delta=2$, computed by  integrating over energy $E$ the sum of Eqs.\ \ref{eq:j00taupara} and \ref{eq:j0tautaupara} multiplied by $\rho^{\text{inter}}(E)$, plus the corresponding contribution from the mixed process. Here we assume a gaussian pulse $A(t)=\bar{A}e^{-t^2/(2\tau_p^2)}\cos(\Omega t)$, by setting $\Omega=\Delta$ and using the values of $\tau_p\Omega$ reported in each column. We set $\eta/(2\Delta)=0.05/(2\pi)$ for the analytic continuation. Panels (d)-(f): 2D map for $\bar{t}/\Delta=0.1$.}}
	\label{fig:semipara}
	\end{figure}
    
    {We stress here that the equivalence between the two-level case and the current lattice system needs to be properly framed. To be consistent with Ref.\ \cite{mahmood_Nat.Phys.21}, in the two-level analysis we have shown the 2D map produced by a frequency independent \textit{electric field} $E(\omega)=\bar{E}$, which would correspond to a frequency-dependent gauge potential $A(\omega)\sim\bar{E}/\omega$ in Eqs.\ \ref{eq:j00taupara} and \ref{eq:j0tautaupara}. In the following we want to analyze the 2D response to a short-in-time (broadband) and long-lasting (narrowband) electric field. Given the relation $E(t)\sim\partial_tA(t)$, we can realize such an effect by varying the envelope of our gaussian $A(t)$ via $\tau_p\Omega$, even if the impulsive limit would not correspond to a $\delta(t)$-like electric field. Within the gaussian approximation for $A(t)$, but keeping a finite duration of the pulse,} the functions $F^{\text{inter}}$ can be reduced to an {\em analytical} expression, given by a combination of Faddeeva functions, as shown in the SM \cite{fiore_}. In practice, this implies a reduction in the numerical effort to calculate the 2D maps, whose result for the full paramagnetic contribution is shown in Fig.\ \ref{fig:semipara}. 
    
    We simulate the signal for three values of $\tau_p\Omega$, which correspond to a incoming narrowband, intermediate, or broadband pulse, while setting the central frequency of the pump inside the gap $\Omega=\Delta$. As expected, for decreasing values of $\tau_p\Omega$, the 2D maps evolve from the six spots at multiples of $\Omega$ expected in the monochromatic limit, see panels (a) and (d), to a more complex spectral distribution, in panels (c) and (f), dominated by the features of the kernel. 
    
    Since the latter strictly depends on the continuum of excitations, we consider two values of $\bar{t}/\Delta$. In Fig.\ \ref{fig:semipara}(a)-(c) we show the results setting {$\bar{t}/\Delta=2$}, while in Fig.\ \ref{fig:semipara}(d)-(f) we consider $\bar{t}/\Delta=0.1$, which intuitively corresponds to a less dispersive {band structure}. In the first case, panel (c), as the pump width increases, the spectral weight accumulates around four spots located at $(2\Delta,\pm 2\Delta)$, $(2\Delta,0)$ and $(2\Delta,4\Delta)$, i.e.\ the same positions appearing in the two-level simulation of Fig.\ \ref{fig:imp}. However, since the bands now have a large dispersion, the peaks show a significant elongation along the lines that connect each one of them with the origin, namely $\omega_\tau=\pm\omega_t,0,2\omega_t$. On top of that, peaks along the line $\omega_t=0$ are also visible in panel (c). As mentioned before, they purely arise from the mixed interband-intraband scattering mediated by $\bar{K}^{\text{mi}}$. Then it is possible to identify regions on the 2D map that are selectively sensitive to this specific nonlinear pathway, as also discussed in Ref.\ \cite{chen_npjComputMater25}. On the other hand, if $\bar{t}/\Delta=0.1$ as in panel (f), we recover {more similar results to} Fig.\ \ref{fig:imp}. This can be understood by analyzing the behavior of the two functions $\rho^{\text{inter}}(E)$ and $\rho^{\text{mi}}(E)$, see the SM \cite{fiore_}. While the former is singular for $E\to\Delta$, the latter vanishes in the same limit; in addition, there is a factor $t^2/\Delta^2$ in the whole $E$ interval suppressing $\rho^{\text{mi}}(E)$ with respect to $\rho^{\text{inter}}(E)$. Consequently, one can approximate $\rho^{\text{inter}}(E)\sim\delta(E-\Delta)$ and $\rho^{\text{mi}}(E)=0$, so that the mixed scattering is now washed out and the elongation of the remaining four peaks is now concentrated in close proximity to $2\Delta$.
	
	Let us now compute the result for the diamagnetic process. Here the calculation is simplified due to the dependence of the nonlinear kernel only on the sum of two incoming frequencies of the field, i.e.\ from Eq.\ \ref{eq:jdia} one has  $\bar{K}^{\text{dia}}(\omega_{ij})$ as appropriate for two-photon vertices. For the $\bar{J}^{\text{dia}}_{00\tau}$ part of the 2D map, one needs then to write out all the possible combinations of two incoming frequencies that enter the kernel: 
    \begin{align}
	\label{eq:diasym}
	\omega_{ij}=\omega+\omega_\tau,\omega_t-\omega,\omega_t-\omega_\tau.
	\end{align}
	More explicitly, by considering the structure of Eq.\ \ref{eq:kdia}, one finds 
	\begin{widetext}
	\begin{align}
	\label{eq:j00tdia}
		\bar{J}^{\text{dia}}_{00\tau}(E,\omega_t,\omega_\tau)&=2A(\omega_\tau)\left[F^{\text{dia}}_{00\tau}(-2E)-F^{\text{dia}}_{00\tau}(2E)\right]\nonumber\\
		&+A(\omega_\tau)A^2(\omega_t-\omega_\tau)\left(\frac{1}{\omega_t-\omega_{\tau}+i\eta-2E}-\frac{1}{\omega_t-\omega_{\tau}+i\eta+2E}\right),\\
		\label{eq:j0ttdia}
		\bar{J}^{\text{dia}}_{0\tau\tau}(E,\omega_t,\omega_\tau)&=2A(\omega_t-\omega_\tau)\left[F^{\text{dia}}_{0\tau\tau}(-2E)-F^{\text{dia}}_{0\tau\tau}(2E)\right]\nonumber\\
		&+A(\omega_t-\omega_\tau)A^2(\omega_\tau)\left(\frac{1}{\omega_{\tau}+i\eta-2E}-\frac{1}{\omega_{\tau}+i\eta+2E}\right),
	\end{align}
	\end{widetext}
	where the function $F_{00\tau}^{\text{dia}}$ reads
	\be
	\label{eq:fadddia}
	F^{\text{dia}}_{00\tau}(2E)=\int d\omega\,\frac{A(\omega)A(\omega_t-\omega_\tau-\omega)}{\omega_\tau+\omega+i\eta+2E},
	\ee
	and $F^{\text{dia}}_{0\tau\tau}$ is obtained from $F^{\text{dia}}_{00\tau}$ by exchanging $\omega_t-\omega_\tau\leftrightarrow\omega_\tau$. 		
	As before, in the limit {$A(\omega)=\bar{A}$} the $F^{\text{dia}}$ functions are constant and  the combination in the first lines of Eqs.\ \ref{eq:j00tdia} and \ref{eq:j0ttdia} vanishes. The remaining terms then give
	\begin{align}
	\label{eq:jdiatot1}
    \bar{J}^{\text{dia}}_{00\tau}(E,\omega_t,\omega_\tau)&\propto\frac{1}{\omega_t-\omega_{\tau}+i\eta-2E}-(E\to-E),\\
	\label{eq:jdiatot2}
	\bar{J}^{\text{dia}}_{0\tau\tau}(E,\omega_t,\omega_\tau)&\propto\frac{1}{\omega_{\tau}+i\eta-2E}-(E\to-E).
	\end{align}
	From this expression we notice that the spectral weight on the 2D map {in this approximation} is spread along the lines $\omega_t-\omega_\tau=\pm2E$ and $\omega_\tau=\pm2E$. 
	
\begin{figure}
	\centering
	\includegraphics[width=\columnwidth]{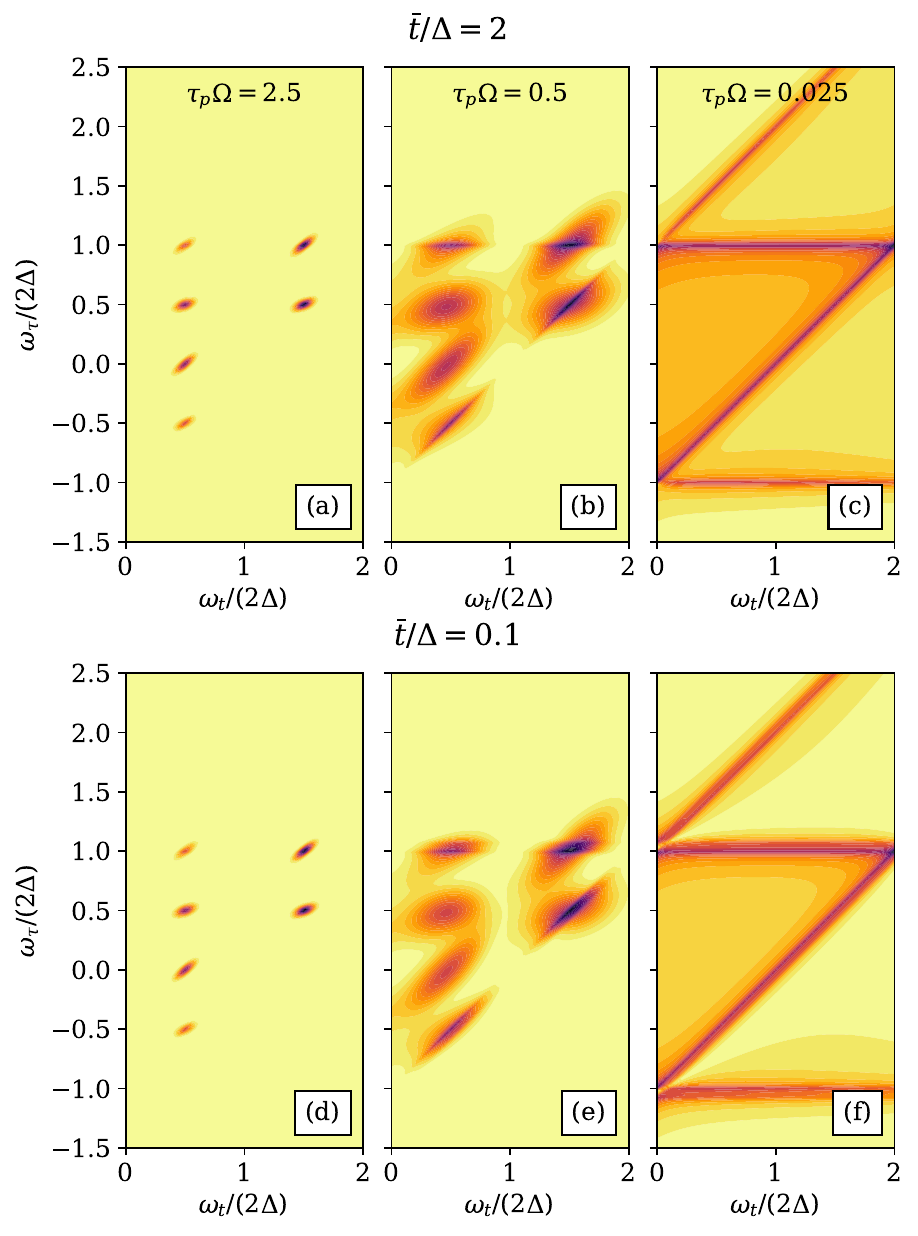}	
	\caption{{Analgous of the previous Fig.\ \ref{fig:semipara} for the fully-diamagnetic process, $|J_{2D}^{\text{dia}}(\omega_t,\omega_\tau)|$, under the same driving conditions. In panels (a)-(c) we consider $\rho^{\text{dia}}(E)$ derived from Eq.\ \ref{rhodia} for $\bar{t}/\Delta=2$, we multiply it by the sum of Eqs.\ \ref{eq:j00tdia} and \ref{eq:j0ttdia} and integrate the result over energy $E$.  In panels (d)-(f) we consider instead $\bar{t}/\Delta=0.1$.}}
	\label{fig:semidia}
	\end{figure}

    The complete computation of the diamagnetic process is shown in Fig.\ \ref{fig:semidia}. Here we used the same resonance condition $\Omega=\Delta$ and values $\tau_p\Omega$ and $t/\Delta$ adopted in Fig.\ \ref{fig:semipara}. In the upper row, panels (a)-(c), we considered the {more dispersive limit $\bar{t}/\Delta=2$}, while in the lower row, panels (d)-(f), we considered the {$\bar{t}/\Delta=0.1$ case}. The diamagnetic process is much less effective in selecting specific spots in the 2D map compared to the paramagnetic ones, as one can observe in panels (c)-(f), where we show the results obtained in the impulsive limit, {which reproduce the stripe-like phenomenology highlighted in Eqs.\ \ref{eq:jdiatot1} and \ref{eq:jdiatot2}.} 
	
\begin{table}
\caption{\label{tab:recap} Summary of the maxima of the 2D $(\omega_t,\omega_\tau)$ map for the $S_{1111}$ (para) and $S_{22}$ (dia) processes studied in this section, in the {broadband} limit for the gauge field.}
\begin{ruledtabular}
\begin{tabular}{ccc}
&
\textrm{Para}&
\textrm{Dia}\\
\colrule
$J_{00\tau}$ & $(\omega_t,\omega_\tau)=(2\Delta,2\Delta)$ & $\omega_\tau=2\Delta$ \\
& $(\omega_t,\omega_\tau)=(2\Delta,-2\Delta)$ & $\omega_\tau=-2\Delta$ \\
$J_{0\tau\tau}$ & $(\omega_t,\omega_\tau)=(2\Delta,0)$ & $\omega_t-\omega_\tau=2\Delta$\\
& $(\omega_t,\omega_\tau)=(2\Delta,4\Delta)$ & $\omega_t-\omega_\tau=-2\Delta$ \\
\end{tabular}
\end{ruledtabular}
\end{table}

The set of Eqs.\ \ref{eq:scheme}-\ref{eq:jdiatot2} represent the central result of the present work: they provide, indeed, a general expression for the computation of the 2D maps, where the details of the excitation spectrum of the system is fully encoded in the density of states $\rho(E)$ of Eq.\ \ref{eq:scheme}. As an example, in the case of the 2D lattice model in Eq.\ \ref{eps}, this is in turn given for paramagnetic-like and diamagnetic-like processes by the analytical expressions in Eqs. \ref{rhointer}-\ref{rhomix} and \ref{rhodia}, respectively. The present approach can be in principle extended to interacting electronic systems, once that the spectrum $E_\a$ of the interacting eigenstates (labeled by a generic index $\a$) and the matrix elements of light-matter couplings are known. In this case, some additional care should be taken to account for all the possible $E_\a-E_\b$ transitions triggered by the possible light-matter couplings in $K^{(3)}_\mathcal{S}$, see e.g.\ Ref.\ \cite{rostami_AnnalsofPhysics21, ono_Phys.Rev.Lett.25}. 

\begin{figure}
	\centering
	\includegraphics[width=\columnwidth]{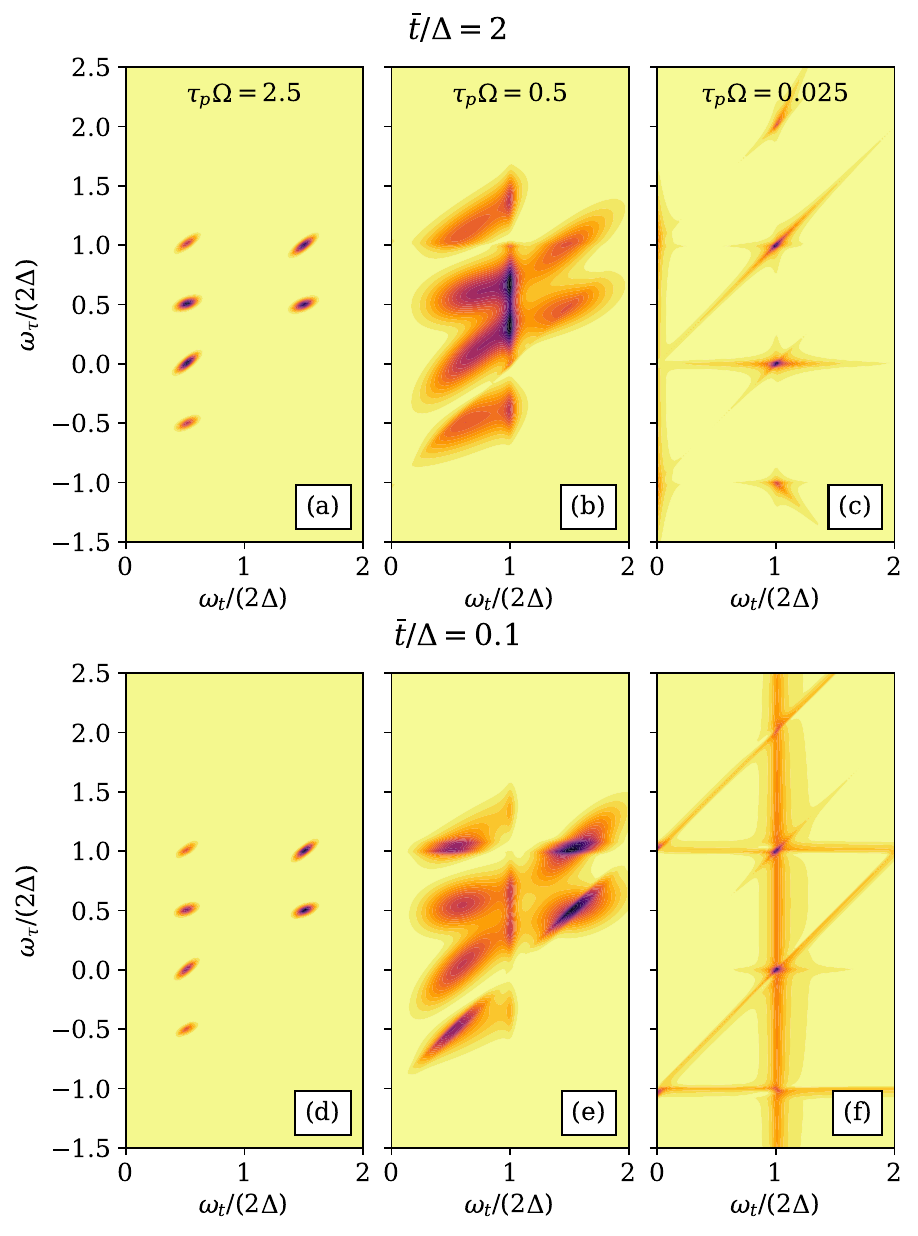}	
	\caption{{Fully gauge-invariant 2D map of $|J_{2D}(\omega_t,\omega_\tau)|$ containing all the processes foreseen by Eq.\ \ref{eq:s4}, under the same driving conditions of Figs.\ \ref{fig:semipara} and \ref{fig:semidia}. In panels (a)-(c) we consider $\bar{t}/\Delta=2$, while in panels (d)-(f) we consider instead $\bar{t}/\Delta=0.1$.}}
	\label{fig:semitot}
	\end{figure}

{In Fig.\ \ref{fig:semitot} we finally show the simulations of the fully gauge-invariant response $|J^{(3)}_{2D}(\omega_t,\omega_\tau)|$ for values $\bar{t}/\Delta=0.1$ and $\bar{t}/\Delta=2$, respectively, obtained by summing all the contributions to the nonlinear response kernel $K^{(3)}$ foreseen by Eq.\ \ref{eq:s4}. In the SM \cite{fiore_}, we also provide separated plots for each class of processes to highlight their relative weight. As one can see, by varying either the spectral envelope, ranging from narrowband in panels (a),(d) to broadband in panels (c),(f), or the system's excitations, ranging from more (top row) to less (bottom row) dispersive bands, one gets quite different 2D maps, where contributions of different subsets can be dominant. As before, in the narrowband limit the spots are always controlled by the pulse frequency. As the pulse become broadband we can observe that for both choices of $\bar t/\Delta$ the fully paramagnetic contribution is always relevant. Signatures coming from two-point responses, arising either from the fully diamagnetic term $S_{22}$ or from the mixed $S_{31}$ one, see the SM \cite{fiore_} for details, are visible in the fully gauge-invariant response as stripe-like features, whose relative strength depends on the ratio $\bar{t}/\Delta$. }

{As a final remark,} we note that in the present discussion the nonlinear current is computed as a function of the pulse {\em inside} the sample. {As mentioned at the beginning of Sec.\ \ref{sec:gen} and explicitly shown in Sec.\ IIB of the SM \cite{fiore_}, this is a reasonable approximation for an experiment in transmission geometry from non-absorbing thin films}. How this has to be connected to the {\em incoming} pulse {in different experimental geometries} is a crucial point that will be discussed in the next section.

\section{Propagation effects in the 2DTS maps: the case of Josephson plasmons}\label{sec:prop}

\subsection{Screening and propagation effects connected to the linear response}
In the previous section we focused on the different signatures obtained in the 2D maps when the nonlinear processes originate from paramagnetic or diamagnetic light-matter coupling terms at microscopic level. In both cases the possibility to disentangle different excitation pathways requires a broad pump spectrum, since in the opposite monochromatic limit the 2D spectrum has only isolated spots where the structure of the kernel cannot be resolved (see panels (a)-(d) of Figs.\ \ref{fig:semipara}, \ref{fig:semidia} and \ref{fig:semitot}).  As a consequence, to really make a bridge between the theoretically-computed $K^{(3)}$, that can be carried out in principle in a controlled way within certain approximations, and the outcomes of the experimental measurements, one must be able to predict the spectral content of the pump field $A_\text{in}$ {\em inside} the sample, where the nonlinear current is generated. In linear response, this is usually a simple problem to be solved via Maxwell's equations: indeed, when the induced current is linear in the gauge field $A$, Maxwell's equations are linear and all the eventual complications of the mixing of electromagnetic waves {with a dipole-active} mode can be accounted for by introducing the frequency-dependent refractive index $n(\omega)$ of the system. In addition, by properly using boundary conditions at the sample/vacuum interfaces, {one} can connect the field inside the material to the one outside. In the presence of nonlinear processes Maxwell's equations themselves become nonlinear and an alternative approach must be followed to answer two questions: (i) what is the spectrum of the THz fields driving the nonlinear current computed via {Eq.\ \ref{eq:jnl}}; (ii) what is the connection between the nonlinear current generated inside the material and the nonlinear electric field in Eq.\ \ref{eq:2dexp} detected experimentally to construct the 2D map. 

To answer these questions we should go back to the definition in Eq.\ \ref{eq:2dexp} of the experimentally measured 2D signal $E_{2D}(t,\tau)$, and discuss the limits of applicability of the approximation in Eq.\ \ref{eq:appr}, where $E_{2D}(t,\tau)$ is identified with the nonlinear {current} inside the material. As one can intuitively understand, the problem arises when the system hosts dipolar modes contributing to a frequency-dependent refractive index $n(\omega)$, that varies considerably on the range of frequencies set by the spectrum of the $A$ pulse. Whenever this happens, we expect that the internal field $A_\text{in}$ gets modified with respect to the  {\em external} one $A_\text{ext}$. This effect can be accounted for macroscopically by the the transmission coefficient
\be
\label{eq:trans}
t(\omega)=\frac{2}{1+n(\omega)},
\ee
whose impact will be relevant whenever optically-active modes overlap with the pump/probe spectrum.

In addition to this, one has to account for the fact that the nonlinear current generated inside the material acts as a source for the Maxwell's equations, modifying the spectrum of the nonlinear field generated inside the system. This problem can be solved using an iterative perturbative approach \cite{zhang_NationalScienceReview23,katsumi_Phys.Rev.B23,gomezsalvador_Phys.Rev.B24,sellati_npjQuantumMater.25}, as detailed in the SM \cite{fiore_} for transmission and reflection geometries of a sample of any thickness. {In summary, in the present case the pipeline of the calculations is the following. Starting from the general assumption that $E_{2D}(t,\tau)$ in Eq.\ \ref{eq:2dexp} represents the experimentally accessible quantity, one first computes the induced nonlinear current inside the system, Eq.\ \ref{eq:2d}, as a $J^{(3)}$ function of the gauge field $A$. Next, one accounts for propagation effects to connect a generic outgoing electric field $E^{\rm out}$ to the $J^{(3)}$ inside the system. This latter part is accomplished by solving the Maxwell's equations inside the system in the presence of a nonlinear current, and accounting for both boundary conditions at the sample/vacuum interfaces, and for phase-matching conditions in the presence of multi-wave mixing. All these steps are detailed in the SM \cite{fiore_}.} Here we just report the final result for the third-order outgoing electric field in a reflection geometry from a bulk sample, i.e.\ the situation appropriate for the case of the Josephson plasmon measured in Ref.\ \cite{liu_Nat.Phys.24} that we will discuss in the following: 
	\begin{widetext}
	\be
	\label{eq:genout}
	E^{\text{out}}(\omega_t)=\frac{4\pi i}{c}\frac{1}{1+n(\omega_t)}\int d\omega_i\,K^{(3)}(\omega_1,\omega_2,\omega_3)A_{\text{ext}}(\omega_1)A_{\text{ext}}(\omega_2)A_{\text{ext}}(\omega_3)\frac{t(\omega_1)t(\omega_2)t(\omega_3)}{n(\omega_t)\omega_t/c+\sum_in(\omega_i)\omega_i/c}\delta\left(\omega_t-\sum_i\omega_i\right).
	\ee
	\end{widetext}
    {In the case of a bulk system considered here, one only retains the generated reflected field on the left of the vacuum/material interface and the transmitted one on its right side, thus neglecting the effect of back-reflected fields from the opposite side of the sample. This is a reasonable approximation for a millimeter-scale superconducting sample probed with THz light, as it is the case for the experimental geometry in Ref.\ \cite{liu_Nat.Phys.24} discussed below. The transmission coefficients in Eq.\ \ref{eq:genout} account then for the boundary conditions at the interface, while the denominator encodes the phase-matching condition ($\sum_i k_i-k_r=0$) for the momenta $k_i=n(\omega_i)\omega_i/c$ of the generating pulses and the momentum $k_r=-n(\omega_t)\omega_t/c$ of the reflected outgoing field, where we also assume a local ($q=0$) dielectric response via $n(\omega)$ appropriate for THz and optical probes. Further details on the derivation of the formulas for the reflected/transmitted $E^{\text{out}}$ in the bulk/thin film limit, along with their validation against known expressions, are provided in Secs.\ IIA-B of the SM \cite{fiore_}, where all the needed conversion factors between the electric field and the gauge potential are carefully taken into account.}
    
	Once established a general expression for the nonlinear field propagating outside {the sample} we can proceed in the evaluation of the 2D map by following the same steps leading to Eq.\ \ref{eq:j00tau} and Eq.\ \ref{eq:j0tautau} in Sec.\ \ref{sec:gen}. {Let us consider e.g.\ the contribution with the double interaction with the $A_0(t)$ field and the single interaction with $A_{\tau}(t)$. One starts from Eq.\ \ref{eq:genout} and replaces one frequency with $\omega_\tau$ and the other two with $\omega$ and $\omega_t-\omega_\tau-\omega$, eliminating one frequency integration via the energy conservation and another one with the $\tau\to\omega_\tau$ Fourier transform. We end up with }
	\begin{widetext}
	\begin{align}
	\label{eq:fre00t}
	E_{00\tau}(\omega_t,\omega_\tau)=\int d\omega\,&K^{(3)}_{\mathcal{S}}(\omega_\tau,\omega,\omega_t-\omega_\tau-\omega)A_{\tau,\text{ext}}(\omega_\tau)A_{0,\text{ext}}(\omega)A_{0,\text{ext}}(\omega_t-\omega_\tau-\omega)\nonumber\\
	\times&\frac{2\pi i t(\omega_t)t(\omega_\tau)t(\omega)t(\omega_t-\omega_\tau-\omega)}{n(\omega_t)\omega_t+n(\omega_\tau)\omega_\tau+n(\omega)\omega+n(\omega_t-\omega_\tau-\omega)(\omega_t-\omega_\tau-\omega)}.
	\end{align}
	\end{widetext}
	{In the spirit of Eq.\ \ref{eq:genfre},} in this expression we have separated on the first line inside the integral the result that one would obtain neglecting the propagation effects inside the material, where the external fields are convoluted with the nonlinear third-order kernel. In the second line, instead, {we see the explicit form of the function $M(\omega_\tau,\omega,\omega_t-\omega_\tau-\omega)$ introduced in Eq.\ \ref{eq:genfre}.} which has its own frequency dependence and can mask the genuine nonlinear effects included in $K^{(3)}$. In this way, we will see that even the simplest instantaneous kernel can show complicated signatures linked to the response due to the propagation of the generated signal at the interface between vacuum and material. An analogous result for $E_{0\tau\tau}(\omega_t,\omega_\tau)$ is obtained, as explained above, by switching $\omega_\tau\leftrightarrow\omega_t-\omega_\tau$ {and $A_\tau\leftrightarrow A_0$}, corresponding to the system interacting once with the field $A_0(t)$ and twice with $A_{\tau}(t)$. 

Eq.\ \ref{eq:fre00t} and its counterpart for $E_{0\tau\tau}$ are general. However, to give a concrete example of the relevance of the propagation effects we apply this approach to a physically relevant case, that is the understanding of recent 2DTS experiments in Josephson plasmons \cite{liu_Nat.Phys.24}. The example is particularly relevant since, in contrast to the toy model of the insulator discussed in the previous Section, in this case the {\em same} collective excitations contribute to screen the linear response to a THz pulse and to generate a nonlinear current. The understanding of 2DTS spectra of Josephson plasmon then offers a perfect benchmark to highlight the necessary steps required to extract from {\em measured} nonlinear electric field the relevant information on the {\em microscopic} processes at play in the system, encoded in the $K^{(3)}$ kernel. 

\subsection{Screening effects from Josephson plasmons}
\label{sec:plasmon1}
\begin{figure}
\includegraphics[width=\columnwidth]{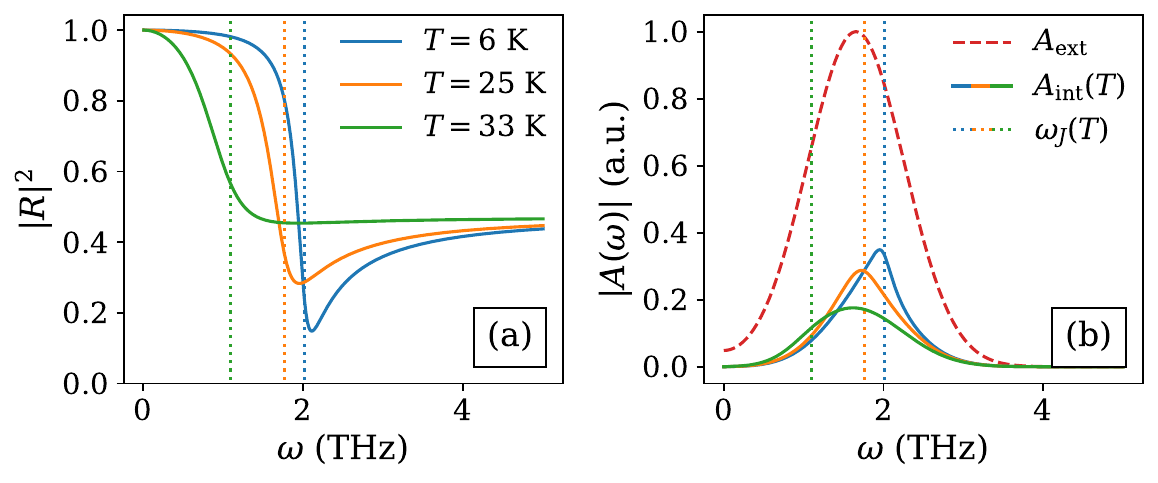}
\caption{Panel (a): simulated reflectance (left panel) of a LSCO sample at doping level $x=0.17$ for three selected temperatures, using the interpolated parameters from Ref.\ \cite{tamasaku_Phys.Rev.Lett.92} and Ref.\ \cite{henn_ThinSolidFilms98} with the two-fluid model in Eq.\ \ref{eq:epsilon}. The dotted vertical lines denote the temperature-dependent value of $\omega_J(T)$. Panel (b): comparison between the spectrum of the gauge field outside the sample $A_\text{ext}$ (dotted line), modeled as the experimental one used in Ref.\ \cite{liu_Nat.Phys.24}, and inside the sample $A_\text{in}(\omega)=t(\omega)A_\text{ext}(\omega)$ (solid lines), where $t(\omega)$ is the transmission coefficient of Eq.\ \ref{eq:trans}.  At low $T$ the plasma edge (dotted vertical lines) produces a dramatic filtering of the incoming field  around $\omega_J$.}
\label{fig:tra}
\end{figure}

As we mentioned in the introduction, soft Josephson plasmons are the natural low-energy collective excitations in layered superconductors that emerge when the phase of the SC order parameter mixes with the electromagnetic waves propagating inside the material. Since layered systems, like SC cuprates, have a weak interlayer coupling, the low energetic cost for pair tunneling among planes leads to a soft (i.e.\ of the order of few THz) plasma mode, affected by a strong Landau damping by quasiparticles above $T_c$. However, below $T_c$ the opening of a SC gap removes damping effects and a well-defined mode appears, whose signature in the linear response is a sharp plasma edge in the reflectivity of the sample at the temperature-dependent value $\omega_J$, see Fig.\ \ref{fig:tra}(a). This is also the case for the nearly optimally-doped ($x=0.17$) La$_{2-x}$Sr$_x$CuO$_4$  sample measured by 2DTS in Ref.\ \cite{liu_Nat.Phys.24}. In this experiment the broadband external pump and probe fields have a central frequency $\Omega$ which matches the $T=0$ value of $\omega_J$, $\Omega\sim \omega_J(T=0)=2$ THz.  As a consequence, the propagation of the light pulse {\em inside} the sample is modified by the formation of a plasmon-polariton, whose presence in encoded in the refractive index of the material. In the specific case of the SC Josephson plasmon a convenient description of the linear response can be achieved \cite{dordevic_Phys.Rev.Lett.03} by modelling the dielectric function for the transport along the stacking direction as the sum of a superfluid response and the normal dissipative term, namely
	\be
	\label{eq:epsilon}
	\epsilon(\omega)=\epsilon_{\infty}\left(1-\frac{\omega_J^2}{\omega^2}+i\frac{\gamma}{\omega}\right).
	\ee
Here $\gamma$ is a phenomenological damping factor that accounts for the experimentally-observed smearing of the plasma edge, that also appears in the two-plasmon propagator in Eq.\ \ref{eq:2pk}. The microscopic origin of this damping has been discussed in the literature in connection with a possible inhomogeneity of the Josephson couplings among different planes \cite{vandermarel_CzechJPhys96}. For the sake of the present discussion, the analysis of Ref.\ \cite{dordevic_Phys.Rev.Lett.03} suggests that the two-fluid model of Eq.\ \ref{eq:epsilon} gives an excellent description of the measured linear response for the doping level of the sample of Ref.\ \cite{liu_Nat.Phys.24}, that is the relevant aspect to quantitatively describe the refractive index $n(\omega)=\sqrt{\epsilon(\omega)}$ and the transmission coefficient in Eq.\ \ref{eq:trans}. 

Here we considered for $A_{\text{ext}}$ the pulse spectrum of Ref.\ \cite{liu_Nat.Phys.24}{, fitting the experimental applied field with $E(\omega)=i\omega A(\omega)$}. We modeled the dielectric function in Eq.\ \ref{eq:epsilon} by fitting the experimental data for LSCO samples at doping levels $x=0.16$ \cite{tamasaku_Phys.Rev.Lett.92} and $x=0.18$ \cite{henn_ThinSolidFilms98} at several temperatures, interpolating then the extracted values of $\omega_J$ and $\gamma$ to estimate the ones appropriate for the $x=0.17$ LSCO sample of \cite{liu_Nat.Phys.24}. As we can see from Fig.\ \ref{fig:tra}(b), at all temperatures the field gets strongly reshaped, with a narrowing of the peak and a shift from the (fixed) central frequency $\Omega$  of the pump towards the temperature-dependent value of $\omega_J(T)$. In the next section, we will discuss how this effect influences the {\em nonlinear} response of the plasmon. 

\subsection{Nonlinear response of Josephson plasmons}
\label{sec:plasmon2}
A possible description of nonlinear effects emerging from plasma waves below $T_c$ is provided by the Josephson model, describing a cosine-like energy dependence on the gauge-invariant phase difference among neighboring planes \cite{bulaevskii_Phys.Rev.B94,machida_PhysicaC:Superconductivity00,savelev_NaturePhys06,savelev_Rep.Prog.Phys.10,laplace_Adv.Phys.X16}
\be
\label{josephson2}
H=J\sum_n \cos \left(\theta_n-\theta_{n+1}-\frac{2e}{\hbar c}A\right),
\ee
where $\theta_n$ is the SC phase in layer $n$, $J$ is the effective Josephson interaction between the phase degrees of freedom in neighbouring layers, connected to the phase rigidity in the SC state. Here $\theta$ is a quantum variable whose conjugate field is the density: one can then show that the spectrum of the phase mode is the same of density fluctuations, which describe plasma waves with a value $\omega^2_J\sim4\pi e^2 J$ in the long-wavelength limit \cite{sun_Phys.Rev.Res.20,michael_Phys.Rev.B20,gabriele_Phys.Rev.Res.22,sellati_Phys.Rev.B23}. In addition, thanks to the minimal coupling of the phase difference to the gauge field, the {\em anharmonic} phase-only model of Eq.\ \ref{josephson2} provide a natural source for optical non linearity in the SC state \cite{savelev_NaturePhys06,savelev_Rep.Prog.Phys.10,laplace_Adv.Phys.X16,gabriele_NatCommun21}. While in the past these effects have been mainly described by solving the equation of motion for the phase variable $\phi_n\equiv \theta_n-\theta_{n+1}$ in the time domain, recent experimental measurements of high-harmonic generation and 2DTS with strong THz field triggered a direct theoretical computation of the nonlinear optical kernel in frequency domain, along the lines of the approach discussed in the previous section. In full analogy with the discussion of the previous Section, see Eq.\ \ref{eq:s4}, the $K^{(3)}$ is obtained by deriving from Eq.\ \ref{josephson2} the terms of order $A^4$.  As one can intuitively understand, there is a first instantaneous (in time) contribution $\sim JA^4$ coming directly from the expansion of the cosine in Eq.\ \ref{josephson2}, that has no structure in frequency domain. In addition, one recovers a {\em diamagnetic-like} coupling term $\sim A^2 \phi_n^2$ where two photons from the pump field are able to excite two plasma waves $\phi_n\equiv \theta_n-\theta_{n+1}$. The resulting contribution to the $K^{(3)}$ has the structure of a Kubo-like response, where two plasmons with opposite momenta are excited and then reabsorbed by the THz field. The computation of this kernel has been the subject of previous work \cite{gabriele_NatCommun21,fiore_Phys.Rev.B24}, and we refer the reader to these paper for additional details in the derivation. For the purposes of the present discussion, we can then assume that the first, instantaneous contribution is represented by a constant kernel in frequency domain, scaling with the Josephson coupling \cite{fiore_Phys.Rev.B24}
\be
\label{instk}
K^{\text{inst}}\sim J,
\ee
while for the  two-plasmon kernel we can use the general expression \cite{fiore_Phys.Rev.B24}
	\be
	\label{eq:2pk}
	K^{2\text{p}}(\omega)\sim J^2\int dz\,\rho^{\text{2p}}(z)\frac{1}{z}\frac{\coth{(\beta z/2)}}{(2z)^2-(\omega+i\gamma)^2},
	\ee
where $\rho^{\text{2p}}(z)$ is the density of state associated with the two-plasmon process. 
To get an intuition on the behavior of the two-plasmon kernel in Eq.\ \ref{eq:2pk}, we observe that if one neglects completely the momentum dispersion of the plasma waves in Eq.\ \ref{eq:2pk} then $\rho^{\text{2p}}(z)\propto \delta(z-\omega_J)$. As a consequence, one immediately recovers a maximum response of the system when the driving frequency $\omega$ matches $2\omega_J$.  When the momentum dependence of the plasma waves is taken into account via the $\rho^{\text{2p}}(z)$ dependence, the resonances of the nonlinear kernel depend on the dispersion and the number of the plasma branches, that is controlled by the number of layers per unit cells \cite{michael_Phys.Rev.B20,sellati_Phys.Rev.B23,fiore_Phys.Rev.B24}. However, for systems with a single layer per unit cell, as the La$_{2-x}$Sr$_x$CuO$_4$ sample measured in Ref.\ \cite{liu_Nat.Phys.24},  the result remains qualitatively the same. Indeed, in the region in proximity of $\omega_J$, the density of states grows linearly as $\rho^{\text{2p}}(z)\sim z\theta(z-\omega_z)$. Moreover, near the critical temperature we can also approximate $\coth{\beta z/2}\sim T/z$, so that the above kernel can be easily integrated and reads, neglecting constant prefactors,
\be
\label{eq:2pkappr}
K^{2\text{p}}(\omega)\sim\frac{T}{(\omega+i\gamma)^2}\log{\left(1-\frac{(\omega+i\gamma)^2}{4\omega_J^2}\right)},
\ee
which almost coincides with the so-called plasmon squeezing contribution in Ref.\ \cite{gomezsalvador_Phys.Rev.B24}. The only effect of the plasmon dispersion is then to weaken the singularity from a delta-like to a log-like divergence for $\omega\sim 2\omega_J$. In the following, we will use the expression in Eq.\ \ref{eq:2pkappr} to simulate the 2D maps given by this process.

In full analogy with the diamagnetic electronic kernel in Eq.\ \ref{eq:k3diasemi}, also Eq.\ \ref{eq:2pk} depends on a single frequency, corresponding in the experiment to the sum of the frequencies of the two photons exciting the two plasmons. To compute the full 2D map we can then symmetrize the diamagnetic kernel in Eq.\ \ref{eq:2pk} exactly as we did in Eq.\ \ref{eq:diasym}. We then obtain the kernel determining the nonlinear current originated from two interactions with $A_0$ and one with $A_\tau$

	\begin{align}
	\label{eq:2d2p}
	K^{(3),2\text{p}}_\mathcal{S}(\omega_\tau,\omega,\omega_t-\omega_\tau-\omega)&\sim K^{2\text{p}}(\omega+\omega_\tau)\nonumber\\
    &+K^{2\text{p}}(\omega_t-\omega)\nonumber\\
	&+K^{2\text{p}}(\omega_t-\omega_\tau).
	\end{align}
	Using again the symmetry in Eq.\ \ref{eq:j00tau} upon replacement $\omega\to\omega_t-\omega_\tau-\omega$ we can show that the first two terms of the previous sum are actually equal. As it happened before, the form of the kernel $K^{(3),2\text{p}}_\mathcal{S}(\omega_t-\omega_\tau,\omega,\omega_\tau-\omega)$ entering the response from one interaction with $A_0$ and two with $A_\tau$ is obtained by switching $\omega_\tau\leftrightarrow\omega_t-\omega_\tau$ in the previous expression.
	
\begin{figure}
	\centering
	\includegraphics[width=\columnwidth]{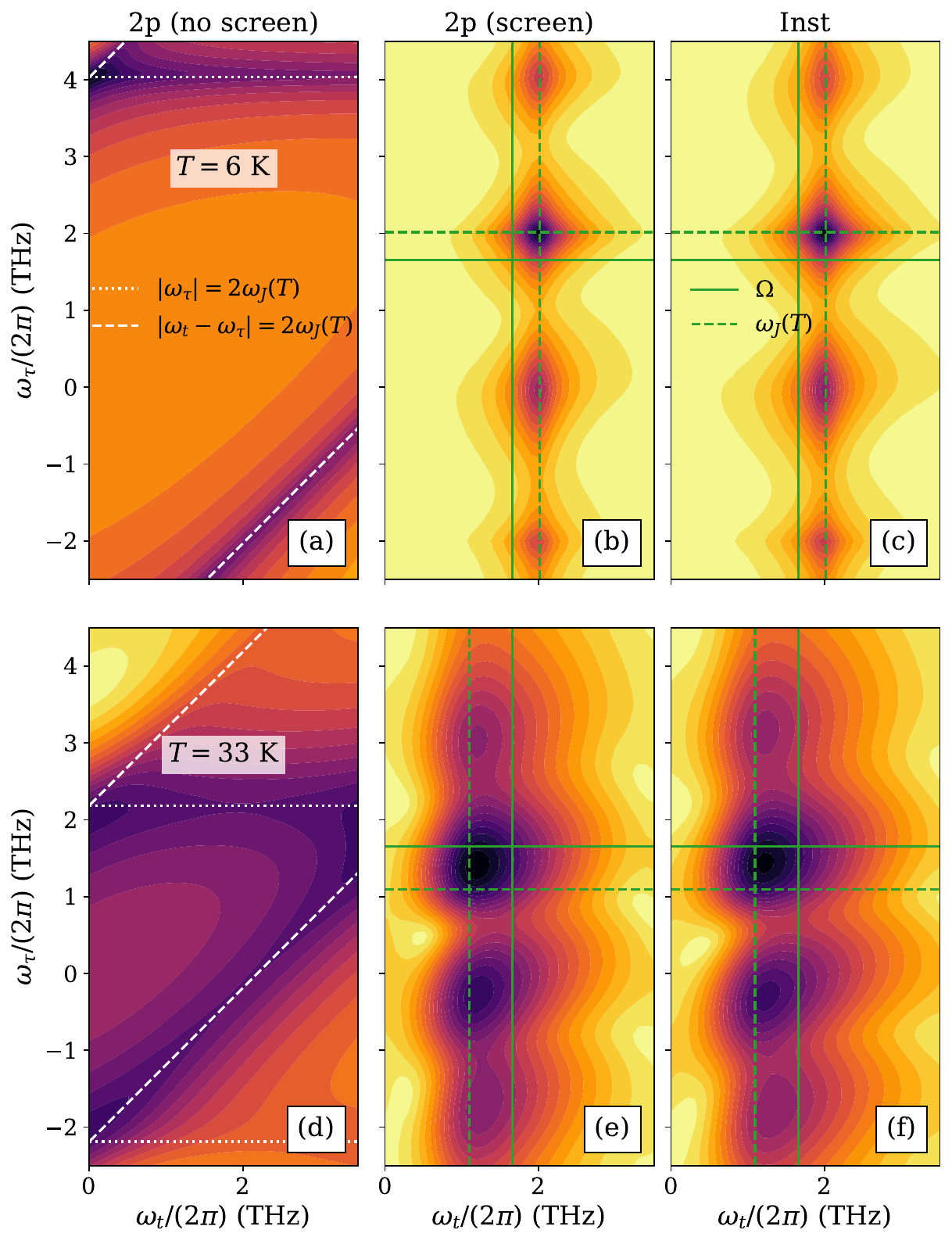}	
	\caption{The simulated 2D-THz response of LSCO ($x=0.17$) probed with $c-$axis polarized electric field, with incoming spectrum similar to the one used in Ref.\ \cite{liu_Nat.Phys.24}, for $T=6$ K (panels a-c) and $T=33$ K (panels d-f). In panels (a) and (d) we plot the behavior of the 2D signal produced by the two-plasmon kernel from Eq.\ \ref{eq:2pk} and Eq.\ \ref{eq:2d2p}, considering the ideal situation of a constant in frequency field and neglecting propagation effects, which shows the typical signatures of the diamagnetic processes discussed in Section \ref{sec:mod}. In panels (b) and (e) we consider the same two-plasmon excitation mechanism but we include propagation effects according to Eq.\ \ref{eq:genout}, while in panels (c) and (f) we compute the signal resulting from an instantaneous kernel. We superimpose as guides to the eye the positions of the central frequency of the pump (green dashed line) and of the plasma frequency at the various temperatures (blue dashed line).}
	\label{fig:inst2p}
	\end{figure}

In analogy with the discussion of the previous Section,  when the $A(\omega)$ spectrum is broad, one would expect that, in the absence of any modification of the pump spectrum inside the system, the processes mediated by the instantaneous kernel in Eq.\ \ref{instk} do not have clear structure in the frequency domain, while diamagnetic-like processes give rise to stripe-like features in the 2DTS maps, corresponding to the resonance frequencies of the kernel, see Fig.\ \ref{fig:semidia}(f). In Ref.\ \cite{liu_Nat.Phys.24} one is exactly in the impulsive limit for the driving $A_{\text{ext}}$ THz field: assuming that this is also the case for $A$ inside the system, the  $K^{2\text{p}}$ kernel leads to the map shown in Fig.\ \ref{fig:inst2p}(a). Here indeed  the signal piles up along the lines $\omega_\tau\sim 2\omega_J$ and $\omega_\tau-\omega_t\sim 2\omega_J$, where $\omega_J=2$ THz is the zero-temperature limit of the long-wavelength out-of-plane Josephson plasmon measured in Ref.\ \cite{liu_Nat.Phys.24}. Since the Josephson frequency $\omega_J(T)$ softens by increasing temperature, these signatures move in the 2D map when the system is warmed up, see Fig.\ \ref{fig:inst2p}(d). 
	
While the numerical results of Fig.\ \ref{fig:inst2p}(a)-(d) are consistent with the general discussion of the previous Section and with the {analytical estimate} of the kernel in Eq.\ \ref{eq:2pk}, they are in striking contrast with the experimental findings on Ref.\ \cite{liu_Nat.Phys.24}. Indeed, the main outcomes of these measurements is the appearance of four well-defined spots in the 2D map located at $(\omega_J,0)$, $(\omega_J, \pm \omega_J)$ and $(\omega_J, 2\omega_J)$. These correspond to the response expected for a generic diamagnetic kernel when the pump field {\em inside} the sample can be approximated as a monochromatic one centered at $\Omega\sim \omega_J$, see Fig.\ \ref{fig:semidia}(a). As we shall see, this is indeed an effect of the plasmon contribution to the {\em linear} response, that acts as a ``filter'' on the external broadband pump, washing out any resonant effect connected to the kernel itself. 

\subsection{2D maps including propagation effects}
\label{sec:plasmon3}
To account for the filtering effect shown in Fig.\ \ref{fig:tra}(b) we must resort to the full expression in Eq.\ \ref{eq:fre00t}, where the second line describes filtering by the $t(\omega)$ factors and propagation effects connected to multi-wave mixing. We then computed the realistic 2D response by using the general expression  in Eq.\ \ref{eq:fre00t} and its related counterpart for $E_{0\tau\tau}(\omega_t,\omega_\tau)$ to the case of the Josephson plasmon, by adopting the previously discussed modeling of $n(\omega)$. The results are shown in Fig.\ \ref{fig:inst2p}(c)-(f) for what concerns an instantaneous kernel, that just reduces to a constant, and in panels \ref{fig:inst2p}(b)-(e) for what concerns the two-plasmon kernel in Eq.\ \ref{eq:2pk}. Notice that, consistently with the physical origin of the $\epsilon(\omega)$, the same damping $\gamma(T)$ enters both the dielectric function via Eq.\ \ref{eq:epsilon} and the two-plasmon kernel via Eq.\ \ref{eq:2pk}. {As a further remark, consistently with Ref.\ \cite{liu_Nat.Phys.24}, we also want to only consider contributions from $t,\tau>0$ in the 2D frequency map. To do so, we compute the the inverse Fourier transform of Eq.\ \ref{eq:fre00t} and its related $A_0\leftrightarrow A_\tau$ counterpart, we select the relevant time quadrant and then we Fourier-transform back the signal to frequency space.}

Let us first consider the $T=6$ K case, reported in Fig.\ \ref{fig:inst2p}(b)-(c) for the instantaneous $K^{\text{inst}}$ and two-plasmon $K^{2\text{p}}$ kernel, respectively.  We clearly observe the four spots located at the positions $(\omega_J,\pm \omega_J)$, $(\omega_J,0)$, $(\omega_J,2\omega_J)$, with $\omega_J\sim2$ THz, consistent with the experiments \cite{liu_Nat.Phys.24}. This result simply follows from the fact that at low temperature, where the plasma edge is particularly sharp, the filtering effect encoded in the $t(\omega_i)$ factors of Eq.\ \ref{eq:genout} effectively convert the maps expected for a broadband pulse to the ones of a narrow-band pulse, whose spectral content is controlled by the linear response. By increasing the temperature the peaks redshift along the $\omega_t$ axis following the softening of the $\omega_J(T)$  plasma edge, see the dashed guiding lines in panels (e)-(f), while along the $\omega_\tau$ axis the peak maxima are still sensitive to the central frequency of the pump. This behavior can be understood by close inspection of Eq.\ \ref{eq:fre00t}. Indeed, the overall $t(\omega_t)$ factor, which is only related to the propagation of the outgoing radiation, fixes the detection frequency of the 2D signal at the temperature-dependent plasma edge $\omega_J(T)$, while the remaining transmission factors integrated over $\omega$ contribute to spread out the $\omega_\tau$ response in a range given by the overlap between $A_{\text{ext}}$, centered at $\Omega$, and the transmission factor $t(\omega)$, centered at $\omega_J(T)$.  A direct consequence of these findings is that {it is unlikely that an experiment could resolve the difference between the instantaneous and diamagnetic-like process. Indeed, both at $T=6$ K (Fig.\ \ref{fig:inst2p}(b) and (c)) and at $T=33$ K (Fig.\ \ref{fig:inst2p}(e) and (f)) the computed maps display essentially the same spectral features, due to a dominant effect of propagation, making it hard to distinguish minor differences in an experimental set-up.} 
	
	\begin{figure}
		\centering
		\includegraphics[width=\columnwidth]{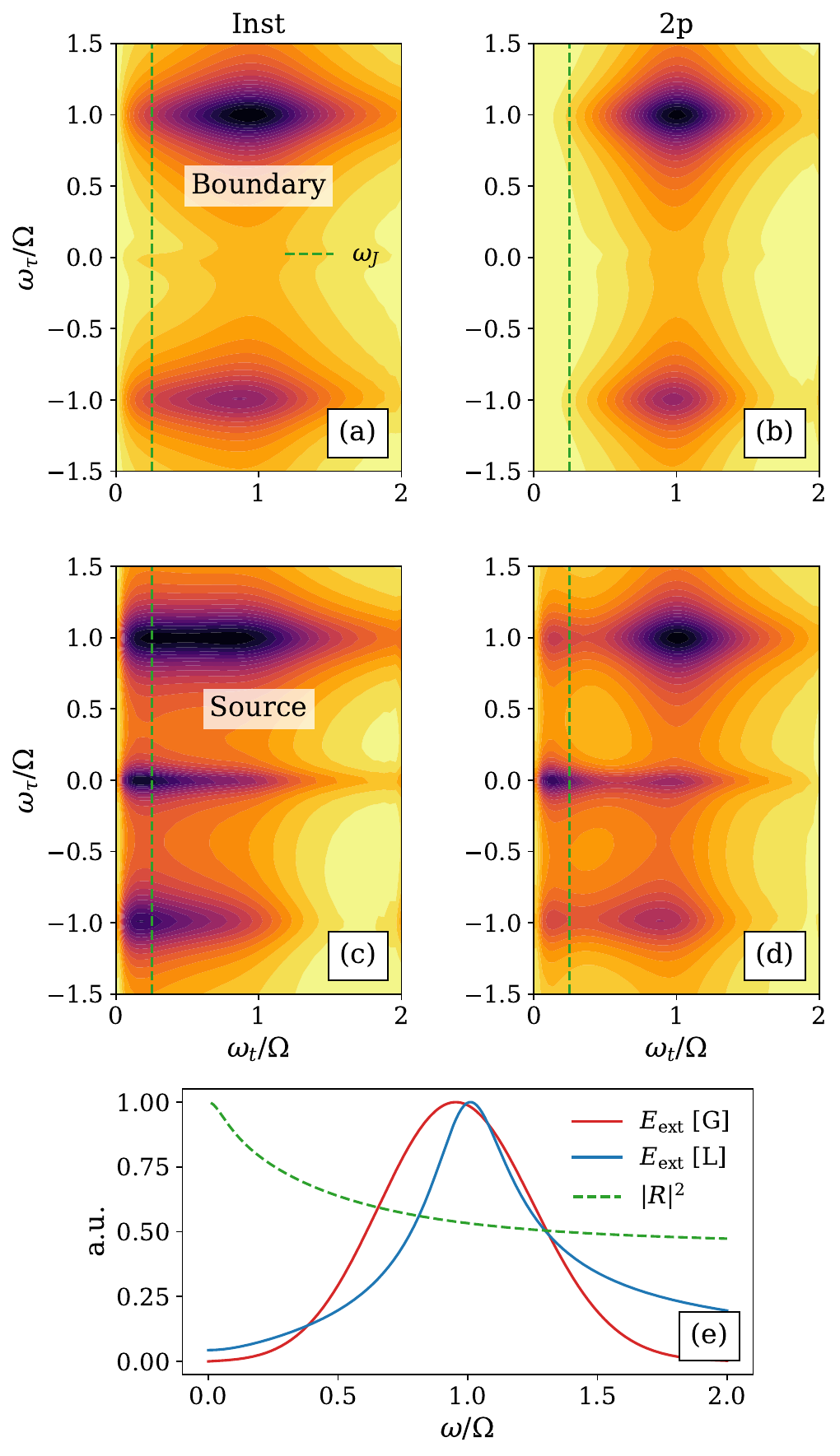}
		\caption{2D contour maps for the $E_{00\tau}$ signal, containing only the rephasing and non-rephasing contributions, obtained for the instantaneous (panels (a) and (c)) and two-plasmon (panels (b) and (d)) contributions using the screening procedure presented in this work (panels (a) and (b)) and the one in Ref.\ \cite{gomezsalvador_Phys.Rev.B24} (panels (c) and (d)). In the latter case a somehow qualitative difference between the two processes can be appreciated. (e) Comparison between the lorentzian external pulse (blue line) used for the 2D maps of panels (a)-(d), that coincides with the model adopted in Ref.\ \cite{gomezsalvador_Phys.Rev.B24}, and the gaussian one (red line) used in the simulations of  Fig.\ \ref{fig:inst2p} (red). We also superimposed the reflectance (green dashed line) corresponding to a choice of parameters in Eq.\ \ref{eq:epsilon} consistent with the one of Ref.\ \cite{gomezsalvador_Phys.Rev.B24}, i.e.\ $\omega_J/\Omega=0.25$, $\gamma/\Omega=1$ and $\epsilon_\infty=25$. {Additional simulations for reasonable variations of damping and pump spectrum are also shown in the SM \cite{fiore_}.}}
		\label{fig:gomez}
	\end{figure}

A somehow different conclusion on this respect has been reached recently in Ref.\ \cite{gomezsalvador_Phys.Rev.B24}, where the authors argue that a different behavior between the instantaneous (named mean-field) and the two-plasmon (named squeezing) process can observed for $T\sim T_c$. {There are two differences between our approach and the one of Ref.\ \cite{gomezsalvador_Phys.Rev.B24}. First, the authors model the external pulse with a Lorentzian instead of a Gaussian profile, see Fig.\ \ref{fig:gomez}(e). To make  the comparison with their result meaningful, we then repeat the computation of the 2D maps using the same pulse profile of Ref.\ \cite{gomezsalvador_Phys.Rev.B24}. The results are shown in Fig.\ \ref{fig:gomez}(a) and (b). Apart from a sharpening of the features with respect to Fig.\ \ref{fig:inst2p} due to the different choice of the pulse profile, the qualitative findings are the same. More specifically, the instantaneous and two-plasmon responses remain almost spectroscopically indistinguishable, in contrast to what stated in Ref.\ \cite{gomezsalvador_Phys.Rev.B24}. }

{In addition to a different modeling of the external pulse, the authors of Ref.\ \cite{gomezsalvador_Phys.Rev.B24}  propose a different description of the propagation effects, which are detailed in the Sec.\ IIC of the SM \cite{fiore_}. In short, the assumption there is to compute how the system responds to a homogeneous external current perturbation $J^{\text{ext}}(\omega)$ representing the applied electric field $E^{\text{ext}}(\omega)$, instead of solving the boundary-value problem leading to Eq.\ \ref{eq:genout}. This leads to a modified expression which does not contain the phase-matching denominator in Eq.\ \ref{eq:genout} and screens the internal fields by $1/\epsilon(\omega)$ rather than $1/(1+n(\omega))$. To test the impact of this approximation as compared to the expression in Eq.\ \ref{eq:genout} we then implement it and reproduce in Fig.\ \ref{fig:gomez}(c)-(d) the 2D maps of Ref.\ \cite{gomezsalvador_Phys.Rev.B24}. Let us first focus on the comparison between panels (a) and (b), where the computation is done with the featureless instantaneous kernel, so that different spectral signatures can only be ascribed to propagation effects. In the approach based on the general formula $\ref{eq:genout}$ the spectral weight of the THz pulse inside the system remains localized around the central pump frequency, see Fig.\ \ref{fig:gomez}(a). In contrast, when one approximate screening effects as done in Ref.\ \cite{gomezsalvador_Phys.Rev.B24} the spectral weight of the internal field extends towards $\omega_t \sim 0$, see Fig.\ \ref{fig:gomez}(c). On the other hand, the two approaches to the screening give more similar results for what concerns the two-plasmon process, shown in Fig.\ \ref{fig:gomez}(b) in the case of the simulation based on Eq.\ \ref{eq:genout}, and in Fig.\ \ref{fig:gomez}(d) for the approximation used in Ref.\ \cite{gomezsalvador_Phys.Rev.B24}. Indeed, for the two-plasmon process the $K^{2\text{p}}$ kernel filters out the spectral components of the driving fields around the resonance at $2\omega_J$, making the final result less sensitive to the modeling of the internal field.}

{The outcome of the comparison shown in Fig.\ \ref{fig:gomez} is that when screening effects are particularly pronounced some extra care should be taken in their modeling. Indeed, by following the approach of Ref.\ \cite{gomezsalvador_Phys.Rev.B24} one can in principle identify spectroscopically different features between the instantaneous and two-plasmon process, since the latter, in Fig.\ \ref{fig:gomez}(d) shows well defined spots, absent in the former, in Fig.\ \ref{fig:gomez}(c). In other words, the experiments could discriminate among the two, as indeed suggested by the authors of Ref.\ \cite{gomezsalvador_Phys.Rev.B24}. However, this conclusion is at odd with the results obtained with the formula based on the boundary conditions in Eq.\ \ref{eq:genout}, where both processes lead to extremely similar maps, see Fig.\ \ref{fig:gomez}(a) and (b). At present, the most promising experimental route to distinguish the two processes relies on the identification of the frequency resonance of the two-plasmon kernel  as a temperature enhancement\cite{fiore_Phys.Rev.B24}, as observed e.g. in Ref.\ \cite{kaj_Phys.Rev.B23} by THG measurement with multicycle pulses. }
\section{Discussion and Conclusions}\label{sec:con}

The general formalism presented in Sec.\ \ref{sec:mod} and Sec.\ \ref{sec:prop} aims to provide a theoretical framework that allows us to extract from the experimental 2DTS maps the crucial physical ingredients behind them.  These include two complementary aspects: the first one is the theoretical computation of the nonlinear $\chi^{(3)}$ optical kernel, which for electronic systems amounts to computing correlation functions that in the diagrammatic language correspond to bubbles with multiple (up to four) photon vertices. These have been classified as paramagnetic-like and diamagnetic-like, referring to microscopic light-matter processes where the particle-hole excitations are driven by a single photon or two photons, respectively. As we have seen, when the pump pulse is short enough in time, so that its frequency spectrum is relatively broad on the energy scales relevant for microscopic excitations, the two classes of processes lead to specific and distinct signatures in the 2DTS. These general features, summarized in Table \ref{tab:recap}, can be understood from the set of equations Eqs.\ \ref{eq:scheme}-\ref{eq:jdiatot2}, where we formulate the 2D problem by disentangling the detection structure, carried out by the dependence in $(\omega_t, \omega_\tau)$,  from the energy spectrum of the excitations, encoded in the form of the energy prefactors in Eqs.\ \ref{rhointer}-\ref{rhomix} and \ref{rhodia}. In the present manuscript we applied the formalism to a {toy model reproducing a semiconducting-like band structure.} We worked out several semi-analytical expressions for the optical kernel and for the convolution with gaussian light pulses. {This approximation allowed us to make the numerical computation significantly faster, especially in the present case where the density of states of the fermionic excitations admits an analytical representation. Even though the experimental pulses are not necessarily gaussian, and realistic band structures can be more complex, the present results provided us with a clear {\em interpretation} of the numerically-obtained 2D maps,} which had been already attempted by some authors \cite{mootz_CommunPhys22,mootz_Phys.Rev.B24,valmispild_Nat.Photon.24,chen_npjComputMater25,ono_Phys.Rev.Lett.25,chen_25,tsuji_25}. An example of future applications of these results to realistic cases is provided, e.g.\, by the recent measurements of the nonlinear response of superconducting NbN in Ref.\ \cite{katsumi_Phys.Rev.Lett.24}. In that work the possible predominance of paramagnetic processes over the diamagnetic ones, that is expected on the basis of theoretical analysis of the nonlinear kernel, has been deduced from the temperature dependence of the nonlinear first-harmonic measurements, rather than from the structure of the 2D maps. The present formalism offers instead the possibility to infer this information directly from the analysis of the 2D maps at a given temperature, and possibly to distinguish {\em spectroscopically} pure electronic processes from those mediating collective modes, as, e.g.\, the Higgs mode. In fact, despite the fact that this issue has been extensively addressed theoretically, a clear experimental signature of the different excitation pathways is still lacking in single-pulse experiments, while recent experimental findings \cite{kim_Sci.Adv.24,katsumi_Phys.Rev.Lett.24,yuan_24,huang_Sci.Adv.25,katsumi_Phys.Rev.Lett.25} suggest that it could be captured by two-pulse protocols. {We observe that the present formalism can also be applied in presence of the lower-order $\chi^{(2)}$ susceptibility, that is the leading nonlinear process in system lacking inversion symmetry, due to lattice or magnetic effects. Also in this case the response will be built upon paramagnetic and diamagnetic light-matter vertices, leading to potentially different spectroscopic signatures. Another interesting consequence is the possibility of extracting the \textit{full} frequency dependence of $\chi^{(2)}(\omega_1,\omega_2)$ from the 2D map, since the procedure described in Sec.\ \ref{sec:gen} would now saturate all the internal frequency integrals, while with $\chi^{(3)}$ there is always a residual $\omega-$integral to perform in Eqs.\ \ref{eq:j00tau} and \ref{eq:j0tautau}.}

A second aspect concerns a proper treatment of the possible mixing of the electromagnetic waves with dipolar modes of the material. This is a well-known effect already present in linear response, that is usually described at the level of Maxwell's equations and Fresnel boundary conditions, which condense all the complications of the polariton formation in a frequency-dependent refractive index $n(\omega)$. However, for nonlinear response the same exact treatment is not possible, since Maxwell's equations themselves are not linear. To overcome this limitation we discussed the outcomes of an iterative procedure, based again on a perturbative approach to the linear response. The generic 2DTS map must then be computed using Eq.\ \ref{eq:genout}, or its generalization to transmission geometry. Such a tedious procedure turns out to be necessary, since propagation effects can dominate the features of the experimental response, masking microscopic mechanisms manifest via the $\chi^{(3)}$ optical kernel. As a case study we applied the formalism to the 2DTS of soft Josephson plasmons in layered superconductors, and we showed that in this case we cannot find a way to distinguish experimentally the two nonlinear contributions expected theoretically, corresponding to the presence or absence of a two-plasmon intermediate process. Such a conclusions is somehow at odd with previous theoretical results of Ref.\ \cite{gomezsalvador_Phys.Rev.B24}, due to a different approach to the problem of the screening. 

\begin{figure}
\centering
\includegraphics[width=\columnwidth]{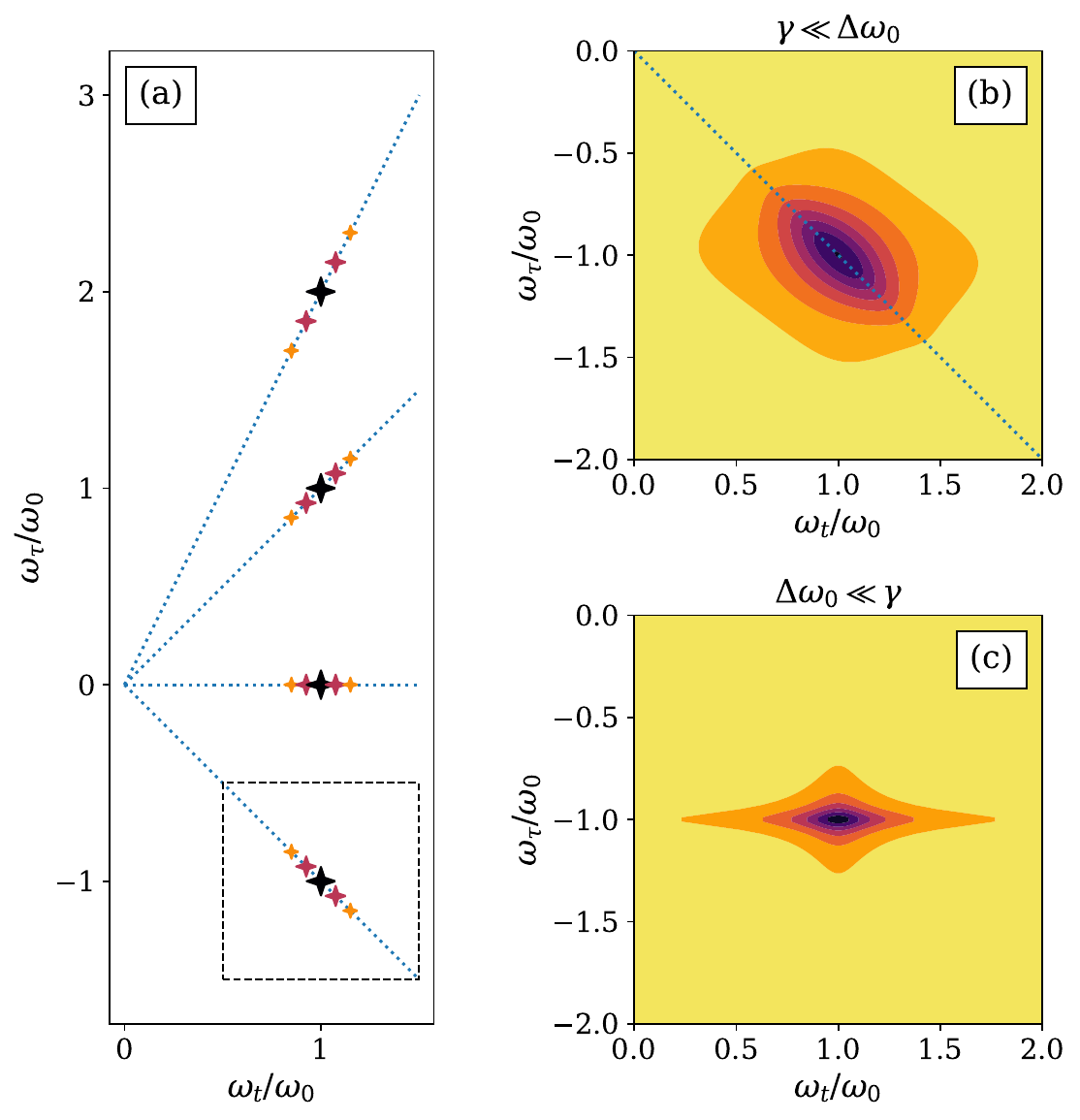}	
\caption{A sketch showing the effect inhomogeneous broadening. Panel (a): a collection of non-interacting dipoles (orange stars) with an energy distribution centered around $\omega_0$ produces a sliding of the four peaks foreseen in Fig.\ \ref{fig:imp} along specific lines (blue dotted lines), with an envelope sensitive to the width of the distribution $\Delta\omega_0$. In the case where the intrinsic mode damping $\gamma$ is small with respect to $\Delta\omega_0$, panel (b), an elongation of the rephasing peak along $\omega_\t=-\omega_t$ is observed. In the opposite limit $\Delta\omega_0\ll\gamma$, panel (c), the mode has a star-like shape \cite{siemens_Opt.ExpressOE10}.}
\label{fig:inho}
\end{figure}	

As we discussed in Sec.\ \ref{sec:imp}, the present diagrammatic approach can be used also to address the well-studied case of 2DTS in two-level systems, by reproducing the results usually obtained  via a time-dependent density-matrix formalism. An interesting question is then how to incorporate in the many-body language the description of typical effects emerging within this context, the primary one being the role of {\em inhomogeneity}. Indeed, for molecular systems 2DTS can distinguish spectroscopically the signatures connected to the intrinsic lifetime $1/\gamma$ of a single mode from the ones due to the existence of a finite distribution $\rho$ of possible values for the mode excitation energy $\omega_0$. As discussed in Ref.\ \cite{siemens_Opt.ExpressOE10,liu_npjQuantumMater.25} the two effects appear as a broadening of the rephasing peak in different directions: while $\Delta \omega_0$ mostly controls the peak width along the diagonal of the lower half plane, $\omega_\tau=-\omega_t$, $\gamma$ is determines its width in the perpendicular direction see Fig.\ \ref{fig:inho}. This effect can be reproduced by a phenomenological formula for the spectrum of the rephasing peak:
\be
P^{\text{rep}}(\omega_t,\omega_\tau)\sim\int d\omega_0\, \rho{(\omega_0)}\frac{1}{\omega_t-\omega_0+i\gamma}\frac{1}{\omega_\t+\omega_0+i\gamma},
\ee
where the choice of a gaussian envelope for $\rho(\omega_0)$ leads to the same findings of Ref.\ \cite{siemens_Opt.ExpressOE10}. A natural question is how much of this phenomenology, that can be easily understood within a single-mode picture, survives and can be observed in dispersive electronic systems and in realistic situations where also propagation effects must be taken into account. A somehow similar signature is indeed already present in our non interacting electronic model considered in Section \ref{sec:mod}. If we examine the difference between the nondispersive and dispersive limits of the paramagnetic process in Fig.\ \ref{fig:semipara}(c) and (f), we see that the tensorial density of states $\rho^{\text{para}}$ in Eq.\ \ref{eq:tenspara} acts as an effective inhomogeneous broadening of the $2\Delta$ resonance in momentum space, producing an elongation of the various peaks related to the finite {bandwidth set by $\bar{t}$}. In particular, the rephasing peak acquires an elongation in the antidiagonal direction analogous to the one of Fig.\ \ref{fig:inho}(b).

The renewed interest in this problem has been triggered by the 2DTS measurements of the Josephson plasmon discussed in Sec\ \ref{sec:prop}: indeed, the measured spectra show an elongation of the non-rephasing peak at $(\omega_J,-\omega_J)$ in the diagonal direction, which is not reproduced by the simulations of Fig.\ \ref{fig:inst2p}. At the level of the formalism of the present manuscript, one could easily account for this effect by phenomenologically mediating the 2DTS spectra for a single realization of the $\omega_J$ value over a gaussian distribution of possible $\omega_J$ values with variance $\Delta \omega_J$. However, such a procedure completely misses the answer to a rather profound question: how a microscopic distribution of local plasma values $\omega_J$ impacts on the macroscopic propagation of electromagnetic waves in the medium, leading to signatures in the linear and nonlinear response. Valuable attempts have already been made to answer this question for the linear \cite{vandermarel_CzechJPhys96,dordevic_Phys.Rev.Lett.03} and nonlinear \cite{salvador_25} responses. Still, a clear understanding of the additional effects due to propagation in a dipolar inhomogeneous system is lacking, leaving open the question of the real potential of 2DTS to distinguish spectroscopically intrinsic mode lifetimes from emergent space inhomogeneities. 

A related open question concerns the recent proposal \cite{barbalas_Phys.Rev.Lett.25,chaudhuri_25} concerning the ability of 2DTS to identify energy-relaxation channels, absent in the linear optical response, which is only sensitive to momentum-relaxation processes. So far, the persistence of such slow-relaxation processes has been reported for a variety of conventional and unconventional metals, but a clear formulation of the problem in terms of microscopic processes relevant for this specific nonlinear optical effect is still in its infancy \cite{kryhin_Phys.Rev.Lett.25}. The results of the present manuscript represent a first step in this direction, offering a novel route to bridge the experimental results based on multidimensional THz spectroscopy to the theoretical investigation of interaction processes not accessible in linear response, with the final goal of understanding the microscopic mechanisms responsible for the observed physical effects. 

\begin{acknowledgments}

We are grateful to N. P. Armitage for illuminating exchanges and a critical reading of the manuscript. We thank C. Castellani, A. G\'{o}mez-Salvador, K. Katsumi, A. Liu and G. Seibold for useful discussions and suggestions. This work has been supported by the EU under the project MORE-TEM ERC-SYN (No.\ 951215), by ICSC–Centro Nazionale di Ricerca in High Performance Computing, Big Data and Quantum Computing, funded by European Union – NextGenerationEU, by Sapienza University under the projects Ateneo (RM123188E357C540, RP124190A63FAA97 and AR224190752E4212) and by the Italian MIUR under the project PRIN2022-CoInEx (2022WS9MS4).

\end{acknowledgments}


%

\clearpage
\onecolumngrid
\setcounter{secnumdepth}{3}
\setcounter{section}{0}\setcounter{equation}{0}\setcounter{figure}{0}\setcounter{table}{0}
\renewcommand{\thesection}{S\arabic{section}}
\renewcommand{\thesubsection}{S\arabic{section}\Alph{subsection}}
\renewcommand{\theequation}{S\arabic{equation}}
\renewcommand{\thefigure}{S\arabic{figure}}
\renewcommand{\thetable}{S\arabic{table}}
\begin{center}
  {\large\bfseries Supplementary Material:\\[2pt]
  Two-dimensional THz spectroscopy in electronic systems:\\ a many-body diagrammatic approach}
\end{center}
\vspace{1em}
\section{Full $K^{(3)}$ response}

    In this section, we provide the derivation for all the processes contributing to the third-order response kernel of the CDW model in the main text. For each of them, we will provide a symmetrized expression over the internal frequencies and the corresponding behavior as a function of $(\omega_t,\omega_\tau)$ in the 2D plane. In the following we use Gaussian atomic units.

    The key quantity to be derived is the partition function
    \be
    Z=\int\mathcal{D}\psi\mathcal{D}\bar{\psi}\,e^{-S_0[\psi,\bar{\psi}]},
    \ee
	written in terms of the effective action
    \be
    S_0[\psi,\bar{\psi}]=\int_0^\beta d\tau\, \sum_\mathbf{k}\bar{\psi}_\mathbf{k}(\tau)
    \begin{pmatrix}
    \partial_\tau+\epsilon_\mathbf{k}&\Delta\\
    \Delta & \partial_\tau+\epsilon_{\mathbf{k}+\mathbf{Q}}
    \end{pmatrix}
    \psi_\mathbf{k}(\tau),
    \ee
    where we promoted the operators $\Psi_\mathbf{k}^\dagger,\Psi_\mathbf{k}$, used in the main text to write the CDW hamiltonian, to Grassmann variables $\bar{\psi}_\mathbf{k}(\tau)=
\begin{pmatrix}
\bar{c}_{\mathbf{k}}(\tau)&\bar{c}_{\mathbf{k}+\mathbf{Q}}(\tau)\\
\end{pmatrix}
$ and $\psi_\mathbf{k}(\tau)$, respectively. In terms of fermionic Matsubara frequencies\footnote{The conventions for transforming $\psi(\tau)\leftrightarrow\psi(i\omega_n)$ foresee a $1/\sqrt{\beta}$ in both directions. For the gauge field $A$ and the current $J$, instead, there is a $1/\beta$ when transforming from $A(i\omega_n)$ to $A(\tau)$. This ensures dimensional consistency when the sum over bosonic Matsubara frequencies is extended to a continuous variable \cite{kopnin_01SMREF}.} $i\omega_n=(2n+1)\pi T$ one then gets, defining $k=(i\omega_n,\mathbf{k})$
    \be
    S_0[\psi,\bar{\psi}]=-\sum_k\bar{\psi}_kG_0^{-1}(k)\psi_k,
    \ee
    with $G_0(k)$ defined as in the main text:
    \be
    G_0(i\omega_n,\mathbf{k})=
    \begin{pmatrix}
    i\omega_n-\epsilon_\mathbf{k}&-\Delta\\
    -\Delta&i\omega_n+\epsilon_\mathbf{k}
    \end{pmatrix}^{-1}.
    \ee
    
    For uniform perturbations, the Peierls substitution amounts to replacing $\epsilon_{\mathbf{k}}\to\epsilon_{\mathbf{k}-e/c\mathbf{A}(\tau)}$, where $e=-1$ is the electron charge and $\mathbf{A}(\tau)$ is the applied time-dependent vector potential. The effective action now depends on the vector potential and can be expanded in powers of $\mathbf{A}$. One has, setting $k^\prime=(i\omega_m,\mathbf{k}^\prime)$,
    \be
    \label{eq:sa}
    S[\psi,\bar{\psi},\mathbf{A}]=S_0[\psi,\bar{\psi}]+\sum_{kk^\prime}\bar{\psi}_{k^\prime}\sigma_3\Sigma_A(k^\prime-k)\psi_k,
    \ee
    where $\Sigma_A(k^\prime-k)=\delta_{\mathbf{k}\mathbf{k}^\prime}\Sigma_A(i\Omega=i\omega_m-i\omega_n,\mathbf{k})$, with $i\Omega=i\omega_m-i\omega_n=2(m-n)\pi T$ is now a bosonic Matsubara frequency. Being $S$ still quadratic for the $\psi$ variables, it is now possible to integrate out the fermionic degrees of freedom exactly, getting 
    \be
    \label{eq:expa}
    S[\mathbf{A}]=\Tr\ln{(1-G_0\Sigma_A\sigma_3)}=\sum_n\frac{1}{n}\Tr_k{(G_0\sigma_3\Sigma_A)^n},
    \ee
    where we denoted by $\Tr$ the trace over momenta, Matsubara frequencies and Nambu degrees of freedom. When we take the trace of the term of $n$-th order in Eq.\ \ref{eq:expa}, we effectively contract the product of $n$ fermionic Green’s functions with the matrix $\sigma_3$. This contraction generates closed fermionic loops with insertions of $n$ vertices, each carrying a different power of the gauge field. The resulting expression can be organized as an expansion in powers of $\mathbf{A}$, whose coefficients correspond to fermionic susceptibilities. Since we are interested in the current up to the third order in the gauge field, we need to expand $S$ up to the fourth order in $\mathbf{A}$. Consequently, the relevant terms in the expansion of the self-energy $\Sigma_A$ in Eq.\ \eqref{eq:sa} are
	\be
	\label{eq:sesemi-sm}
	\begin{split}
		\beta\Sigma_A(i\Omega,\mathbf{k})=&-\frac{e}{c}\sum_{\alpha_1}\frac{\partial\epsilon_{\mathbf{k}}}{\partial k_{\alpha_1}}A_{\alpha_1}(i\Omega)+\frac{1}{2}\left(\frac{e}{c}\right)^2\sum_{\alpha_1,\alpha_2}\frac{\partial^2\epsilon_{\mathbf{k}}}{\partial k_{\alpha_1}\partial k_{\alpha_2}}A^2_{\alpha_1\alpha_2}(i\Omega)\\
		&-\frac{1}{6}\left(\frac{e}{c}\right)^3\sum_{\alpha_1,\alpha_2,\alpha_3}\frac{\partial^3\epsilon_{\mathbf{k}}}{\partial k_{\alpha_1}\partial k_{\alpha_2}\partial k_{\alpha_3}}A^3_{\alpha_1\alpha_2\alpha_3}(i\Omega)\\
		&+\frac{1}{24}\left(\frac{e}{c}\right)^4\sum_{\alpha_1,\alpha_2,\alpha_3,\alpha_4}\frac{\partial^4\epsilon_{\mathbf{k}}}{\partial k_{\alpha_1}\partial k_{\alpha_2}\partial k_{\alpha_3}\partial k_{\alpha_4}}A^4_{\alpha_1\alpha_2\alpha_3\alpha_4}(i\Omega),
	\end{split}
	\ee
	with the same notation adopted in the main text for $A^n_{\alpha_1\dots\alpha_n}(i\Omega)$. Notice that in full generality one can write 
    \be
    \label{eq:funa}
    A^n_{\alpha_1\dots\alpha_n}(i\Omega)=\frac{1}{\beta^{n-1}}\sum_{i\Omega_1\dots i\Omega_n}A_{\alpha_1}(i\Omega_1)\cdots A_{\alpha_n}(i\Omega_n)\delta\left(i\Omega-i\Omega_{1\dots n}\right),
    \ee
    where $i\Omega_{1\dots n}=i\Omega_1+\dots+i\Omega_n$. 

    \begin{figure}
		\centering
		\includegraphics[width=0.5\textwidth]{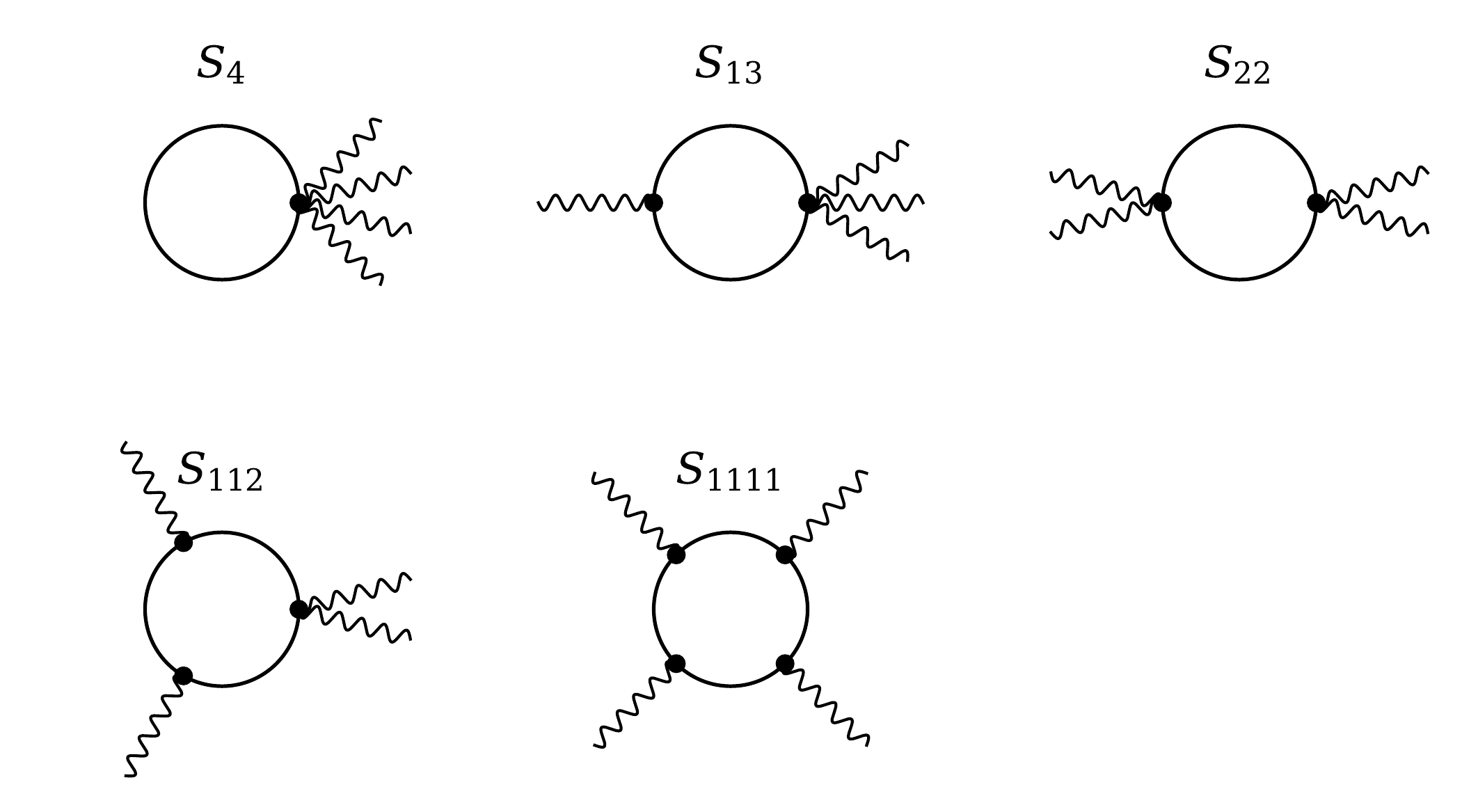}
		\caption{The set of diagrams presented in Eq.\ \ref{eq:sesemi-sm}. Wavy lines denote the gauge field, bullets denote the $\sigma_1$ vertex connecting the electronic propagators in Nambu space, represented by solid lines.}
		\label{fig:diasemi-sm}
	\end{figure}
    
    Denoting the previous four contributions to the self-energy as $\Sigma_A^{(1)},\dots, \Sigma_A^{(4)}$, one can generate the contributions to the fourth order $S[A^4]$. They can be classified according to the frequency structure of the corresponding Feynman diagram, see Fig.\ \ref{fig:diasemi-sm}. There is the instantaneous contribution
	\be
    \label{eq:sinst}
	S_4=\sum_{k}\Tr\left[G_0(i\omega_n,\mathbf{k})\sigma_3\right]\Sigma_A^{(4)}(0,\mathbf{k}),
	\ee
    which has the structure of a vacuum polarization. Then we find the two-point, Kubo-like responses $S_{13}$ and $S_{22}$ (fully-diamagnetic). They read
    \begin{align}
    \label{eq:dia}
    S_{22}&=\frac{1}{2}\sum_{k}\sum_{i\Omega}\Tr\left[G_0(i\omega_n,\mathbf{k})\sigma_3G_0(i\omega_n+i\Omega,\mathbf{k})\sigma_3\right]\Sigma_A^{(2)}(i\Omega,\mathbf{k})\Sigma_A^{(2)}(-i\Omega,\mathbf{k}),\\
    \label{eq:s13}
    S_{13}&=\frac{1}{2}\cdot 2 \sum_{k}\sum_{i\Omega}\Tr\left[G_0(i\omega_n,\mathbf{k})\sigma_3G_0(i\omega_n+i\Omega,\mathbf{k})\sigma_3\right]\Sigma_A^{(1)}(i\Omega,\mathbf{k})\Sigma_A^{(3)}(-i\Omega,\mathbf{k}).
	\end{align}
    Both contributions contain the factor $1/2$ coming from the $n=2$ term in Eq.\ \ref{eq:expa} that generates two-point responses; in addition, $S_{13}$ contains a factor $2$ due to the equivalent contribution where $\Sigma_A^{(3)}$ and $\Sigma_A^{(1)}$ are swapped. Next in line, we have the three-point response $S_{112}$ $(i\Omega_{12}=i\Omega_1+i\Omega_2)$
	\begin{align}
	S_{112}=\frac{1}{3}\cdot3\sum_{k}\sum_{i\Omega_1,i\Omega_2}&\Tr\left[G_0(i\omega_n,\mathbf{k})\sigma_3G_0(i\omega_n+i\Omega_1,\mathbf{k})\sigma_3G_0(i\omega_n+i\Omega_{12},\mathbf{k})\sigma_3\right]\nonumber\\
    &\times\Sigma_A^{(1)}(i\Omega_1,\mathbf{k})\Sigma_A^{(1)}(i\Omega_2,\mathbf{k})\Sigma_A^{(2)}(-i\Omega_{12},\mathbf{k}),
	\end{align}
    where the factor $1/3$ comes again from Eq.\ \ref{eq:expa}  and the multiplicity $3$ is due to the three distinct but equivalent ways to obtain a term with two $\Sigma_A^{(1)}$ and one $\Sigma_A^{(2)}$ insertion. Finally, we have the four-point (fully-paramagnetic) response which is written $(i\Omega_{123}=i\Omega_1+i\Omega_2+i\Omega_3)$
	\begin{align}
		S_{1111}=\frac{1}{4}\sum_{k}\sum_{i\Omega_1,i\Omega_2,i\Omega_3}&\Tr\left[G_0(i\omega_n,\mathbf{k})\sigma_3G_0(i\omega_n+i\Omega_1,\mathbf{k})\sigma_3G_0(i\omega_n+i\Omega_{12},\mathbf{k})\sigma_3G_0(i\omega_n+i\Omega_{123},\mathbf{k})\sigma_3\right]\nonumber\\
        &\times\Sigma_A^{(1)}(i\Omega_1,\mathbf{k})\Sigma_A^{(1)}(i\Omega_2,\mathbf{k})\Sigma_A^{(1)}(i\Omega_3,\mathbf{k})\Sigma_A^{(1)}(-i\Omega_{123}).
	\end{align}
    As we did in the main text, we also introduce the $n$-point correlator that is needed to express the various contributions to $K_{\alpha\beta\gamma\delta}^{(3)}$ obtained above by expanding the effective action \eqref{eq:sa}. Since each gauge-field vertex always carries a $\sigma_3$ insertion in Nambu space, we define:
    \be
    \label{eq:corr}
    X^{(n+1)}(i\Omega_1,\dots,i\Omega_n,\mathbf{k})=\frac{1}{\beta}\sum_{i\omega_m}\Tr{\left[G_0(i\omega_m,\mathbf{k})\sigma_3G_0(i\omega_m+i\Omega_1,\mathbf{k})\sigma_3\dots G_0(i\omega_m+i\Omega_{1\dots n},\mathbf{k})\sigma_3\right]}. 
    \ee
    Let us now analyze how the various pieces generate a corresponding contribution to the nonlinear kernel and 2D map. To get the current density one has to perform the following derivative\footnote{Strictly speaking, this is not a functional but an ordinary derivative since $i\Omega$ are discrete. Nevertheless, we keep the same symbol for notational consistency with the corresponding expressions in real time. Notice also that the presence of the $\beta$ factor is reabsorbed in the analytical continuation.}
    \be
    \label{eq:derj}
    J_\alpha(i\Omega_t)=\frac{\beta}{V}\left.\frac{\delta S[\mathbf{A}]}{\delta A_\alpha(-i\Omega_t)}\right|_{A=0}.
    \ee
    The third-order nonlinear kernel can analogously be defined as
    \be
    \label{eq:der4}
    \bar{K}_{\alpha\beta\gamma\delta}^{(3)}(i\Omega_t;i\Omega_1,i\Omega_2,i\Omega_3)=\frac{\beta^4}{V}\left.\frac{\delta^4 S[\mathbf{A}]}{\delta A_\alpha(-i\Omega_t)\delta A_\beta(i\Omega_1)\delta A_\gamma(i\Omega_2)\delta A_\delta(i\Omega_3)}\right|_{A=0},
    \ee
    where we assume that in each fermionic loop we have three incoming frequencies $i\Omega_{i}$, with $i=1,2,3$, and one outgoing frequency $i\Omega_t$. The nonlinear current can then be expressed as a function of the third-order kernel for an arbitrary field spectrum after analytical continuation $i\Omega_i\to\omega_i+i\eta$ \cite{rostami_AnnalsofPhysics21SMREF,kopnin_01SMREF}. In particular, $i\Omega_t$ will be analytically continued to the detection frequency $\omega_t+i\eta$ of the 2DCS protocol, while each incoming frequency will be continued as $i\Omega_i\to\omega_i+i\eta$. 
    
    We notice that the kernel obtained from the derivative in Eq.\ \ref{eq:der4} contains an overall Dirac delta $\beta\delta(i\Omega-i\Omega_{123})\to\delta(\omega-\omega_{123})$ enforcing energy conservation. In our general treatment in the first section of the manuscript we singled out that Dirac delta from the third-order response function to understand the behavior of the 2D maps. For this reason, it is useful to define
    \be
    \label{eq:kbar}
    \bar{K}_{\alpha\beta\gamma\delta}^{(3)}(\omega_t;\omega_1,\omega_2,\omega_3)=6K^{(3)}_{\alpha\beta\gamma\delta}(\omega_1,\omega_2,\omega_3)\delta\left(\omega_t-\omega_{123}\right),
    \ee
    that allows us to express the relation between current and gauge field in the way that was introduced in the main text:
    \be
	\label{eq:gen}
	J^{(3)}_\alpha(\omega_t)=\int d\omega_i\,K^{(3)}_{\alpha\beta\gamma\delta}(\omega_1,\omega_2,\omega_3)A_\beta(\omega_1)A_\gamma(\omega_2)A_\delta(\omega_3)\delta\left(\omega_t-\omega_{123}\right).
	\ee
    Note also that the multiplicity $3!=6$ in the exchange of the internal tensorial and frequency indices is reabsorbed in the definition of $K^{(3)}$.

    

    For simplicity, we will now stick to the one-dimensional chain and neglect the tensorial nature of the response, the generalization for the square lattice being straightforward and provided later. 
    To construct the 2D map we should further make the dependence on $\tau$ of the $A(\omega_i)$ fields in Eq.\ \eqref{eq:gen} explicit, and Fourier transform with respect to that variable. Using $A(\omega_,\tau)=A_0(\omega)+e^{-i\omega\tau}A_\tau(\omega)$, as we explained in the main text, the contributions to the 2D map can be conveniently classified as: 
    \be
    \label{eq:all}
    J^{(3)}_{2D}=J^{(3)}_{00\tau}+J^{(3)}_{0\tau\tau}.
    \ee
    The first term corresponds to the nonlinear current generated by two interactions with the field $A_0$ and one with the field $A_{\tau}$, which we will denote by $J^{(3)}_{00\tau}$. The Fourier transform with respect to $\omega_\tau$ leads to the assignment $\omega_i\to \omega_\tau$, leaving only one integration left in Eq.\ \eqref{eq:gen}. The second term of Eq.\ \eqref{eq:all} corresponds to two interactions with field $A_{\tau}$ and one with field $A_0$, leading to the assignment $\omega_i\to \omega_t-\omega_\tau$, which we will denote accordingly as $J^{(3)}_{0\tau\tau}$. Their expression is, as a function of $\omega_t$ and $\omega_\tau$,
    \begin{align}
    \label{eq:j00tau-sm}
    J^{(3)}_{00\tau}(\omega_t,\omega_\tau)&=A_\tau(\omega_\tau)\int d\omega K^{(3)}_{\mathcal{S}}(\omega_\tau,\omega,\omega_t-\omega_\tau-\omega)A_0(\omega)A_0(\omega_t-\omega_\tau-\omega),\\
    \label{eq:j0tautau-sm}
    J^{(3)}_{0\tau\tau}(\omega_t,\omega_\tau)&=A_0(\omega_t-\omega_\tau)\int d\omega\, K^{(3)}_{\mathcal{S}}(\omega_t-\omega_\tau,\omega,\omega_\tau-\omega)A_\tau(\omega)A_\tau(\omega_\tau-\omega).
    \end{align}
    We assume as in the main text equal field envelopes $A_0(\omega)=A_\tau(\omega)=A(\omega)$, and we introduce the symmetrized form of nonlinear kernel, namely
    \be
    \label{eq:chi3sy-sm}
    K^{(3)}_{\mathcal{S}}(\omega_\tau,\omega,\omega_t-\omega_\tau-\omega)=K^{(3)}(\omega_\tau,\omega,\omega_t-\omega_\tau-\omega)+K^{(3)}(\omega,\omega_\tau,\omega_t-\omega_\tau-\omega)+K^{(3)}(\omega,\omega_t-\omega_\tau-\omega,\omega_\tau).
    \ee
    We also notice that by construction $K^{(3)}(\omega_1,\omega_2,\omega_3)$ obtained from Eq.\ \ref{eq:kbar} will be symmetric under any permutation of $\omega_1,\omega_2,\omega_3$, which means that $K_\mathcal{S}^{(3)}=3K^{(3)}$.
    
    We can now proceed with the evaluation of the diagrams. We introduce the compact notation $J^{[m;n\dots]}$ and $K^{[m;n\dots]}$ to separate the various contributions to the nonlinear third-order current and kernel, respectively, where $m$ denotes the vertex $\Sigma_A^{(m)}$ with respect to which the functional derivative $A(-i\Omega_t)$ is performed and $n\dots$ is a string containing the remaining vertices. This notation will be made evident in the following Figs.\ \ref{fig:k13}-\ref{fig:k112} by denoting by a dashed line the $A(-i\O_t)$ line carrying the outgoing detection frequency $\o_t$ and by solid lines the remaining ones. 

    \subsection{\label{sec:k4}$S_4$ diagram}

    The easiest contribution is the instantaneous one obtained from Eq.\ \eqref{eq:sinst}. Here, the $\Sigma_A^{(4)}$ structure forces the total frequency flowing in the loop to be zero since  
    \be
    A^{4}(i\Omega=0)=\frac{1}{\beta^3}\sum_{i\Omega_1,i\Omega_2,i\Omega_3}A(i\Omega_1)A(i\Omega_2)A(i\Omega_3)A(-i\Omega_{123}).
    \ee
    We are then left with the vacuum bubble, see Eq.\ \ref{eq:corr} for $n=0$
    \be
    X^{(1)}(0,\mathbf{k})=\frac{1}{\beta}\sum_{i\omega_n}\Tr{\left[G_0(i\omega_n,\mathbf{k})\sigma_3\right]}=-\frac{\epsilon_\mathbf{k}}{E_\mathbf{k}}\left(1-2f_\mathbf{k}\right).
    \ee
    The corresponding contribution to the third-order nonlinear kernel will be frequency independent, as appropriate for this kind of process. It reads explicitly
    \be
	K^{[4;]}(\omega_1,\omega_2,\omega_3)=-\frac{1}{6V}\sum_\mathbf{k}\frac{\partial^4 \epsilon_\mathbf{k}}{\partial k^4}\frac{\epsilon_\mathbf{k}}{E_\mathbf{k}}\left(1-2f_\mathbf{k}\right),
	\ee
	where $1/24$ from the expansion of the Peierls phase in $\Sigma_A^{(4)}$ is canceled out by the functional derivative with respect to the gauge field $A$ in the expression for $A^{4}(i\Omega=0)$, see Eq.\ \ref{eq:der4}. The factor $1/6$ comes from the definition of Eq.\ \ref{eq:kbar} and will be present as an overall factor in any contribution to $K^{(3)}$. When plugged into Eqs.\ \ref{eq:j00tau-sm} and \ref{eq:j0tautau-sm} to get the behavior of the 2D map, we get the pieces
	\begin{align} 
	J^{[4;]}_{00\tau}(\omega_t,\omega_\tau)&=-\frac{1}{2V}\sum_\mathbf{k}\frac{\partial^4 \epsilon_\mathbf{k}}{\partial k^4}\frac{\epsilon_\mathbf{k}}{E_\mathbf{k}}\left(1-2f_\mathbf{k}\right)A(\omega_\tau)A^2(\omega_t-\omega_\tau),\\
	J^{[4;]}_{0\tau\tau}(\omega_t,\omega_\tau)&=-\frac{1}{2V}\sum_\mathbf{k}\frac{\partial^4 \epsilon_\mathbf{k}}{\partial k^4}\frac{\epsilon_\mathbf{k}}{E_\mathbf{k}}\left(1-2f_\mathbf{k}\right)A(\omega_t-\omega_\tau)A^2(\omega_\tau).
	\end{align}
    The factor $1/2=3\cdot1/6$ comes from the further factor $3$ introduced by the symmetrization in Eq.\ \ref{eq:chi3sy-sm}.
	
	\subsection{$S_{13}$ diagrams}

    \begin{figure}
		\centering
		\includegraphics[width=0.375\textwidth]{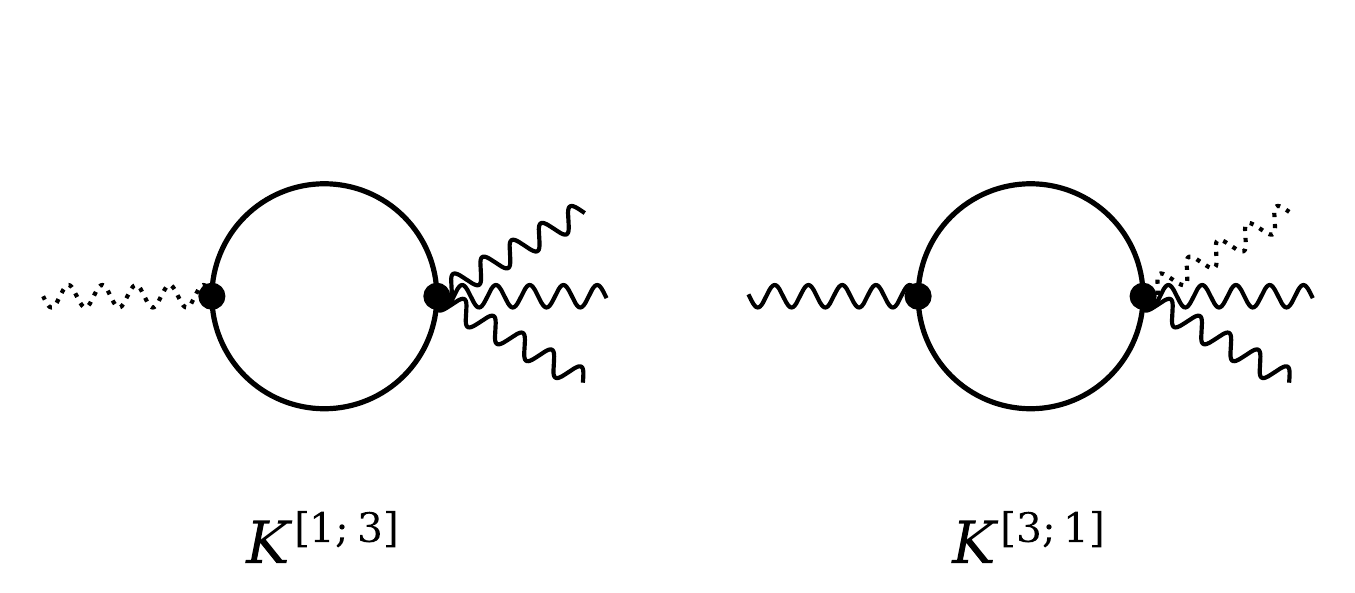}
		\caption{The two possible contributions stemming from deriving $S_{13}$. On the left $K^{[1;3]}$, see Eq.\ \ref{eq:k13}, on the right $K^{[3;1]}$, see Eq.\ \ref{eq:k31}.}
		\label{fig:k13}
	\end{figure}
    
	To compute this diagram, we need 
    \be
    X^{(2)}(i\Omega,\mathbf{k})=\frac{1}{\beta}\sum_{i\omega_n}\Tr{\left[G_0(i\omega_n,\mathbf{k})\sigma_3G_0(i\omega_n+i\Omega,\mathbf{k})\sigma_3\right]}=\frac{\Delta^2}{E_\mathbf{k}^2} \left(1-2f_\mathbf{k}\right)\left(\frac{1}{i\Omega-2E_\mathbf{k}}-\frac{1}{i\Omega+2E_\mathbf{k}}\right),
    \ee
    In this case, we need to make a distinction when the derivative with respect to $A$ in Eq.\ \eqref{eq:derj} is performed on the $\Sigma_A^{(1)}$ or $\Sigma_A^{(3)}$ vertex, see Fig.\ \ref{fig:k13}. Since there is only one field line attached to $\Sigma_A^{(1)}$, when we perform the functional derivative on that one we have
	\be
    \label{eq:k13}
	K^{[1;3]}(\omega_1,\omega_2,\omega_3)=\frac{1}{6V}\sum_\mathbf{k}\frac{\partial^3 \epsilon_\mathbf{k}}{\partial k^3}\frac{\partial \epsilon_\mathbf{k}}{\partial k}\frac{\Delta^2}{E_\mathbf{k}^2} \left(1-2f_\mathbf{k}\right)\left(\frac{1}{\omega_{123}+i\eta-2E_\mathbf{k}}-\frac{1}{\omega_{123}+i\eta+2E_\mathbf{k}}\right),
	\ee
    where the prefactor $1/6$ associated with the $\Sigma_A^{(3)}$ vertex is canceled out by the possible ways of performing the remaining three functional derivatives in Eq.\ \ref{eq:der4} at the same vertex. This process produces the following signatures on the 2D map:
    \begin{align}
	J^{[1;3]}_{00\tau}(\omega_t,\omega_\tau)&=\frac{1}{2V}\sum_\mathbf{k}\frac{\partial^3 \epsilon_\mathbf{k}}{\partial k^3}\frac{\partial \epsilon_\mathbf{k}}{\partial k}\frac{\Delta^2}{E_\mathbf{k}^2} \left(1-2f_\mathbf{k}\right)\left(\frac{1}{\omega_t+i\eta-2E_\mathbf{k}}-\frac{1}{\omega_t+i\eta+2E_\mathbf{k}}\right)A(\omega_\tau)A^2(\omega_t-\omega_\tau),\\
	J^{[1;3]}_{0\tau\tau}(\omega_t,\omega_\tau)&=\frac{1}{2V}\sum_\mathbf{k}\frac{\partial^3 \epsilon_\mathbf{k}}{\partial k^3}\frac{\partial \epsilon_\mathbf{k}}{\partial k}\frac{\Delta^2}{E_\mathbf{k}^2} \left(1-2f_\mathbf{k}\right)\left(\frac{1}{\omega_t-i\eta-2E_\mathbf{k}}-\frac{1}{\omega_t+i\eta+2E_\mathbf{k}}\right)A(\omega_t-\omega_\tau)A^2(\omega_\tau).
	\end{align}

    In contrast, the situation for $K^{[3;1]}$ is slightly more complicated. There are three ways to assign the derivative with respect to $A(-i\Omega_t)$ to the three-photon vertex, while the other three derivatives need to be distributed among the remaining two photon lines in $\Sigma_A^{(3)}$ and the one in $\Sigma_A^{(1)}$. Each possible choice for the single-photon vertex will determine the corresponding $\omega_i$ running in the two-point bubble $X^{(2)}$, consequently there are two ways left to assign the remaining two derivatives to $\Sigma_A^{(3)}$. Collecting the multiplicities ($3\cdot2$), we cancel out the factor $1/6$ from the vertex and we also get a sum over the possible frequencies $\omega_i$:
	\be
    \label{eq:k31}
	K^{[3;1]}(\omega_1,\omega_2,\omega_3)=\frac{1}{6}\sum_i\frac{1}{V}\sum_\mathbf{k}\frac{\partial^3 \epsilon_\mathbf{k}}{\partial k^3}\frac{\partial \epsilon_\mathbf{k}}{\partial k}\frac{\Delta^2}{E_\mathbf{k}^2} \left(1-2f_\mathbf{k}\right)\left(\frac{1}{\omega_{i}+i\eta-2E_\mathbf{k}}-\frac{1}{\omega_{i}+i\eta+2E_\mathbf{k}}\right).
	\ee
    However, when we now include the $\tau$ dependence of the fields we do not always succeed in factorizing the frequency dependence of the bubble from the one of the fields. For example, in the $J^{[3;1]}_{00\tau}$ combination we assign only one $\omega_i$ to $\omega_\tau$. But then only one term in the summation over $\omega_i$ in Eq.\ \ref{eq:k31} factorizes out, while the remaining two terms still depend on $\omega$ and $\omega_t-\omega_\tau-\o$, where $\omega$ is the frequency of the integration in Eq.\ \eqref{eq:j00tau-sm}. Since the response \eqref{eq:k31} is symmetric with respect to the incoming frequencies, this structure will repeat three times, as anticipated above. Taking this multiplicity into account, we can then proceed as in the main text by introducing a suitable density of states for each process. We rewrite the two contributions to the nonlinear current as
    \begin{align}
    \label{eq:j31}
    J^{[3;1]}_{00\tau}(\omega_t,\omega_\tau)&=\frac{1}{2}\int dE\rho^{(13)}(E)\bar{J}^{[3;1]}_{00\tau}(E,\omega_t,\omega_\tau),\nonumber\\ 
    J^{[3;1]}_{0\tau\tau}(\omega_t,\omega_\tau)&=\frac{1}{2}\int dE\rho^{(13)}(E)\bar{J}^{[3;1]}_{0\tau\tau}(E,\omega_t,\omega_\tau),
    \end{align}
    where we introduced 
    \be
    \rho^{(13)}(E)=\frac{E}{\sqrt{E^2-\Delta^2}}\frac{\Delta^2}{E^2}\left(1-2f(E)\right)\tilde{\rho}^{(13)}\left(\sqrt{E^2-\Delta^2}\right),
    \ee
    with
    \be
    \tilde{\rho}^{(13)}(\epsilon)=\frac{1}{V}\sum_{\mathbf{k}}\frac{\partial^3 \epsilon_\mathbf{k}}{\partial k^3}\frac{\partial \epsilon_\mathbf{k}}{\partial k}\delta(\epsilon-\epsilon_\mathbf{k}),
    \ee
    and we defined the two quantities
	\begin{align}
    \label{eq:j3100t}
	\bar{J}^{[3;1]}_{00\tau}(E,\omega_t,\omega_\tau)&=2A(\omega_\tau)\left[F^\prime_{00\tau}(-2E)-F^\prime_{00\tau}(2E)\right]+A(\omega_\tau)A^2(\omega_t-\omega_\tau)\left(\frac{1}{\omega_{\tau}+i\eta-2E}-\frac{1}{\omega_{\tau}+i\eta+2E}\right),\\
    \label{eq:j310tt}
	\bar{J}^{[3;1]}_{0\tau\tau}(E,\omega_t,\omega_\tau)&=2A(\omega_t-\omega_\tau)\left[F^\prime_{0\tau\tau}(-2E)-F^\prime_{0\tau\tau}(2E)\right]\nonumber\\
    &+A(\omega_t-\omega_\tau)A^2(\omega_\tau)\left(\frac{1}{\omega_t-\omega_{\tau}+i\eta-2E}-\frac{1}{\omega_t-\omega_{\tau}+i\eta+2E}\right).
	\end{align}
    In the previous expression, we introduced the functions
	\begin{align}
	F^\prime_{00\tau}(2E)&=\int d\omega\,\frac{A(\omega)A(\omega_t-\omega_\tau-\omega)}{\omega+i\eta+2E},\nonumber\\
	F^\prime_{0\tau\tau}(2E)&=\int d\omega\,\frac{A(\omega)A(\omega_\tau-\omega)}{\omega+i\eta+2E}.
	\end{align}
    In the presence of gaussian field envelopes, $A(\omega)=e^{-(\omega-\Omega)^2\tau^2/2}+e^{-(\omega+\Omega)^2\tau^2/2}$, the previous integral can be expressed in terms of a sum of Faddeeva functions $w(z)$. Knowing that, for $\text{Im}(z)>0$,
    \be
    w(z)=\frac{i}{\pi}\int_{-\infty}^{+\infty}\frac{e^{-t^2}}{z+t},
    \ee
    one can show immediately that
    \begin{align}
    \int d\omega^\prime\,\frac{A(\omega^{\prime})A(\omega-\omega^\prime)}{\omega^\prime+i\eta+2E}&=-i\pi w{\left[\left(\omega+4E+2i\eta\right)\tau/2\right]}\left(e^{-(\omega-2\Omega)^2\tau^2/4}+e^{-(\omega+2\Omega)^2\tau^2/4}\right)\nonumber\\
    &-i\pi\left(w{\left[\left(\omega+2\Omega+4E+2i\eta\right)\tau/2\right]}+w{\left[\left(\omega-2\Omega+4E+2i\eta\right)\tau/2\right]}\right)e^{-\omega^2\tau^2/4}.
    \end{align}

    \subsection{$S_{22}$ diagrams}

    The fully-diamagnetic contribution has been derived in the main text, we report here the results for the sake of completeness and to account for multiplicities. Similarly to $K^{[1;3]}$ and $K^{[3;1]}$, it has a two-point structure where the running frequency is now the sum of two incoming ones:
	\be
	K^{[2;2]}(\omega_1,\omega_2,\omega_3)=\frac{1}{6}\cdot\frac{1}{2}\sum_{i\neq j}\frac{1}{V}\sum_\mathbf{k}\left(\frac{\partial^2\epsilon_{\mathbf{k}}}{\partial k^2}\right)^2\frac{\Delta^2}{E_\mathbf{k}^2} \left(1-2f_\mathbf{k}\right)\left(\frac{1}{\omega_{ij}+i\eta-2E_\mathbf{k}}-\frac{1}{\omega_{ij}+i\eta+2E_\mathbf{k}}\right),
	\ee
	where by $i\neq j$ we mean the six \textit{ordered} pairs of frequencies $\omega_i,\omega_j$. In this case, there is also an overall factor $(1/2)$ that is inherited from the difference between $S_{22}$ and $S_{13}$, see Eq.\ \ref{eq:dia} and \ref{eq:s13}, respectively. The derivation goes on as before and one introduces $\rho^{(22)}$, $\bar{J}^{[2;2]}_{00\tau}$ and $\bar{J}^{[2;2]}_{00\tau}$ exactly as done for Eq.\ \ref{eq:j31}. They read:
    \be
    \label{eq:rhodia}
    \rho^{(22)}(E)=\frac{E}{\sqrt{E^2-\Delta^2}}\frac{\Delta^2}{E^2}\left(1-2f(E)\right)\tilde{\rho}^{(22)}\left(\sqrt{E^2-\Delta^2}\right),
    \ee
    with
    \be
    \tilde{\rho}^{(22)}(\epsilon)=\frac{1}{V}\sum_{\mathbf{k}}\left(\frac{\partial^2 \epsilon_\mathbf{k}}{\partial k^2}\right)^2\delta(\epsilon-\epsilon_\mathbf{k}),
    \ee
    while we have
    \begin{align}
		\bar{J}^{[2;2]}_{00\tau}(E,\omega_t,\omega_\tau)&=2A(\omega_\tau)\left[F^{\prime\prime}_{00\tau}(-2E)-F^{\prime\prime}_{00\tau}(2E)\right]\nonumber\\
		&+A(\omega_\tau)A^2(\omega_t-\omega_\tau)\left(\frac{1}{\omega_t-\omega_{\tau}+i\eta-2E}-\frac{1}{\omega_t-\omega_{\tau}+i\eta+2E}\right),\\
		\bar{J}^{[2;2]}_{0\tau\tau}(E,\omega_t,\omega_\tau)&=2A(\omega_t-\omega_\tau)\left[F^{\prime\prime}_{0\tau\tau}(-2E)-F^{\prime\prime}_{0\tau\tau}(2E)\right]+A(\omega_t-\omega_\tau)A^2(\omega_\tau)\left(\frac{1}{\omega_{\tau}+i\eta-2E}-\frac{1}{\omega_{\tau}+i\eta+2E}\right).
	\end{align}
    Notice that the structure is very similar to the one presented in Eqs.\ \ref{eq:j3100t} and \ref{eq:j310tt}, with the important difference that in the propagators now the energies corresponding to the sum of two photons appear. This is also reflected in the new definitions of the functions $F^{\prime\prime}$ which replace the previous $F^{\prime}$:
    \begin{align}
	F^{\prime\prime}_{00\tau}(2E)&=\int d\omega\,\frac{A(\omega)A(\omega_t-\omega_\tau-\omega)}{\omega+\omega_\tau+i\eta+2E},\nonumber\\
	F^{\prime\prime}_{0\tau\tau}(2E)&=\int d\omega\,\frac{A(\omega)A(\omega_\tau-\omega)}{\omega_t-\omega+i\eta+2E}.
	\end{align}
    
	\subsection{$S_{112}$ diagrams}

    \begin{figure}
		\centering
		\includegraphics[width=0.375\textwidth]{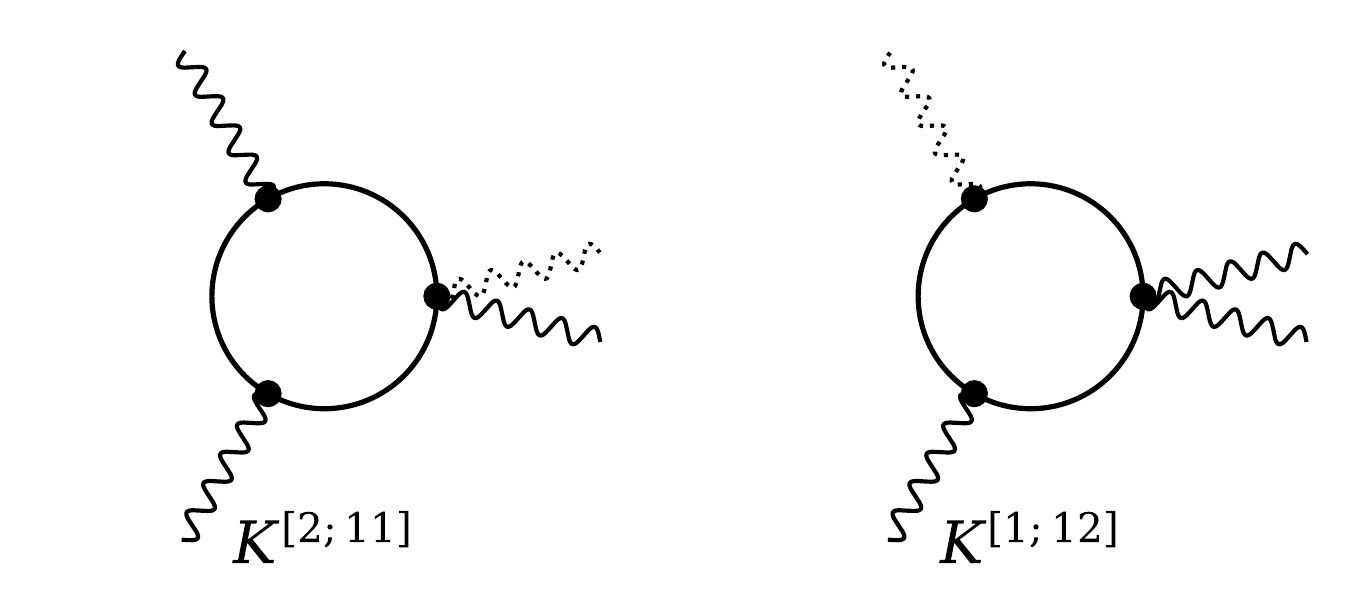}
		\caption{The two possible contributions stemming from deriving $S_{112}$. On the left $K^{[2;11]}$, see Eq.\ \ref{eq:k211}, on the right $K^{[1;12]}$, see Eq.\ \ref{eq:k112}.}
		\label{fig:k112}
	\end{figure}
    
    To compute this diagram, we need to evaluate
    \begin{align}
    X^{(3)}(i\Omega_i,i\Omega_j,\mathbf{k})&=\frac{1}{\beta}\sum_{i\omega_n}\Tr{\left[G_0(i\omega_n,\mathbf{k})\sigma_3G_0(i\omega_n+i\Omega_i,\mathbf{k})\sigma_3G_0(i\omega_n+i\Omega_{ij},\mathbf{k})\sigma_3\right]}\nonumber\\
    &=\frac{\epsilon_\mathbf{k}\Delta^2}{E_\mathbf{k}^3}(1-2f_\mathbf{k})\left[\left(\frac{1}{i\Omega_{ij}-2E_\mathbf{k}}
    -\frac{1}{i\Omega_{ij}}\right)\left(\frac{1}{i\Omega_{i}-2E_\mathbf{k}}+\frac{1}{i\Omega_{j}-2E_\mathbf{k}}\right)+(E_\mathbf{k}\to-E_\mathbf{k})\right]\nonumber\\
    &=\frac{\epsilon_\mathbf{k}\Delta^2}{E_\mathbf{k}^3}(1-2f_\mathbf{k})\left[\frac{1}{i\Omega_{ij}-2E_\mathbf{k}}\left(\frac{1}{i\Omega_{i}-2E_\mathbf{k}}+\frac{1}{i\Omega_{j}-2E_\mathbf{k}}\right)-\frac{1}{i\Omega_{i}-2E_\mathbf{k}}\frac{1}{i\Omega_{j}+2E_\mathbf{k}}+(E_\mathbf{k}\to-E_\mathbf{k})\right].
    \end{align}
	Here we provided two equivalent expressions; the latter is explicitly divergence-free in the static limit, so it can be used for gauge-invariance checks. As we did for the $S_{13}$ processes, we have to be careful in performing the derivative with respect to $A(-i\Omega_t)$, see Fig.\ \ref{fig:k112}. If we assign it to the diamagnetic vertex, we have
	\begin{align}
    \label{eq:k211}
	K^{[2;11]}(\omega_1,\omega_2,\omega_3)&=\frac{1}{6}\sum_{i\neq j}\frac{1}{V}\sum_\mathbf{k}\frac{\epsilon_\mathbf{k}\Delta^2}{E_\mathbf{k}^3}\frac{\partial^2 \epsilon_\mathbf{k}}{\partial k^2}\left(\frac{\partial \epsilon_\mathbf{k}}{\partial k}\right)^2 \left(1-2f_\mathbf{k}\right)\nonumber\\
    &\times\left(\frac{1}{\omega_{ij}+i\eta-2E_\mathbf{k}}-\frac{1}{\omega_{ij}+i\eta}\right)\left(\frac{1}{\omega_{i}+i\eta-2E_\mathbf{k}}+\frac{1}{\omega_{j}+i\eta-2E_\mathbf{k}}\right)+(E_\mathbf{k}\to-E_\mathbf{k}).
	\end{align}
    The multiplicity coming from the $\Sigma_A^{(2)}$ vertex is canceled out by the possible ways of assigning the derivative $A(-i\Omega_t)$ to the two-photon vertex, whereas the other derivatives are accounted for by the sum over the possible ordered pairs. We also introduce here the density of states 
    \be
    \rho^{(112)}(E)=\frac{E}{\sqrt{E^2-\Delta^2}}\frac{\sqrt{E^2-\Delta^2}\Delta^2}{E^3}\left(1-2f(E)\right)\tilde{\rho}^{(112)}\left(\sqrt{E^2-\Delta^2}\right),
    \ee
    with
    \be
    \tilde{\rho}^{(112)}(\epsilon)=\frac{1}{V}\sum_{\mathbf{k}}\frac{\partial^2 \epsilon_\mathbf{k}}{\partial k^2}\left(\frac{\partial \epsilon_\mathbf{k}}{\partial k}\right)^2\delta(\epsilon-\epsilon_\mathbf{k}),
    \ee
    such that we can again define $J^{[2;11]}_{00\tau}(E,\omega_t,\omega_\tau)$ analogously to Eq.\ \ref{eq:j31} to express the contribution $J^{[2;11]}_{00\tau}(\omega_t,\omega_\tau)$. We can finally rewrite
	\be
	\begin{split}
	\bar{J}^{[2;11]}_{00\tau}(E,\omega_t,\omega_\tau)&=4A(\omega_\tau)\left(\frac{1}{\omega_t-\omega_\tau+i\eta-2E}-\frac{1}{\omega_t-\omega_\tau+i\eta}\right)F^{\prime}_{00\tau}(-2E)\\
	&+4A(\omega_\tau)\left[\frac{F^{\prime}_{00\tau}(-2E)-F^{\prime\prime}_{00\tau}(-2E)}{\omega_\tau+i\eta}+\frac{F^{\prime\prime}_{00\tau}(-2E)-F^{\prime
}_{00\tau}(2E)}{\omega_\tau+i\eta-2E}\right]+(E\to-E).
	\end{split}
	\ee
    
	Let us now report the piece coming from the derivative of $\Sigma_A^{(1)}$. It reads, within the same conventions introduced before,
    \begin{align}
	\label{eq:k112}
    K^{[1;12]}(\omega_1,\omega_2,\omega_3)&=\frac{1}{6}\sum_{i\neq j}\frac{1}{V}\sum_\mathbf{k}\frac{\epsilon_\mathbf{k}\Delta^2}{E_\mathbf{k}^3}\frac{\partial^2 \epsilon_\mathbf{k}}{\partial k^2}\left(\frac{\partial \epsilon_\mathbf{k}}{\partial k}\right)^2(1-2f_\mathbf{k})\nonumber\\
    &\times\left(\frac{1}{\omega_{123}+i\eta-2E_\mathbf{k}}-\frac{1}{\omega_{123}+i\eta}\right)\left(\frac{1}{\omega_{ij}+i\eta-2E_\mathbf{k}}+\frac{1}{\omega_{123}-\omega_{ij}+i\eta-2E_\mathbf{k}}\right)+(E_\mathbf{k}\to-E_\mathbf{k}).
	\end{align}
    The structure of the corresponding $J^{[1;12]}_{00\tau}(E,\omega_t,\omega_\tau)$ in $(\omega_t,\omega_\tau)$ is now different and reads
	\be
	\begin{split}
	\bar{J}^{[1;12]}_{00\tau}(E,\omega_t,\omega_\tau)&=2A(\omega_\tau)A^2(\omega_t-\omega_\tau)\left(\frac{1}{\omega_t+i\eta-2E}-\frac{1}{\omega_t+i\eta}\right)\left(\frac{1}{\omega_\tau+i\eta-2E}+\frac{1}{\omega_t-\omega_\tau+i\eta-2E}\right)\\
	&+4A(\omega_\tau)\left(\frac{1}{\omega_t+i\eta-2E}-\frac{1}{\omega_t+i\eta}\right)\left[F^\prime_{00\tau}(-2E)+F^{\prime\prime}_{00\tau}(-2E)\right]+(E\to-E).
	\end{split}
	\ee
    The contributions $J^{[2;11]}_{0\tau\tau}$ and $J^{[1;12]}_{0\tau\tau}$ can be obtained as usual by sending $\omega_\tau\to\omega_t-\omega_\tau$ in the expressions for $J^{[2;11]}_{00\tau}$ and $J^{[1;12]}_{00\tau}$.

    \subsection{\label{sec:k1111} $S_{1111}$ diagram}
    
    For the fully-paramagnetic process we need to 
    evaluate the function
	\begin{align}
    X^{(4)}(i\Omega_1,i\Omega_2,i\Omega_3,\mathbf{k})&=\frac{1}{\beta}\sum_{i\omega_n}\Tr{\left[G_0(i\omega_n,\mathbf{k})\sigma_3G_0(i\omega_n+i\Omega_1,\mathbf{k})\sigma_3G_0(i\omega_n+i\Omega_{12},\mathbf{k})\sigma_3G_0(i\omega_n+i\Omega_{123},\mathbf{k})\sigma_3\right]}\nonumber\\
    &\equiv(1-2f_\mathbf{k})\left[\frac{\Delta^4}{E_\mathbf{k}^4}X^{(4),\text{inter}}(i\Omega_1,i\Omega_2,i\Omega_3,\mathbf{k})+\frac{2\epsilon_\mathbf{k}^2\Delta^2}{E_\mathbf{k}^4}X^{(4),\text{mi}}(i\Omega_1,i\Omega_2,i\Omega_3,\mathbf{k})\right],
    \end{align}
    where we explicitly separated the fully interband, containing four $\sigma_1$ Pauli matrices, from the mixed intraband-interband scattering, containing two $\sigma_1$ and two $\sigma_3$, as discussed in the text. This bubble, differently from the previous ones, shows a non-trivial dependence on the order of the incoming frequencies. However, the functional derivatives with respect to the gauge fields automatically symmetrize the response over the possible permutations of $i\Omega_i$. For this reason, we provide an expression for $X^{(4),\text{inter}}$ and $X^{(4),\text{mi}}$ which has already been symmetrized over the six permutations of the frequencies 
    \be
    \label{eq:x4si}
    X^{(4),\text{inter}}_\mathcal{S}(i\Omega_1,i\Omega_2,i\Omega_3,\mathbf{k})=\frac{1}{i\Omega_{123}-2E_\mathbf{k}}\sum_{i\neq j}\left(\frac{1}{i\Omega_{i}-2E_\mathbf{k}}\frac{1}{i\Omega_{j}+2E_\mathbf{k}}+\frac{1}{i\Omega_{i}+2E_\mathbf{k}}\frac{1}{i\Omega_{j}-2E_\mathbf{k}}\right)-(E_\mathbf{k}\to-E_\mathbf{k}),
    \ee
    which is explicitly finite for the frequencies going to zero, and
    \begin{align}
    \label{eq:x4sm}
    X^{(4),\text{mi}}_\mathcal{S}(i\Omega_1,i\Omega_2,i\Omega_3,\mathbf{k})&=\left(\frac{1}{i\Omega_{123}-2E_\mathbf{k}}-\frac{1}{i\Omega_{123}}\right)\sum_{i\neq j}\frac{1}{i\Omega_{ij}-2E_\mathbf{k}}\left(\frac{1}{i\Omega_{i}-2E_\mathbf{k}}+\frac{1}{i\Omega_{j}-2E_\mathbf{k}}\right)-(E_\mathbf{k}\to-E_\mathbf{k})\nonumber\\
    &=\frac{1}{i\Omega_{123}-2E_\mathbf{k}}\sum_{i\neq j}\frac{1}{i\Omega_{ij}-2E_\mathbf{k}}\left(\frac{1}{i\Omega_{i}-2E_\mathbf{k}}+\frac{1}{i\Omega_{j}-2E_\mathbf{k}}\right)\nonumber\\
    &-\sum_{i\neq j}\left(\frac{1}{i\Omega_{123}-i\Omega_j-2E_\mathbf{k}}-\frac{1}{i\Omega_{123}-i\Omega_i+2E_\mathbf{k}}\right)\frac{1}{i\Omega_i-2E_\mathbf{k}}\frac{1}{i\Omega_j+2E_\mathbf{k}}-(E_\mathbf{k}\to-E_\mathbf{k}),
    \end{align}
    where also in this case we provided the last expression for an easy gauge-invariance check. The total kernel reads
    \be
    K^{[1;111]}(\omega_1,\omega_2,\omega_3)=\frac{1}{6V}\sum_\mathbf{k}(1-2f_\mathbf{k})\left(\frac{\partial \epsilon_\mathbf{k}}{\partial k}\right)^4\left[\frac{\Delta^4}{E_\mathbf{k}^4}X^{(4),\text{inter}}_\mathcal{S}(\omega_1,\omega_2,\omega_3,\mathbf{k})+\frac{2\epsilon_\mathbf{k}^2\Delta^2}{E_\mathbf{k}^4}X^{(4),\text{mi}}_\mathcal{S}(\omega_1,\omega_2,\omega_3,\mathbf{k})\right],
    \ee
    where $1/4$ of the expansion of the logarithm in Eq.\ \ref{eq:expa} is canceled out by the four possible ways of assigning the functional derivative with respect to $A(-i\Omega_t)$, while the further three derivatives provide the symmetrization of the response needed in the previous Eqs.\ \ref{eq:x4si} and \ref{eq:x4sm}. Introducing the corresponding density of states as presented in the main text, 
    \begin{align}
    \label{eq:rhointer}
    \rho^{\text{inter}}(E)&=\frac{E}{\sqrt{E^2-\Delta^2}}\left(1-2f(E)\right)\frac{\Delta^4}{E^4}\tilde{\rho}^{[1;111]}\left(\sqrt{E^2-\Delta^2}\right),\\
    \label{eq:rhomix}
    \rho^{\text{mi}}(E)&=\frac{E}{\sqrt{E^2-\Delta^2}}\left(1-2f(E)\right)\frac{2(E^2-\Delta^2)\Delta^2}{E^4}\tilde{\rho}^{[1;111]}\left(\sqrt{E^2-\Delta^2}\right),
    \end{align}
    with
    \be
    \label{eq:tenspara-sm}
	\tilde{\rho}^{[1;111]}(\epsilon)\equiv\frac{1}{V}\sum_{\mathbf{k}}\left(\frac{\partial\epsilon_\mathbf{k}}{\partial k}\right)^4\delta(\epsilon-\epsilon_{\mathbf{k}}),
	\ee
    we get to the functions that allow to express the contribution of the interband and mixed processes:
    \begin{align}
    \bar{J}^{[1;111],\text{inter}}_{00\tau}(E,\omega_t,\omega_\tau)&=4A(\omega_\tau)\left(\frac{1}{\omega_t+i\eta-2E}-\frac{1}{\omega_t+i\eta+2E}\right)\nonumber\\
		&\times\left[\frac{F^{\prime}_{00\tau}(2E)}{\omega_\tau+i\eta-2E}+\frac{F^{\prime}_{00\tau}(-2E)}{\omega_\tau+i\eta+2E}+\frac{F^{\prime}_{00\tau}(2E)+F^{\prime}_{00\tau}(-2E)}{\omega_t-\omega_\tau+i\eta}\right],\\
    \bar{J}^{[1;111],\text{mi}}_{00\tau}(E,\omega_t,\omega_\tau)&=4A(\omega_\tau)\left(\frac{1}{\omega_t+i\eta-2E}-\frac{1}{\omega_t+i\eta}\right)\nonumber\\
    &\times\left[\frac{F^{\prime}_{00\tau}(2E)-F^{\prime\prime}_{00\tau}(2E)}{\omega_\tau+i\eta}+\frac{F^{\prime\prime}_{00\tau}(2E)}{\omega_\tau+i\eta-2E}+\frac{F^{\prime}_{00\tau}(2E)}{\omega_t-\omega_\tau+i\eta-2E}\right]-(E\to-E).
    \end{align}
    In this case, also, $\bar{J}^{[1;111],\text{inter}}_{0\tau\tau}$ and $\bar{J}^{[1;111],\text{mi}}_{0\tau\tau}$  can be obtained by sending $\omega_\tau\to\omega_t-\omega_\tau$ in the expression for $\bar{J}^{[1;111],\text{inter}}_{00\tau}$ and $\bar{J}^{[1;111],\text{mi}}_{00\tau}$, respectively.

    \subsection{Gauge-invariance check}

    \begin{figure}
		\centering
		\includegraphics[width=0.5\textwidth]{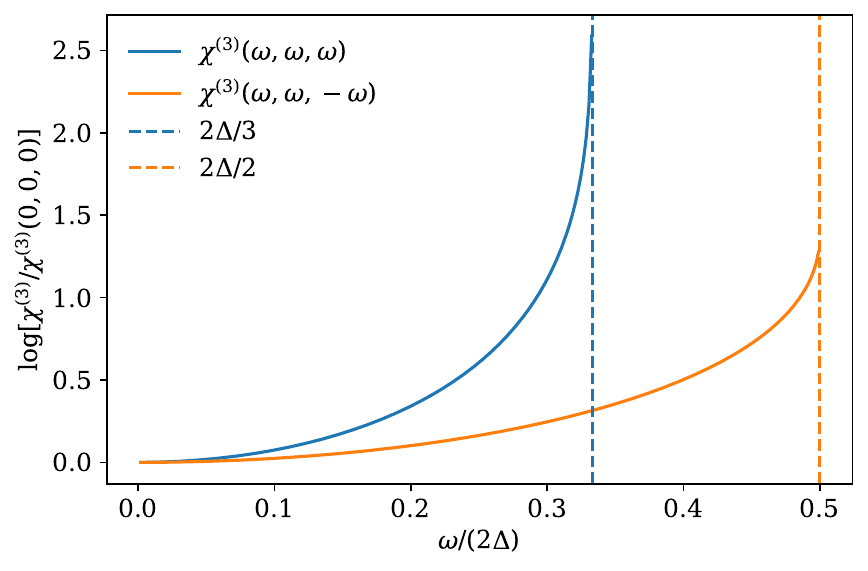}
		\caption{Gauge-invariant susceptibilities for some common nonlinear processes: third harmonic generation (THG) $\chi^{(3)}(\omega,\omega,\omega)$ (blue) and first harmonic generation (FHG) $\chi^{(3)}(\omega,\omega,-\omega)$ (orange). The susceptibilities are computed in the corresponding off-resonant regions $3\omega<2\Delta$ (THG) and $2\omega<2\Delta$ (FHG) and are normalized to their common value $\chi^{(3)}(0,0,0)$. We can observe that below resonance the behavior is perfectly regular, as expected for an insulator.}
		\label{fig:gcheck}
	\end{figure}

    Having evaluated all the contributions, we can construct the total third-order kernel to check the gauge invariance of the response. We have
    \be
    \label{eq:fr}
    K^{(3)}=K^{[4;]}+K^{[1;3]}+K^{[3;1]}+K^{[2;2]}+K^{[2;11]}+K^{[1;12]}+K^{[1;111]}.
    \ee
    We check that up to numerical precision the relation $K^{(3)}(0,0,0)=K^{(3)}(\omega,0,0)=K^{(3)}(\omega,\omega^{\prime},0)=0$ holds for any value of the frequencies $\omega,\omega^\prime<2\Delta$, since in this interval we can use the expressions presented above in their divergence-free form by explicitly setting $\eta\to0$. The above relation is a sufficient condition to ensure that 
    \be
    K^{(3)}(\omega_1,\omega_2,\omega_3)\propto\omega_1\omega_2\omega_3\sigma^{(3)}(\omega_1,\omega_2,\omega_3),
    \ee
    which allows us to connect $K^{(3)}$ to the manifestly gauge-invariant third-order conductivity $\sigma^{(3)}$. From the conductivity, we can obtain the corresponding susceptibility using the constitutive relation $J=\partial P/\partial t$, which gives
    \be
    \chi^{(3)}(\omega_1,\omega_2,\omega_3)=\frac{K^{(3)}(\omega_1,\omega_2,\omega_3)}{\omega_{123}\omega_1\omega_2\omega_3}.
    \ee
    As expected for an insulator at zero temperature, the static limit of the susceptibility $\chi^{(3)}(0,0,0)$ is finite and can be calculated from the above relation for arbitrarily small frequencies. We present in Fig.\ \ref{fig:gcheck} the behavior of the third-order susceptibilities for the one-dimensional chain which allows us to compute some common experimental observables: third harmonic generation $\chi^{(3)}(\omega,\omega,\omega)$ and first harmonic generation $\chi^{(3)}(\omega,\omega,-\omega)$. We can immediately notice that in the corresponding nonresonant regions the susceptibility does not show spurious divergencies, confirming the gauge-invariance of the total response computed from $K^{(3)}$.

    \subsection{Symmetry analysis for square lattice. Density of states.}

     \begin{figure}
		\centering
		\includegraphics[width=0.75\textwidth]{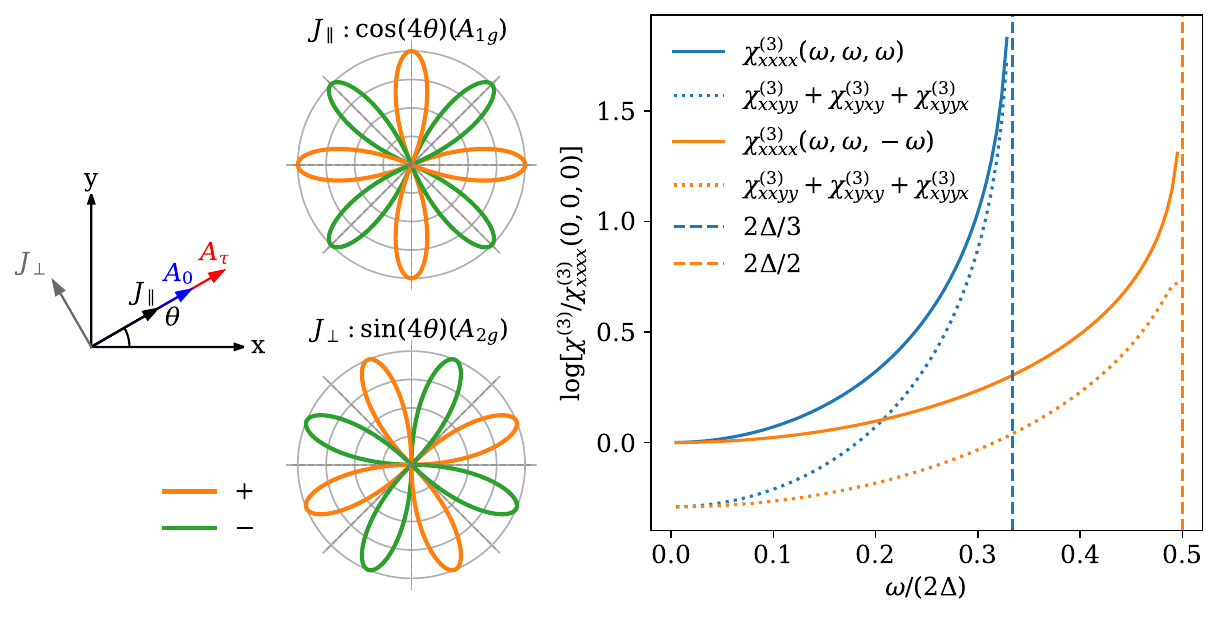}
		\caption{Schematics of the 2D experiment modeled in the main text, where the incoming fields $A_0$ and $A_\tau$ are parallel and one detects the nonlinearly generated current either in the same $(J_\parallel)$ or perpendicular $(J_\perp)$ direction. The parallel (perpendicular) components of the current transform as the $A_{1g}$ ($A_{2g}$) irreducible representation of the $D_{4h}$ point group according to Eq.\ \ref{eq:jpar} (Eq.\ \ref{eq:jper}). We show here also the behavior of the relevant combinations of third-order susceptibility elements entering $J_\parallel$ and perpendicular $J_\perp$, focusing on THG (blue) and FHG (orange).}
		\label{fig:geom2d}
	\end{figure}

    In this subsection we provide a symmetry analysis of the third-order response for the case of the two-dimensional square lattice, to derive the relevant density of states used in the main text to compute the 2D maps. We assume that the experimental geometry is the one in Fig.\ \ref{fig:geom2d}, with $A_0$ and $A_\tau$ fields parallel and oriented along the versor $\hat{\theta}=(\cos\theta,\sin\theta)$, while the nonlinearly generated current is decomposed in the parallel and perpendicular components:
    \begin{align}
    J_\parallel&=\mathbf{J}\cdot\hat{\theta}= J_x\cos\theta+J_y\sin{\theta}\\
    J_\perp&=\mathbf{J}\cdot(\hat{z}\times\hat{\theta})=-J_x\sin\theta+J_y\cos\theta.
    \end{align}
    For the $D_{4h}$ point group, there are only four independent nonzero elements of the frequency-dependent thrid-order response tensor, namely $K^{(3)}_{xxxx}$, $K^{(3)}_{xxyy}$, $K^{(3)}_{xyxy}$, $K^{(3)}_{xyyx}$. Omitting frequency dependencies and convolutions, we have
    \begin{align}
    \label{eq:jpar}
    J_\parallel&=\left[K^{(3)}_{xxxx}(\cos^4\theta+\sin^4\theta)+(K^{(3)}_{xxyy}+K^{(3)}_{xyxy}+K^{(3)}_{xyyx})2\cos^2\theta\sin^2\theta\right]A^3\nonumber\\
    &=\left[\frac{1}{4}(3K^{(3)}_{xxxx}+K^{(3)}_{xxyy}+K^{(3)}_{xyxy}+K^{(3)}_{xyyx})+\frac{1}{4}(K_{xxxx}^{(3)}-K^{(3)}_{xxyy}-K^{(3)}_{xyxy}-K^{(3)}_{xyyx})\cos{4\theta}\right]A^3;\\
    \label{eq:jper}
    J_\perp&=\left[K_{xxxx}^{(3)}(-\sin\theta\cos^3\theta+\sin^3\theta\cos\theta)+(K_{xxyy}^{(3)}+K_{xyxy}^{(3)}+K_{xyyx}^{(3)})(-\sin^3\theta\cos\theta+\sin\theta\cos^3\theta)\right]A^3\nonumber\\
    &=-\frac{1}{4}(K_{xxxx}^{(3)}-K_{xxyy}^{(3)}-K_{xyxy}^{(3)}-K_{xyyx}^{(3)})\sin{4\theta}A^3.
    \end{align}
    This decomposition shows that both components of the nonlinear current are modulated by varying the angle $\theta$. In particular, the parallel component contains an angle-indipendent contribution plus a $\cos4\theta$ modulation, both transforming as the $A_{1g}$ irreducible representation of $D_{4h}$, while the perpendicular component is proportional to $\sin4\theta$, transforming as $A_{2g}$. This behavior is highlighted in Fig.\ \ref{fig:geom2d}, where we represent $\lvert\cos4\theta\rvert$ and $\lvert\sin4\theta\rvert$ in a polar plot with the lobes colored according to the sign of $\cos4\theta$ and $\sin4\theta$, respectively. It is evident that while the former is invariant under all the symmetry operations of the $D_{4h}$ point group, the latter is odd with respect to $C_2$ rotations around the $x,y$ axis and mirrors with respect to the $x,y$ axis and diagonals, as expected for $A_{2g}$. 

    In the main text, we presented the results for the angle-independent contribution to $J_{\parallel}$ coming from the fully diamagnetic ($K^{[2;2]}$)and fully paramagnetic ($K^{[1;111]}$) diagrams. Due to the symmetry of the model, we can compute analytically the density of states associated with the various light-matter vertices. We define
    \be
    \tilde{\rho}_{\alpha\beta\gamma\delta}^{(\dots)}(\epsilon)=\frac{1}{V}\sum_{\mathbf{k}}\delta(\epsilon-\epsilon_\mathbf{k})w^{(\dots)}_{\alpha\beta\gamma\delta}(\mathbf{k}).
    \ee
    In our minimal tight-binding model, we easily notice that $w^{(\dots)}_{\alpha\beta\gamma\delta}(\mathbf{k})$ is an algebraic combination of $\sin k_x$, $\cos k_x$, $\sin k_x$ and $\sin k_y$ related to the structure of the diagrams. In our case, indeed, the four independent tensor elements allowed by the symmetry of the point group can be expressed in terms of two relevant classes of combinations listed in Table \ref{tab:tensdos}, depending on whether diagonal $(xxxx)-$like or off-diagonal $(xxyy)-$like combinations are involved. In Fig.\ \ref{fig:geom2d}, we present the subgap behavior of the third-order susceptibilities entering the computation of the FHG and THG, as done for the 1D case, separating the various tensor elements. In this case, as well, the susceptibilities show a regular behavior for $\omega\ra0$.

    \begin{table}
    \caption{\label{tab:tensdos} Analytical expressions used to compute the various contributions to the nonlinear response kernel for the case of a square lattice, in terms of the complete elliptic integrals of first kind $\mathcal{K}(\sqrt{1-\epsilon^2/4})\equiv\mathcal{K}$ and second kind $\mathcal{E}(\sqrt{1-\epsilon^2/4})\equiv \mathcal{E}$. Here $\epsilon=\epsilon/(2t).$}
    \begin{ruledtabular}
    \begin{tabular}{ccccc}
    \textrm{Diagram} & \textrm{Weight} ($xxxx$) &
    \textrm{Expression}
    & \textrm{Weight} ($xxyy$) &
    \textrm{Expression}\\
    \colrule
    $K^{[4;]}$ & $\cos k_x$ & $\epsilon \mathcal{K}/2$ & - & - \\
    $K^{[3;1]}, K^{[1;3]}$ & $-\sin^2 k_x$ & $(\epsilon^2 \mathcal{K}-4\mathcal{E})/2$ & - & - \\
    $K^{[2;2]}$ & $\cos^2 k_x$ & $((2+\epsilon^2)\mathcal{K}-4\mathcal{E})/2$ & $\cos k_x\cos k_y$ & $2\mathcal{E}-\mathcal{K}$ \\
    $K^{[2;11]},K^{[1;12]}$ & $-\sin^2k_x\cos k_x$ & $\epsilon((2+\epsilon^2)\mathcal{K}-6\mathcal{E})/2$ & $-\sin^2k_x\cos k_y$ & $\epsilon(\mathcal{E}-\mathcal{K})$ \\
    $K^{[1;111]}$ & $\sin^4k_x$ & $(\epsilon^2(8+3\epsilon^2)\mathcal{K}+(8-22\epsilon^2)\mathcal{E})/6$ & $\sin^2k_x\sin^2 k_y$ & $((4+\epsilon^2)\mathcal{E}-2\epsilon^2\mathcal{K})/3$ \\
    \end{tabular}
    \end{ruledtabular}
    \end{table}

\begin{figure}
		\centering
		\includegraphics[width=0.875\textwidth]{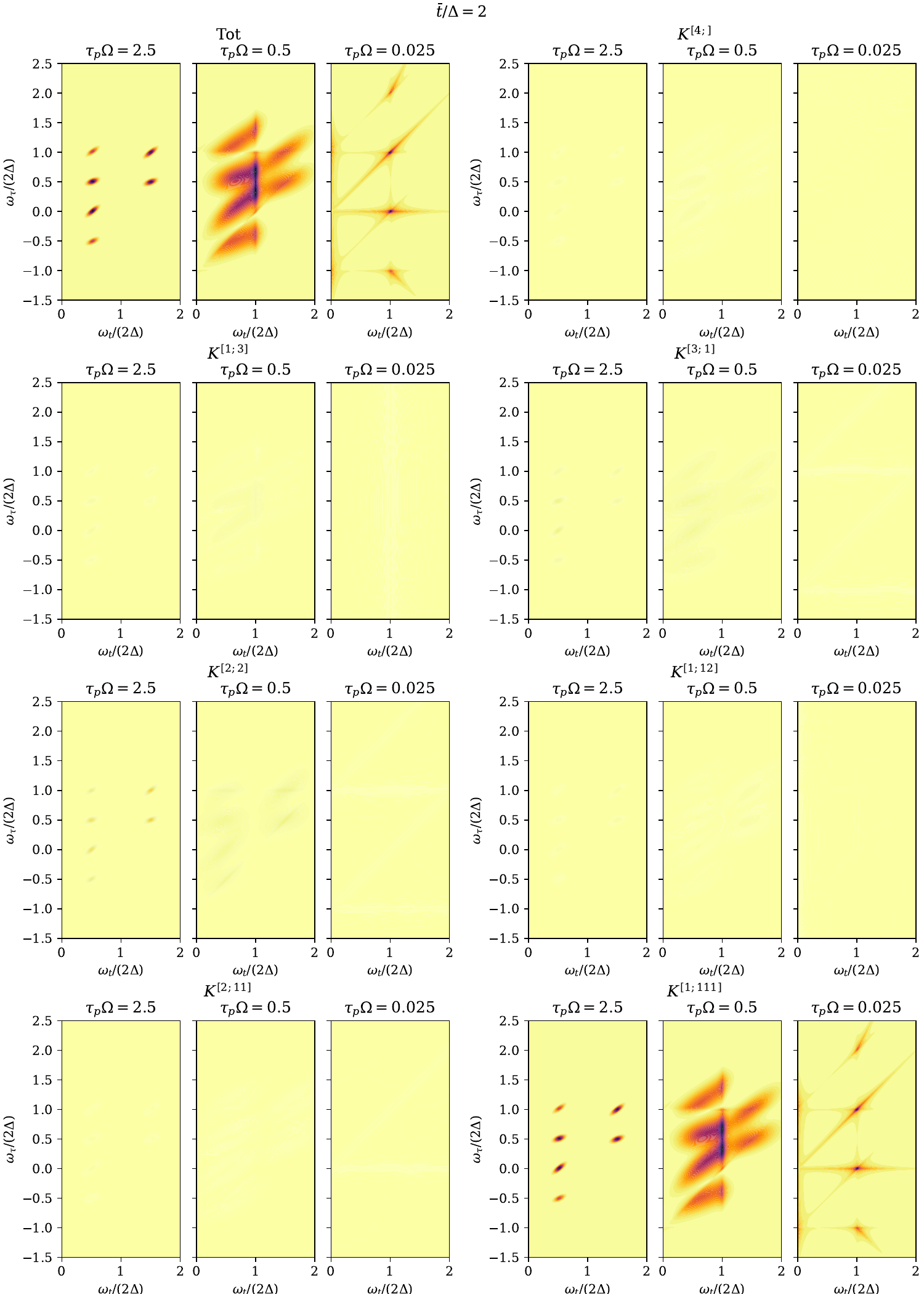}
		\caption{{Fully gauge-invariant 2D map of $|J_{2D}(\omega_t,\omega_\tau)|$ containing all the processes foreseen by Eq.\ \ref{eq:fr}, under the same driving conditions of the Figures in the main text. Each diagrammatic contribution is normalized by the maximum of the total $|J_{2D}(\omega_t,\omega_\tau)|$ at a given value of $\tau_p\Omega$. Here we consider $\bar{t}/\Delta=2$.}}
		\label{fig:td2}
	\end{figure}

\begin{figure}
		\centering
		\includegraphics[width=0.875\textwidth]{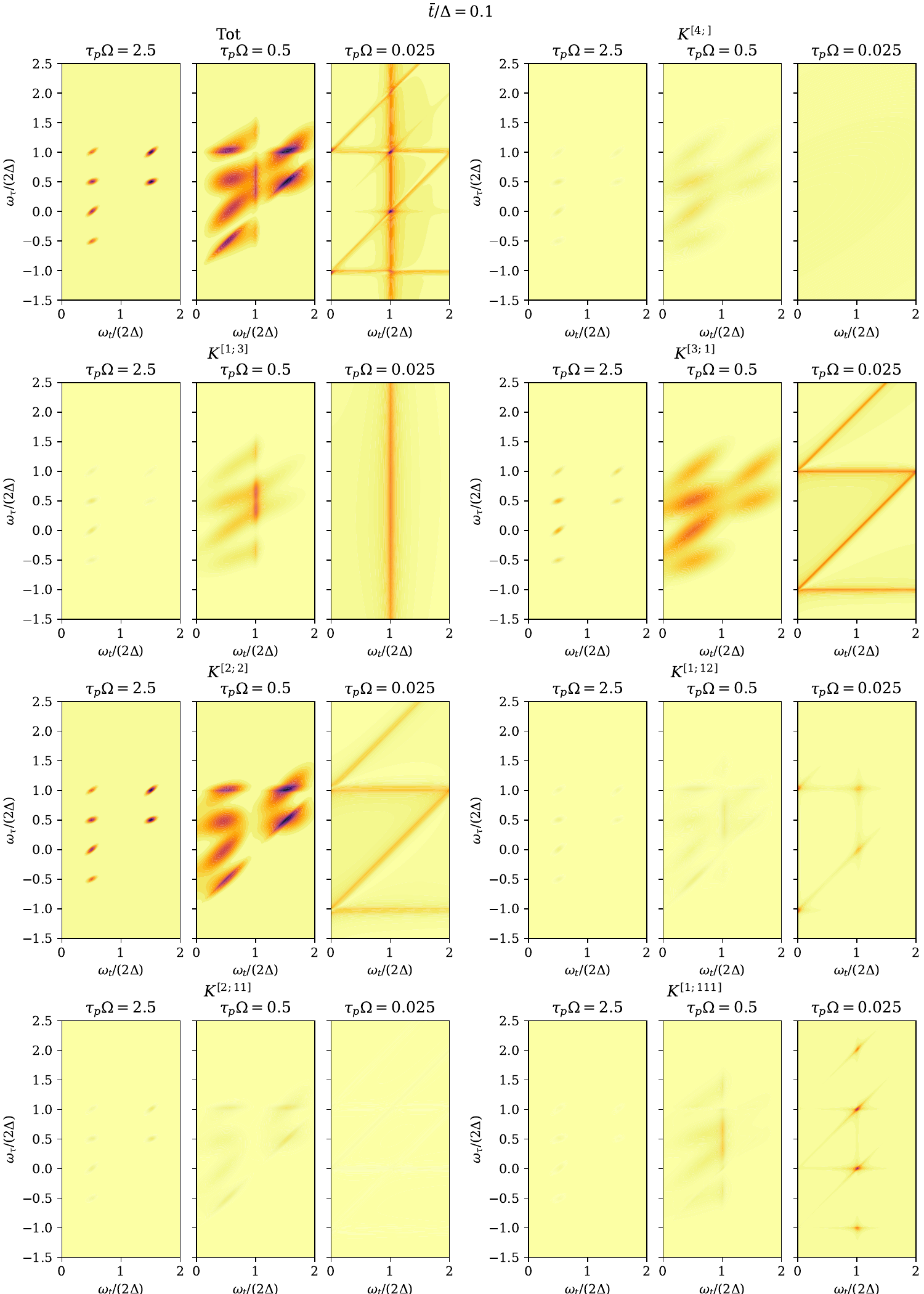}
		\caption{{Fully gauge-invariant 2D map of $|J_{2D}(\omega_t,\omega_\tau)|$ containing all the processes foreseen by Eq.\ \ref{eq:fr}, under the same driving conditions of the Figures in the main text. Each diagrammatic contribution is normalized by the maximum of the total $|J_{2D}(\omega_t,\omega_\tau)|$ at a given value of $\tau_p\Omega$. Here we consider $\bar{t}/\Delta=0.1$.}}
		\label{fig:td01}
	\end{figure}

   {To conclude this section we provide the plot of the fully gauge-invariant sum in the 2D plane $J_{2D}^{\text{(iso)}}(\omega_t,\omega_\tau)$ and of the the separate terms entering it via Eq.\ \ref{eq:fr}. We consider specifically the isotropic, angle-independent component of the 2D response derived above using the tensorial density of states in Table \ref{tab:tensdos} for each process. We employ the same parameters used for generating the various Figures in Sec.\ III of the main text: the gauge potential is taken as $A(t)=\bar{A}e^{-t^2/(2\tau_p^2)}\cos(\Omega t)$, with $\Omega=\Delta$ and $\tau_p\Omega$ fixed at the three values indicated in the panels. Since the contributions entering Eq.\ \ref{eq:fr} are homogeneous, it is reasonable to fix as an overall normalization scale the maximum value of $|J_{2D}^{\text{(iso)}}(\omega_t,\omega_\tau)|$ for the two cases of $\bar{t}/\Delta=2$, see Fig.\ \ref{fig:td2} and $\bar{t}/\Delta=0.1$, see Fig.\ \ref{fig:td01}. We then plot the absolute value of each diagrammatic contribution normalized by the maximum defined above as a proxy of its importance in the gauge-invariant 2D map. As one can notice, each term in the broadband limit brings a distinct signature in the 2D map, with two-point responses producing stripes and three-point responses producing either stripes or spots. This directly reflects the presence of one or more resonant denominators in the frequency structures foreseen in Sec.\ \ref{sec:k4}-\ref{sec:k1111} above, before integration over the density of states. The latter is crucial as well in determining the relative weight of the various terms, according to the nature of the discontinuities encoded in the elliptic integrals in Table \ref{tab:tensdos}.}

    \begin{figure}
		\centering
		\includegraphics[width=0.75\textwidth]{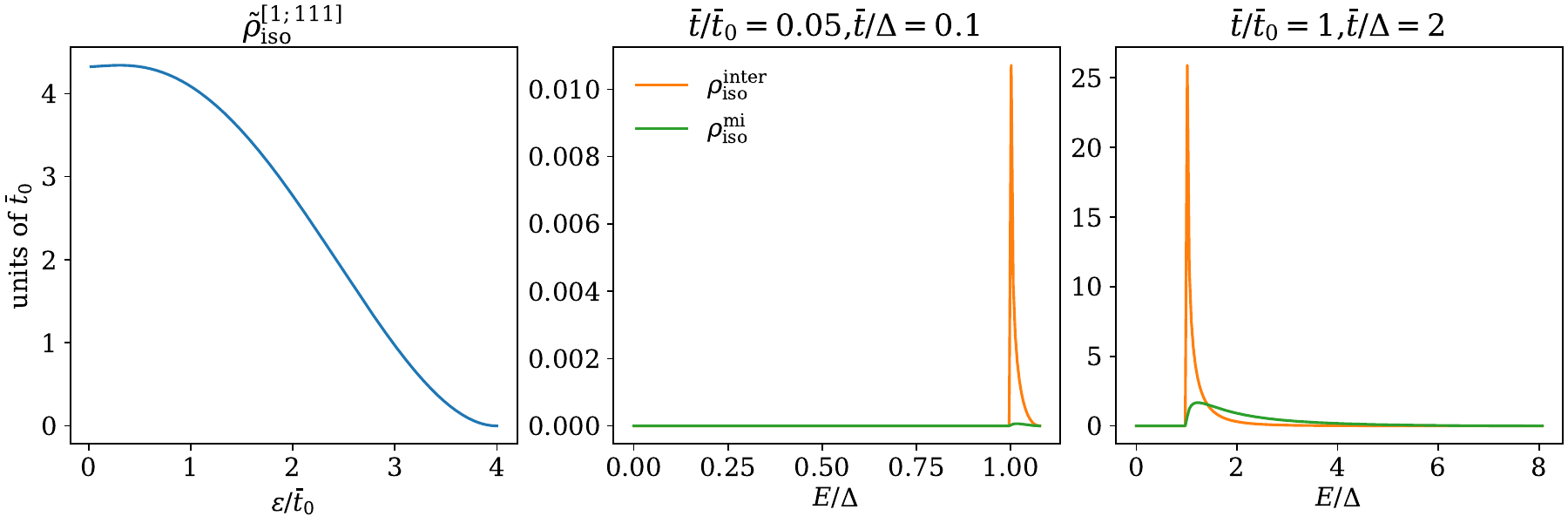}
		\caption{Behavior of the densities of states $\rho^{[1;111]}_{\text{iso}}(\epsilon)$, $\rho^{\text{inter}}_{\text{iso}}(E)$ and $\rho^{\text{mi}}_{\text{iso}}(E)$ used to generate the isotropic contributions to the 2D map coming from the fully-paramagnetic processes. For comparison we set a common energetic unit $\bar{t}_0$ and we have changed $t$ and $\Delta$ as in the main text. One can see that the mixed contribution is negligible in the less dispersive case, while the interband contribution is always diverging for $E\to\Delta$ to the finiteness of $\rho^{[1;111]}_{\text{iso}}(\epsilon)$ at low energy.}
		\label{fig:dosses}
	\end{figure}
    
    Before concluding, we present plots for the relevant densities of states used to generate the angle-independent contributions of the fully-diamagnetic and fully-paramagnetic processes presented in the manuscript. We need to compute
    \begin{align}
    \tilde{\rho}^{[4;]}_{\text{iso}}(\epsilon)&=\frac{1}{V}\sum_\mathbf{k}\delta(\epsilon-\epsilon_\mathbf{k})3w^{[4;]}_{xxxx}(\mathbf{k})\\
    \tilde{\rho}^{[3;1]}_{\text{iso}}(\epsilon)=\tilde{\rho}^{[1;3]}_{\text{iso}}(\epsilon)&=\frac{1}{V}\sum_\mathbf{k}\delta(\epsilon-\epsilon_\mathbf{k})3w^{[3;1]}_{xxxx}(\mathbf{k})\\
    \tilde{\rho}^{[2;2]}_{\text{iso}}(\epsilon)&=\frac{1}{V}\sum_\mathbf{k}\delta(\epsilon-\epsilon_\mathbf{k})\left(3w^{[2;2]}_{xxxx}(\mathbf{k})+w^{[2;2]}_{xxyy}(\mathbf{k})\right)\\
    \tilde{\rho}^{[2;11]}_{\text{iso}}(\epsilon)=\tilde{\rho}^{[2;11]}_{\text{iso}}(\epsilon)&=\frac{1}{V}\sum_\mathbf{k}\delta(\epsilon-\epsilon_\mathbf{k})\left(3w^{[2;11]}_{xxxx}(\mathbf{k})+w^{[2;11]}_{xxyy}(\mathbf{k})\right)\\
    \tilde{\rho}^{[1;111]}_{\text{iso}}(\epsilon)&=\frac{1}{V}\sum_\mathbf{k}\delta(\epsilon-\epsilon_\mathbf{k})\left(3w^{[1;111]}_{xxxx}(\mathbf{k})+3w^{[1;111]}_{xxyy}(\mathbf{k})\right).
    \end{align}
    {We notice that the different prefactor in front of $w^{[2;2]}_{xxyy}$ and $w^{[2;11]}_{xxyy}$ with respect to $w^{[1;111]}_{xxyy}$ is inherited from the nearest neighbors band structure that we have adopted, which implies that the mixed derivatives $\partial_{k_x}\partial_{k_y}\epsilon_\mathbf{k}$ are zero. As a consequence, when computing the contribution $K^{(3)}_{xxyy}+K^{(3)}_{xyxy}+K^{(3)}_{xyyx}$ which enters the angle independent part of the nonlinear current, the processes containing diamagnetic vertices only contribute to one term out of three.} Then, using Eq.\ \ref{eq:rhodia} above we obtain $\rho^{\text{dia}}_{\text{iso}}(E)$, whereas from Eqs.\ \ref{eq:rhointer} and \ref{eq:rhomix} we get $\rho^{\text{inter}}_{\text{iso}}(E)$ and $\rho^{\text{mi}}_{\text{iso}}(E)$, which are reproduced in Fig.\ \ref{fig:dosses}.
    
    \newpage

	\section{Perturbative solution of Maxwell's equations}

    {\subsection{General expressions for the  transmitted/reflected field}}
    In this section, we provide the perturbative solution of Maxwell's equations in presence of a nonlinear source current generated by a $K^{(3)}$ response function. We consider a nonlinear material in the region of space $0<z<d$ characterized by a uniform ($z-$independent) nonlinear third order susceptibility $K^{(3)}(\omega_1,\omega_2,\omega_3)$ sandwiched by vacuum, on which we shine the transverse field $\mathbf{A}=(A(z,\omega),0,0)$ The nonlinearly induced current reads in the most general case
	\be
	\label{eq:pnl}
	J^{(3)}(z,\omega)=\int d\omega_i\,K^{(3)}(\omega_1,\omega_2,\omega_3)A(z,\omega_1)A(z,\omega_2)A(z,\omega_3)\delta\left(\omega-\sum_i\omega_i\right).
	\ee
	This source term will enter Maxwell's equations, resulting in the following Helmholtz equations
	\be
	\label{eq:hel}
	\begin{cases}
		\partial_z^2A(z,\omega)+\frac{\omega^2}{c^2}A(z,\omega)=0&(z<0,z>d)\\
		\partial_z^2A(z,\omega)+\frac{n^2(\omega)\omega^2}{c^2}A(z,\omega)=-\frac{4\pi}{c}J^{(3)}(z,\omega)&(0<z<d),
	\end{cases}
	\ee
	In principle, one has to solve self-consistently the two equations since there is a dependence on the vector potential in the source term of the right hand side. As nonlinear effects are usually smaller with respect to linear ones, which are already taken into account by $n(\omega)$, it is possible to solve the previous differential equation with a perturbative approach. Let us introduce a fictitious small parameter $\eta$ to keep track of this perturbative expansion, such that
	\be
	K^{(3)}(\omega_1,\omega_2,\omega_3)\equiv\eta\tilde{K}^{(3)}(\omega_1,\omega_2,\omega_3).
	\ee
	Then we suppose
	\be
	A(z,\omega)=\sum_{m}\eta^{m}A^{(m)}(z,\omega),
	\ee
	and replace these expressions in Eq.\ \ref{eq:pnl} 
	\be
	\begin{split}
	J^{(3)}(z,\omega)=\sum_{p,q,r}\eta^{p+q+r+1}\int d\omega_i\,K^{(3)}(\omega_1,\omega_2,\omega_3)A^{(p)}(z,\omega_1)A^{(q)}(z,\omega_2)A^{(r)}(z,\omega_3)\delta\left(\omega-\sum_i\omega_i\right).
	\end{split}
	\ee
	We can solve Eq.\ \ref{eq:hel} order by order in $\eta$, collecting the corresponding terms on the two sides of the equation. For $\eta^0$ we have
	\be
	\begin{cases}
		\partial_z^2A^{(0)}(z,\omega)+\frac{\omega^2}{c^2}A^{(0)}(z,\omega)=0&(z<0,z>d)\\
		\partial_z^2A^{(0)}(z,\omega)+\frac{n^2(\omega)\omega^2}{c^2}A^{(0)}(z,\omega)=0&(0<z<d),\\
	\end{cases}
	\ee
	which is the usual propagation of waves in a slab of thickness $d$ in the linear regime. The solution to the system for the linear response in the region $0<z<d$ is
	\be
	\label{eq:inf}
	A^{(0)}(z,\omega)=A_{t}(\omega)e^{i\omega n(\omega)z/c}+A_{r}(\omega)e^{-i\omega n(\omega)z/c}.
	\ee
	Here we have
	\be
    \label{eq:tin}
	A_{t}(\omega)=A_{\text{ext}}(\omega)t(\omega)f(\omega)\equiv A_{\text{ext}}(\omega)t^{\text{in}}(\omega),
	\ee
	\be
    \label{eq:rin}
	A_{r}(\omega)=A_{\text{ext}}(\omega)t(\omega)e^{i2\omega n(\omega)d/c}r(\omega)f(\omega)\equiv A_{\text{ext}}(\omega)r^{\text{in}}(\omega),
	\ee
	in terms of the spectrum of the incident field $A_{\text{ext}}(\omega)$, the usual transmission and reflection coefficients from a vacuum-material interface
	\be
	t(\omega)=\frac{2}{n(\omega)+1}, 
	\ee
	\be
	r(\omega)=\frac{n(\omega)-1}{n(\omega)+1},
	\ee
	and the Fabry-Pérot factor 
	\be
	f(\omega)=\frac{1}{1-r^2(\omega)e^{i2\omega n(\omega)d/c}}.
	\ee 
	
	\begin{figure}
		\centering
		\includegraphics[width=0.5\textwidth]{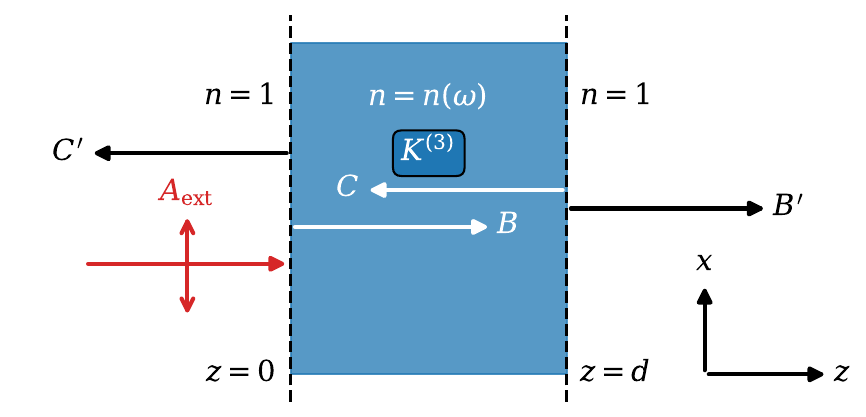}
		\caption{A film of thickness $d$ with refractive index $n(\omega)$ and nonlinear response kernel $K^{(3)}$ is sandwiched by vacuum at $z=0$ and $z=d$. The gauge potential $A_{\text{ext}}(\omega)$ impinges from $z<0$ and generates the fields $A^{(0)}(z,\omega)$, containing linear effects, and $A^{(1)}(z,\omega)$, containing nonlinear effects. The latter depends on the coefficients $B,C$ for $0<z<d$, and $B^{\prime},C^{\prime}$ for $z>d$ and $z<0$, respectively, which enter
			 the system for the boundary conditions in Eq.\ \ref{eq:sistemone}.}
		\label{fig:films}
	\end{figure}
	
	Due to the linearity of Maxwell's equations in the absence of source terms, one can in principle extend by superposition the previous solution for arbitrary frequency spectrum of the incident field. For $\eta^1$ we obtain
	\be
	\label{eq:genmax}
	\begin{cases}
		\partial_z^2A^{(1)}(z,\omega)+\frac{\omega^2}{c^2}A^{(1)}(z,\omega)=0&(z<0,z>d)\\
		\partial_z^2A^{(1)}(z,\omega)+\frac{n^2(\omega)\omega^2}{c^2}A^{(1)}(z,\omega)=-\frac{4\pi}{c}J^{(3)}_{\eta^1}(z,\omega)&(0<z<d),\\
	\end{cases}
	\ee
	where 
	\be
	\label{eq:je1}
	J^{(3)}_{\eta^1}(z,\omega)=\int d\omega_i\,K^{(3)}(\omega_1,\omega_2,\omega_3)A^{(0)}(z,\omega_1)A^{(0)}(z,\omega_2)A^{(0)}(z,\omega_3)\delta\left(\omega-\sum_i\omega_i\right).
	\ee
	The solution to the last equation of the previous system reads as usual
	\be
	A^{(1)}(z,\omega)=A_p(z)+Be^{i\omega n(\omega)z/c}+Ce^{-i\omega n(\omega)z/c},		
	\ee
	where we denote by $A_p(z)$ the particular solution. Taking into account the outgoing nonlinear signal in free space, $A^{(1)}(z<0,\omega)=C^{\prime}e^{-i\omega z/c}$ and $A^{(1)}(z>d,\omega)=B^{\prime}e^{i\omega z/c}$, the boundary conditions read 
	\be
	\label{eq:sistemone}
	\begin{cases}
		C^{\prime}&=A_p(0)+B+C\\
		-\frac{i\omega}{c}C^{\prime}&=A_p^{\prime}(0)+\frac{in(\omega)\omega}{c}(B-C)\\
		B^{\prime}e^{i\omega d/c}&=A_p(d)+Be^{in(\omega)\omega d/c}+Ce^{-in(\omega)\omega d/c}\\
		\frac{i\omega}{c}B^{\prime}e^{i\omega d/c}&=A_p^{\prime}(d)+\frac{in(\omega)\omega}{c}(Be^{in(\omega)\omega d/c}-Ce^{-in(\omega)\omega d/c}).
	\end{cases}
	\ee
	
	If one is interested in finding the nonlinear signal detected in the reflection configuration, the all important coefficient is $C^{\prime}$, while in transmission the relevant quantity is $B^{\prime}$; both can be found by inverting the linear system written before in terms of the values of the particular solution $A_p$ and its first derivative $A_p^{\prime}$ at the two ends of the slab. 
	
	We now focus on the expression for $B^{\prime}$, which is the relevant quantity to derive the expression for the electric field right beyond the sample in the standard THG experiments in transmission configuration. If one uses a reflection geometry, one analogously computes the coefficient $C^{\prime}$, as we will see later in the limit of bulk samples. Redefining now $B^{\prime}e^{i\omega d/c}\to B^{\prime}$ to discard the pure phase factor, which is irrelevant for our purposes, it reads explicitly
	\be
	\begin{split}
	B^{\prime}=-\frac{f(\omega)}{(n(\omega)+1)^2}&\left[2n(\omega)e^{in(\omega)\omega d/c}\left(A_p(0)-\frac{ic}{\omega}A_p^{\prime}(0)\right)+(n(\omega)-1)e^{i2n(\omega)\omega d/c}\left(n(\omega)A_p(d)+\frac{ic}{\omega}A_p^{\prime}(d)\right)\right.\\
	&\left.-(n(\omega)+1)\left(n(\omega)A_p(d)-\frac{ic}{\omega}A_p^{\prime}(d)\right)\right].
	\end{split}
	\ee
	In order to compute the transmission, one should then evaluate the particular solution and its derivative. For instance, one can use Fourier transforms and consider the following expression of the particular solution
	\be
	A_p(z,\omega)=\frac{4\pi}{c}\frac{1}{2\pi}\int dk\,\frac{e^{-ikz}}{k^2-n^2(\omega)\omega^2/c^2}J_{\eta^1}^{(3)}(k,\omega),
	\ee
	where $J_{\eta^1}^{(3)}(k,\omega)$ is the Fourier transform of the nonlinearly induced current, Eq.\ \ref{eq:je1}, that we rewrite as 
	\be
		J_{\eta^1}^{(3)}(k,\omega)=\sum_{\{\alpha_i\}}\int_0^{d}dz\int d\omega_i\,e^{i(k+\sum_i\alpha_in(\omega_i)\omega_i/c)z}K^{(3)}(\omega_1,\omega_2,\omega_3)A_{\alpha_1}(\omega_1)A_{\alpha_2}(\omega_2)A_{\alpha_3}(\omega_3)\delta\left(\omega-\sum_i\omega_i\right).
	\ee
	Here we used the $z$-dependence of the field within the material given by Eq.\ \ref{eq:inf}, separating the (eight) contributions according to the number of forward ($\propto A_t$) or backward ($\propto A_r$) propagating fields appearing in the triple $A^{(0)}$ product. This amounts to a sum of the possible $\{\alpha_i\}=(\alpha_1,\alpha_2,\alpha_3)$ combinations where $\alpha_i=+1$ for the $A_t$ terms and $\alpha_i=-1$ for the $A_r$ terms. The integral over $z$ can be explicitly performed, obtaining
	\be
		J_{\eta^1}^{(3)}(k,\omega)=\sum_{\{\alpha_i\}}\int d\omega_i\,\frac{e^{i(k+\sum_i\alpha_in(\omega_i)\omega_i/c)d}-1}{i(k+\sum_i\alpha_in(\omega_i)\omega_i/c)}K^{(3)}(\omega_1,\omega_2,\omega_3)A_{\alpha_1}(\omega_1)A_{\alpha_2}(\omega_2)A_{\alpha_3}(\omega_3)\delta\left(\omega-\sum_i\omega_i\right).
	\ee	
	Plugging the last expression into $A_p(z)$ we get 
	\be
	\label{eq:apz}
	\begin{split}
		A_p(z,\omega)=\frac{4\pi}{c}\sum_{\{\alpha_i\}}\int d\omega_i\,&K^{(3)}(\omega_1,\omega_2,\omega_3)A_{\alpha_1}(\omega_1)A_{\alpha_2}(\omega_2)A_{\alpha_3}(\omega_3)\delta\left(\omega-\sum_i\omega_i\right)\times\\
		\times&\frac{1}{2\pi}\int dk\,\frac{e^{-ikz}}{k^2-n^2(\omega)\omega^2/c^2}\frac{e^{i(k+\sum_i\alpha_in(\omega_i)\omega_i/c)d}-1}{i(k+\sum_i\alpha_in(\omega_i)\omega_i/c)}.
	\end{split}
	\ee
	
	\begin{figure}
		\centering
		\includegraphics[width=0.5\textwidth]{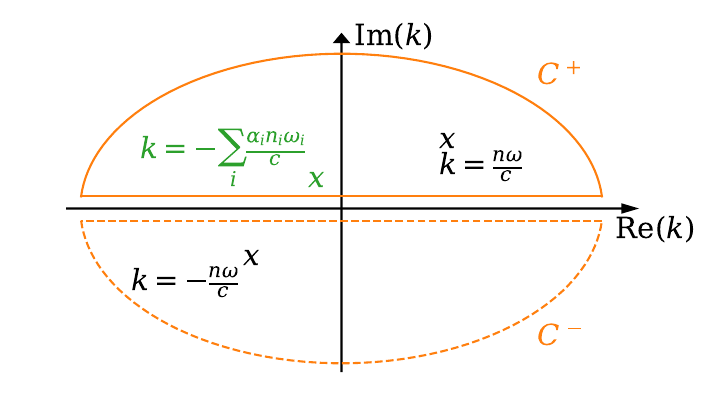}
		\caption{The integration contours used in Eq.\ \ref{eq:contour}. The green pole lies in the upper or lower half-plane according to the signs of $\alpha_i$. Both $\mathcal{C}^{+}$ and $\mathcal{C}^{-}$ are positively oriented.}
		\label{fig:cont}
	\end{figure}
	
	The $k$-integral can then be computed through residues. Since we are interested in the particular solution for $z\to0^+$ and $z\to d^-$, we need to choose the integration contour accordingly. For the first term in the $k$-integral ($\propto e^{-ik(z-d)}$), we should close the contour in the upper half plane, since $z-d<0$, while for the second term ($\propto e^{-ikz}$) we should close the contour in the lower half plane, since $z>0$. We have
	\be
    \label{eq:contour}
	\begin{split}
		\frac{1}{2\pi}\int dk\,\frac{e^{-ikz}}{k^2-n^2(\omega)\omega^2/c^2}\frac{e^{i(k+\sum_i\alpha_in(\omega_i)\omega_i/c)d}-1}{i(k+\sum_i\alpha_in(\omega_i)\omega_i/c)}&=\int_{\mathcal{C}^{+}} dk\,\frac{e^{-ik(z-d)}e^{i\sum_i\alpha_in(\omega_i)\omega_id/c}}{(k^2-n^2(\omega)\omega^2/c^2)(k+\sum_i\alpha_in(\omega_i)\omega_i/c)}\\
		&+\int_{\mathcal{C}^{-}} dk\,\frac{e^{-ikz}}{(k^2-n^2(\omega)\omega^2/c^2)(k+\sum_i\alpha_in(\omega_i)\omega_i/c).}
	\end{split}
	\ee
	In both contours, we will have one pole determining the (forward or backward) Maxwell-like propagation of the nonlinearly generated radiation ($k=\pm n(\omega)\omega/c$) and the one(s) corresponding to the propagation of the internal fields generating the nonlinear signal ($k=-\sum_i\alpha_in(\omega_i)\omega_i/c$). According to the overall sign of $\sum_i\alpha_in^{\prime\prime}(\omega_i)\omega_i/c$ the corresponding pole(s) will lie in the lower or upper half of the complex plane, thus being enclosed in one or the other integration contour. Denoting for brevity $n=n(\omega)$ and $n_i=n(\omega_i)$, for $0<z<d$ we get 
	\be
	\begin{split}
		\frac{1}{2\pi}\int dk\,\frac{e^{-ikz}}{k^2-n^2\omega^2/c^2}\frac{e^{i(k+\sum_i\alpha_in_i\omega_i/c)d}-1}{i(k+\sum_i\alpha_in_i\omega_i/c)}&=\frac{e^{in\omega z/c}}{(-n\omega/c+\sum_i\alpha_in_i\omega_i/c)(2n\omega/c)}-\frac{e^{-in\omega z/c}e^{i\left(n\omega+\sum_i\alpha_in_i\omega_i\right)d/c}}{(n\omega/c+\sum_i\alpha_in_i\omega_i/c)(2n\omega/c)}\\
		&-\frac{e^{i\sum_i\alpha_in_i\omega_iz/c}}{(n\omega/c)^2-\left(\sum_i\alpha_in_i\omega_i/c\right)^2}.
	\end{split}
	\ee
	Using this result in Eq.\ \ref{eq:apz}, we can compute the values of the particular solution and its first derivative at the boundaries of the slab. The coefficient $B^{\prime}$ reads
	\be
	\label{eq:aprime}
	\begin{split}
		B^{\prime}&=\frac{4\pi i}{\omega}\frac{fe^{in\omega d/c}}{n+1}\sum_{\{\alpha_i\}}\int d\omega_i\,K^{(3)}(\omega_1,\omega_2,\omega_3)A_{\alpha_1}(\omega_1)A_{\alpha_2}(\omega_2)A_{\alpha_3}(\omega_3)\times\\
		&\times\left(\frac{n-1}{n+1}\frac{e^{i\left(\sum_i\alpha_in_i\omega_i+n\omega\right)d/c}-1}{i\left(\sum_i\alpha_in_i\omega_i+n\omega\right)/c}+\frac{e^{i\left(\sum_i\alpha_in_i\omega_i-n\omega\right)d/c}-1}{i\left(\sum_i\alpha_in_i\omega_i-n\omega\right)/c}\right)\delta\left(\omega-\sum_i\omega_i\right).
	\end{split}
	\ee
	This result is valid for any value of the thickness of the film and takes into account all the propagation effects inside the material. We recognize in the second line the phase matching conditions between the nonlinearly generated radiation and the internal fields, which appear with both minus and plus signs according to the specific combinations of forward or backward propagating waves that we can consider. As expected, the transmission is maximized when the phase matching condition is satisfied,
    \be
    \sum_i\alpha_in_i\omega_i\pm n\omega=0.
    \ee
    Clearly, in non-dispersive media the phase-matching condition coincides with energy conservation, while here it strongly depends on $n(\omega)$. As a final remark we point out that a similar result has been obtained in Ref.\ \cite{huber_TheJournalofChemicalPhysics21SMREF}, where the effects of phase-mismatch between THz-pump and eV-probe have been studied in the context of ultrafast Kerr effect, i.e.\ considering an instantaneous $K^{(3)}$. A minor difference is just that here we looked for an analytical solution to the full second-order Maxwell's equation in the second line of Eq.\ \ref{eq:genmax}, while in Ref.\ \cite{huber_TheJournalofChemicalPhysics21SMREF} the authors explicilty integrate the $z-$dependence of the nonlinear polarization.

    {We want now to use the coefficient $B^\prime$ to compute the transmitted electric field in a 2D experiment at $z=d^+$. Due to the subtraction procedure, contributions from the direct linear response encoded in $A^{(0)}(z,\omega)$ are canceled out and we just need to take into account those coming from $A^{(1)}(z,\omega)$, which exactly coincides with $B^{\prime}$ at $z=d^+$. Notice that since we are interested to the electric field, we need to consider the relation $E(\omega)=i\omega A(\omega)/c$ and apply it to $B^\prime$. Let us consider e.g.\ the contribution with the double interaction with the $A_0(t)$ field and the single interaction with $A_{\tau}(t)$, which in the notation of the main text reads
    \be
    E_{00\tau}(\omega_t,\omega_\tau)=\int d\omega\,K^{(3)}_{\mathcal{S}}(\omega_\tau,\omega,\omega_t-\omega_\tau-\omega)A_{\tau,\text{ext}}(\omega_\tau)A_{0,\text{ext}}(\omega)A_{0,\text{ext}}(\omega_t-\omega_\tau-\omega)M(\omega_\tau,\omega,\omega_t-\omega_\tau-\omega)
    \ee
    One starts from Eq.\ \ref{eq:aprime} and replaces one frequency with $\omega_\tau$ and the other two with $\omega$ and $\omega_t-\omega_\tau-\omega$, eliminating one frequency integration via the energy conservation and another one with the $\tau\to\omega_\tau$ Fourier transform. We introduce the shortcut notation $n(\omega_t)=n_t$, $n(\omega_\tau)=n_\tau$, $n(\omega)=n$, $n(\omega_t-\omega_\tau-\omega)=n_{t-\tau}$ and we set $\mathcal{I}_{+1}=t^{\text{in}}$ and $\mathcal{I}_{-1}=r^{\text{in}}$ from Eqs.\ \ref{eq:tin} and \ref{eq:rin}, respectively. We obtain finally
    \be
    \label{eq:del}
	\begin{split}
		M(\omega_\tau,\omega,\omega_t-\omega_\tau-\omega)&=-\frac{4\pi}{c}\frac{f_te^{in_t\omega_t d/c}}{n_t+1}\sum_{\alpha_\tau,\alpha,\alpha_{t-\tau}=\pm1}\mathcal{I}_{\alpha_\tau}(\omega_\tau)\mathcal{I}_{\alpha}(\omega)\mathcal{I}_{\alpha_{t-\tau}}(\omega_t-\omega_{\tau}-\omega)\times\\
		&\times\left(\frac{n_t-1}{n_t+1}\frac{e^{i\left(\alpha_\tau n_\tau\omega_\tau+\alpha n\omega+\alpha_{t-\tau}n_{t-\tau}(\omega_t-\omega_{\tau}-\omega)+n_t\omega_t\right)d/c}-1}{i\left(\alpha_\tau n_\tau\omega_\tau+\alpha n\omega+\alpha_{t-\tau}n_{t-\tau}(\omega_t-\omega_{\tau}-\omega)+n_t\omega_t\right)/c}\right.\\
        &\left.+\frac{e^{i\left(\alpha_\tau n_\tau\omega_\tau+\alpha n\omega+\alpha_{t-\tau}n_{t-\tau}(\omega_t-\omega_{\tau}-\omega)-n_t\omega_t\right)d/c}-1}{i\left(\alpha_\tau n_\tau\omega_\tau+\alpha n\omega+\alpha_{t-\tau}n_{t-\tau}(\omega_t-\omega_{\tau}-\omega)-n_t\omega_t\right)/c}\right).
	\end{split}
	\ee
    We stress once again that this expression allows one to simulate the 2D map obtained in a transmission geometry for a film of arbitrary thickness and refractive index.}

\subsection{Approximate expressions for limiting cases}
	
\begin{figure}
		\centering
		\includegraphics[width=\textwidth]{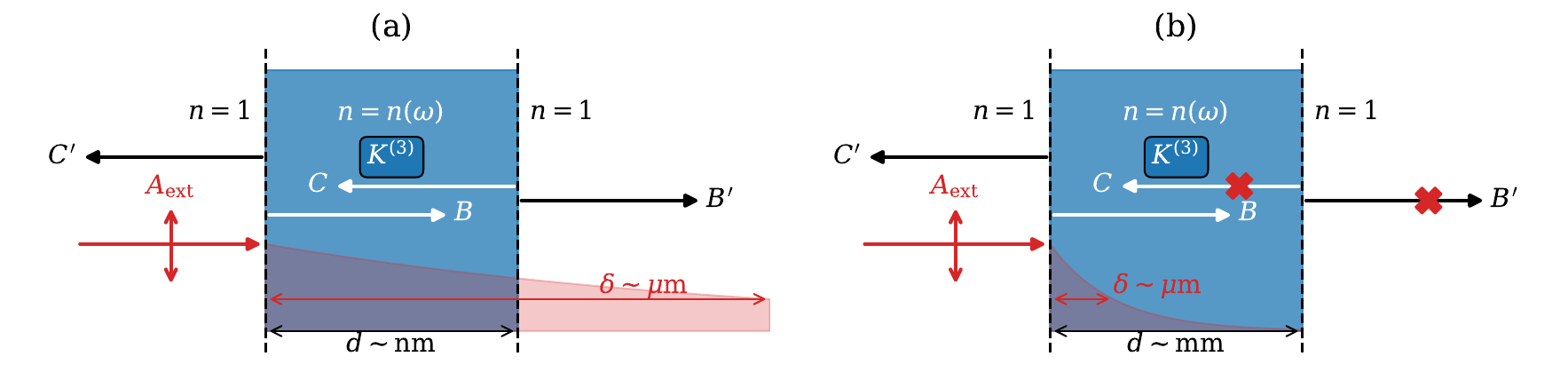}
		\caption{Scheme for the approximation used in trasmission experiments on thin films (a) and reflection from bulk samples (b). In panel (a) the thickness is the smallest lengthscale at play, the crudest approximation in Eq.\ \ref{eq:crude} amounts to consider the profile of the field constant inside the material, while in Eq.\ \ref{eq:cotto} one keeps the effect of screening. In panel (b) the process is approximated as a reflection from a semi-infinite medium. Since the field penetrates just in a small fraction of the sample, one can neglect in Eq.\ \ref{eq:sistemone} both the transmitted field outside the material and the back-propagating internal one.}
		\label{fig:approx}
	\end{figure}

	It is useful to derive an approximate expression of Eq.\ \ref{eq:aprime} to be used in some limiting cases. For very thin films, see Fig.\ \ref{fig:approx}(a), with thickness $d$ much shorter than wavelength $\lambda=c/(n^{\prime}\omega)$ and penetration depth $\delta=c/(n^{\prime\prime}\omega)$ of the radiation inside the material, one can expand the term in round brackets on the second line of Eq.\ \ref{eq:aprime} using $e^{i\alpha}\sim1+i\alpha$ for $\alpha\ll1$, obtaining
	\be
	\frac{n-1}{n+1}\frac{e^{i\alpha d}-1}{i\alpha}+\frac{e^{i\alpha^{\prime} d}-1}{i\alpha^{\prime}}\sim\frac{2nd}{n+1}.
	\ee
	If one is interested in the electric field at leading order in $d$, one must set $d=0$ in all other pieces of Eq.\ \ref{eq:aprime}. Using the relation $E(\omega)=i\omega A(\omega)/c$, we recover the expression
	\be
	\label{eq:crude}
	E(z=d^+,\omega)=E^{(0)}(z=d^+,\omega)-\frac{2\pi d}{c}J^{(3)}_{\eta^1}(z=0,\omega),
	\ee
	where now, since for $d=0$ one has $A^{(0)}(z=0,\omega)=A_{\text{ext}}(\omega)$ in Eq.\ \ref{eq:je1}, the nonlinear current becomes
	\be
	J^{(3)}_{\eta^1}(z=0,\omega)=\int d\omega_i\,K^{(3)}(\omega_1,\omega_2,\omega_3)A_{\text{ext}}(\omega_1)A_{\text{ext}}(\omega_2)A_{\text{ext}}(\omega_3)\delta\left(\omega-\sum_i\omega_i\right).
	\ee
	This result is not surprising: in the approximation $d\to0$, neither the nonlinearly generated current nor the field that generates it is sensitive to any propagation effect. {We have thus provided a basis to justify the approximation employed in Secs.\ II-III of the main text to connect the electric field transmitted through non-absorbing thin films to the nonlinear current generated in response to the \textit{applied} field.}
	
	Instead, a less crude approximation is obtained if one consistently takes into account the screening of the incoming field inside the material. Assuming that $\lvert n\rvert\omega d/c\ll1$, one can rewrite the previous Eq.\ \ref{eq:aprime} as
	\be
	\label{eq:cotto}
	B^{\prime}=\frac{2\pi id}{\omega}T(\omega)\int d\omega_i\,K^{(3)}(\omega_1,\omega_2,\omega_3)A_{\text{in}}(\omega_1)A_{\text{in}}(\omega_2)A_{\text{in}}(\omega_3)\delta\left(\omega-\sum_i\omega_i\right),
	\ee
	where we defined
	\be
	T(\omega)=\frac{2n}{2n\cos(n\omega d/c)-i(n^2+1)\sin(n\omega d/c)}\approx\frac{1}{1-i(n^2+1)\omega d/(2c)},
	\ee
	which is the transmissivity of the vacuum-film-vacuum sandwich, and the internal field $A_{\text{in}}(\omega)=A^{(0)}(z=0,\omega)$ appearing in Eq.\ \ref{eq:je1}
	\be
	A_{\text{in}}(\omega)=A_{\text{ext}}(\omega)\frac{2n\cos(n\omega d/c)-2i\sin(n\omega d/c)}{2n\cos(n\omega d/c)-i(n^2+1)\sin(n\omega d/c)}\approx A_{\text{ext}}(\omega)\frac{1-i\omega d/c}{1-i(n^2+1)\omega d/(2c)}.
	\ee
	Since $\omega d/c\ll1$, by comparing the last two equations, one can also approximate the internal field with the transmitted one, i.e.\ one can take $A_{\text{in}}(\omega)\approx T(\omega)A_{\text{ext}}(\omega)$. This result then validates the common approximation where $E_{\text{in}}(\omega)=T(\omega)E_{\text{out}}(\omega)$.

	On the other hand, we can consider reflection from bulk systems, considering the opposite limit when $d\gg\lambda,\delta$, as usually happens for experiments with $c-$axis polarized electric fields on cuprate samples\cite{zhang_NationalScienceReview23SMREF,katsumi_Phys.Rev.B23SMREF}. An equivalent system of Eq.\ \ref{eq:sistemone}, but including only the boundary conditions at $z=0$, allows us to get
	\be
	\label{eq:bprime}
	C^{\prime}=\frac{1}{n(\omega)+1}\left(n(\omega)A_p(0)+\frac{ic}{\omega}A_p^{\prime}(0)\right).
	\ee
	Considering the expression of the particular solution
	\be
	\label{eq:apb}
	A_p(z,\omega)=\frac{4\pi}{c}\frac{1}{2\pi}\int dk\,\frac{e^{-ikz}}{k^2-n^2(\omega)\omega^2/c^2}J_{\eta^1}^{(3)}(k,\omega),
	\ee
	we should compute the Fourier transform of the nonlinearly induced current. We notice that in the bulk limit, for samples much thicker than the penetration depth of the radiation inside the material, the backward propagating internal fields entering $A^{(0)}(z,\omega)$ in Eq.\ \ref{eq:inf} through $A_r(\omega)$ are exponentially suppressed, see Fig.\ \ref{fig:approx}(b). Thus, only forward propagating internal fields $(\propto A_t^{\prime})$ contribute, giving
	\be
		J_{\eta^1}^{(3)}(k,\omega)=\int_0^{\infty}dz\int d\omega_i\,e^{i(k+\sum_in(\omega_i)\omega_i/c)z}K^{(3)}(\omega_1,\omega_2,\omega_3)A_{\text{ext}}(\omega_1)t(\omega_1)A_{\text{ext}}(\omega_2)t(\omega_2)A_{\text{ext}}(\omega_3)t(\omega_3)\delta\left(\omega-\sum_i\omega_i\right).
	\ee
	Here we extend the upper limit of integration to infinity, since the thickness is the largest length scale at play here. We can nevertheless ensure the convergence of the integral in $z$ since i) $\omega n^{\prime\prime}(\omega)$ is positive-definite for any frequency and ii) the field has a finite cutoff in frequency given by the incident radiation. Then one gets
	\be
		J_{\eta^1}^{(3)}(k,\omega)=i\int d\omega_i\,K^{(3)}(\omega_1,\omega_2,\omega_3)\frac{A_{\text{ext}}(\omega_1)t(\omega_1)A_{\text{ext}}(\omega_2)t(\omega_2)A_{\text{ext}}(\omega_3)t(\omega_3)}{k+\sum_in(\omega_i)\omega_i/c}\delta\left(\omega-\sum_i\omega_i\right).
	\ee
	Plugging the latter expression into Eq.\ \ref{eq:apb} to find the particular solution and computing the coefficient $C^{\prime}$ from Eq.\ \ref{eq:bprime}, one finally gets 
	\be
	C^{\prime}=\frac{4\pi}{\omega}\frac{1}{1+n(\omega)}\int d\omega_i\,K^{(3)}(\omega_1,\omega_2,\omega_3)A_{\text{ext}}(\omega_1)A_{\text{ext}}(\omega_2)A_{\text{ext}}(\omega_3)\frac{t(\omega_1)t(\omega_2)t(\omega_3)}{n(\omega)\omega/c+\sum_in(\omega_i)\omega_i/c}\delta\left(\omega-\sum_i\omega_i\right).
	\ee
	The reflected electric field, again using the relation $E(\omega)=i\omega 
	A(\omega)/c$, reads in this case 
	\be
	\label{eq:eref}
		E(z<0,\omega)=E^{(0)}(z<0,\omega)+2\pi it(\omega)\int d\omega_i\,K^{(3)}(\omega_1,\omega_2,\omega_3)\frac{A_{\text{in}}(\omega_1)A_{\text{in}}(\omega_2)A_{\text{in}}(\omega_3)}{n(\omega)\omega+\sum_in(\omega_i)\omega_i}\delta\left(\omega-\sum_i\omega_i\right),
	\ee
	where the internal field in this case is given by $A_{\text{in}}(\omega)=A_{\text{ext}}(\omega)t(\omega)$. This result, specialized for a monochromatic incoming field and focusing on the third harmonic, has been used in Ref.\ \cite{fiore_Phys.Rev.B24SMREF} to compare the experimental measurements of Ref.\ \cite{katsumi_Phys.Rev.B23SMREF} with the computed two-plasmon kernel. In Ref.\ \cite{zhang_NationalScienceReview23SMREF}, a similar procedure is also presented to simulate the two-dimensional THz map of cuprate LBCO. {To conclude, we recall that from this expression, which we reported in the main text, we can obtain the 2D map for a reflection geometry from a bulk sample following the same steps outlined above to get to Eq.\ \ref{eq:del}.}

   {\subsection{Comparison with the results of Ref.\ \cite{gomezsalvador_Phys.Rev.B24SMREF}}}
	We now comment on the differences between our approach to propagation effects and the one presented in Ref.\ \cite{gomezsalvador_Phys.Rev.B24SMREF}. In the first part of the work, the authors obtain an expression that is completely analogous to our Eq.\ \ref{eq:eref}, which we report here 
    \be
    \label{eq:alex}
		E^{(1)}(z<0,\omega)=\chi_{0}t(\omega)\int d\omega_i\,\frac{E_{\text{ext}}t(\omega_1)E_{\text{ext}}(\omega_2)t(\omega_2)E_{\text{ext}}(\omega_3)t(\omega_3)}{\omega_1\omega_2\omega_3(n(\omega)\omega+\sum_in(\omega_i)\omega_i)}\delta\left(\omega-\sum_i\omega_i\right).
	\ee
    The constant $\chi_0$ is obtained in Ref.\ \cite{gomezsalvador_Phys.Rev.B24SMREF} by a perturbative solution of the Bulaevskii model for Josephson plasmons, derived from a discretized description of Maxwell's equations for a layered superconductor. The reflected electric field is obtained, as in our case, explicitly solving the boundary-value problem set at the interface between vacuum and material. It is no surprise then that the two derivations in Eqs.\ \ref{eq:eref} and \ref{eq:alex} ultimately lead to the same results, although in our case we have separated the computation of the third-order optical kernel, which can be obtained for any other system exhibiting nonlinear response, from the solution of the light-propagation problem.
    
\begin{figure}
\includegraphics[width=0.875\columnwidth]{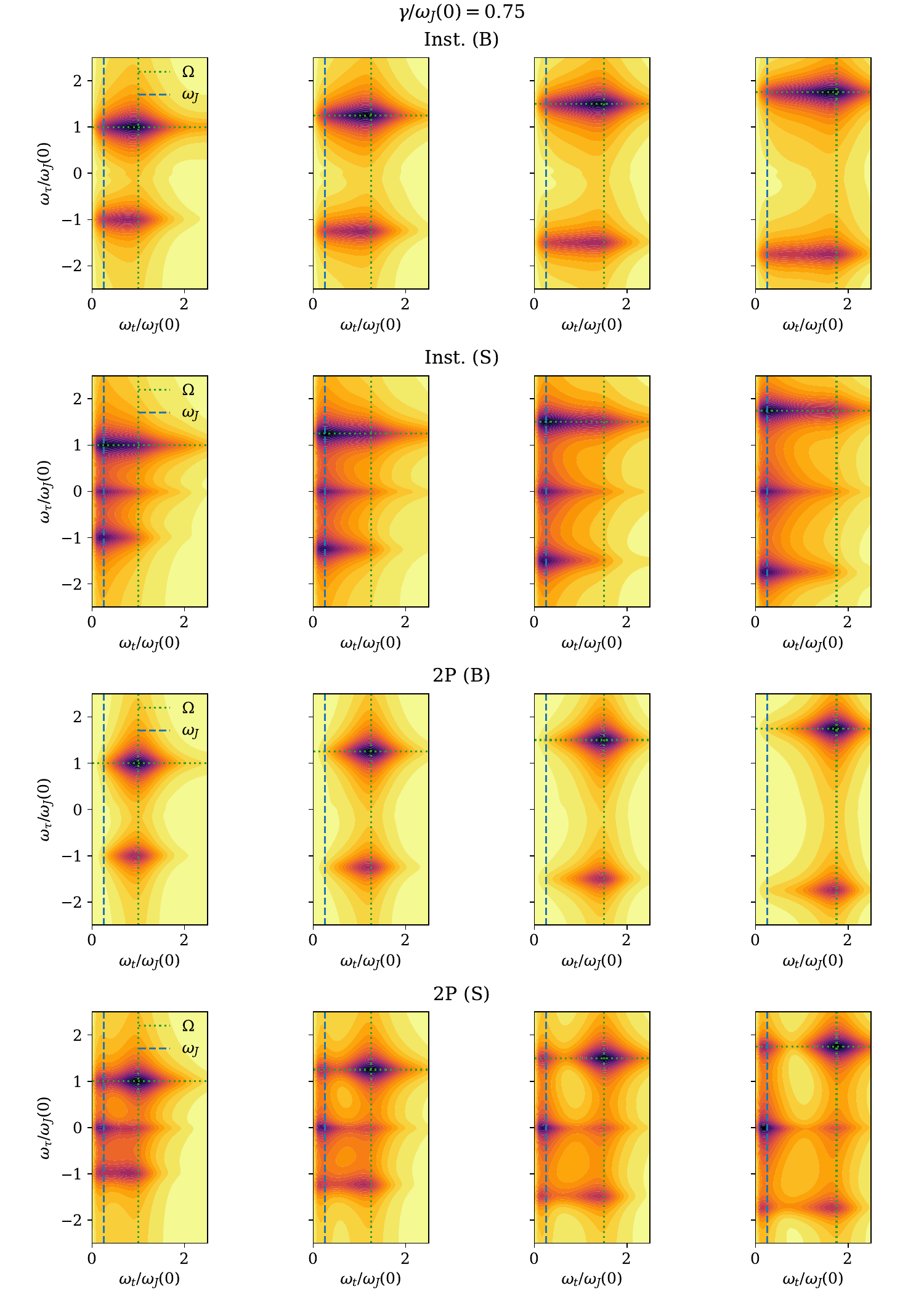}
\caption{\label{fig:c075} {Sensitivity study of the difference between the two-plasmon (2P) and instantaneous (Inst.) kernels to variations of the pump central frequency (left to right, green dotted line) when using the screening procedure of Eq.\ \ref{eq:eref} (B) or Ref.\ \cite{gomezsalvador_Phys.Rev.B24SMREF} (S). In these simulations we set the damping to $\gamma/\omega_J(T=0)=0.75$.}}
\end{figure}

\begin{figure}
\includegraphics[width=0.875\columnwidth]{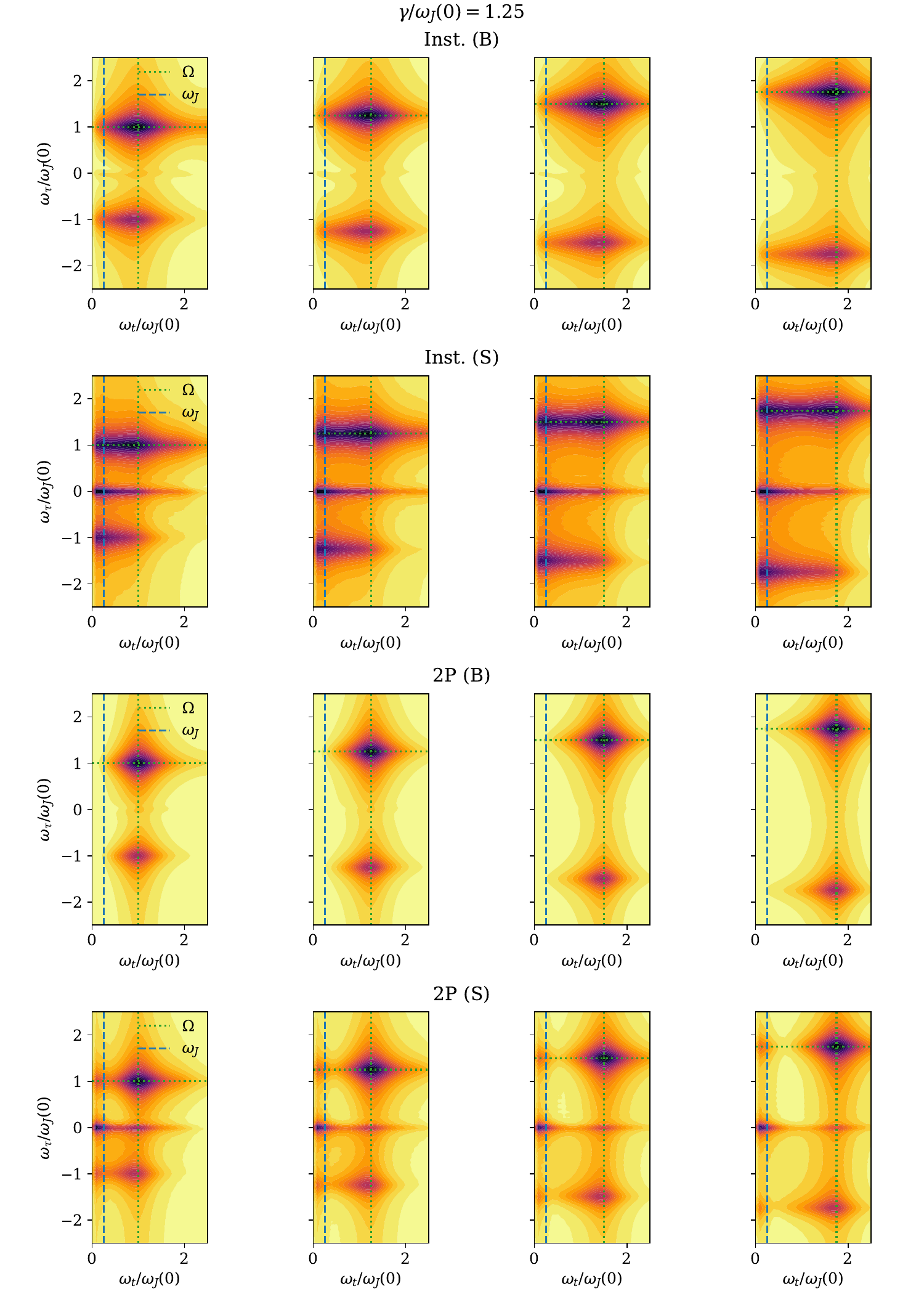}
\caption{\label{fig:c125} {Sensitivity study of the difference between the two-plasmon (2P) and instantaneous (Inst.) kernels to variations of the pump central frequency (left to right, green dotted line) when using the screening procedure of Eq.\ \ref{eq:eref} (B) or Ref.\ \cite{gomezsalvador_Phys.Rev.B24SMREF} (S). In these simulations we set the damping to $\gamma/\omega_J(T=0)=1.25$.}}
\end{figure}

    In the second part of Ref.\ \cite{gomezsalvador_Phys.Rev.B24SMREF}, an alternative approach is used to compute the nonlinear response of the system to an external perturbation. In fact, one can assume to start from a homogeneous situation in which the nonlinear material fills the whole space and introduce a current source term $J_{\text{ext}}(\omega)$ representing the external perturbation, namely
	\be
    \partial_z^2A(z,\omega)+\frac{n^2(\omega)\omega^2}{c^2}A(z,\omega)+\frac{4\pi}{c}J^{(3)}[A(z,\omega)]=-\frac{4\pi}{c}J_{\text{ext}}(\omega),
	\ee
	to be compared with the system of Eq.\ \ref{eq:hel}. It is possible, once again, to solve the equation perturbatively in terms of the nonlinear kernel $K^{(3)}$ entering $J^{(3)}$ above, adopting the same strategy outlined above. We have the solution for $\eta=0$,
	\be
	A^{(0)}(\omega)=-\frac{4\pi c}{n^2(\omega)\omega^2}J_{\text{ext}}(\omega),
	\ee
	which is necessarily $z-$independent since the source is homogeneous. In this case, the absence of a boundary-value problem allows one to take the general solution to coincide with the particular one. For the following iteration, $\eta=1$, the solution is easily found as
	\be
	A^{(1)}(\omega)=-\frac{4\pi c}{n^2(\omega)\omega^2}\int d\omega_i\,K^{(3)}(\omega_1,\omega_2,\omega_3)A^{(0)}(\omega_1)A^{(0)}(\omega_2)A^{(0)}(\omega_3)\delta\left(\omega-\sum_i\omega_i\right).
	\ee
	Using the expression for $\epsilon(\omega)=n^2(\omega)$ appropriate for the response of cuprates along the $c-$axis and \textit{assuming} as in Ref.\ \cite{gomezsalvador_Phys.Rev.B24SMREF} that $J_{\text{ext}}(\omega)\sim E_{\text{ext}}(\omega)$, mimicking the external field, we directly arrive at the nonlinearly generated $E^{(1)}(\omega)$ from the previous expression. For the instantaneous contribution, this reads explicitly
    \be
	E^{(1)}(\omega)\sim\frac{1}{\epsilon(\omega)i\omega}\int d\omega_i\,\frac{J_{\text{ext}}(\omega_1)J_{\text{ext}}(\omega_2)J_{\text{ext}}(\omega_3)}{\epsilon(\omega_1)\omega_1^2\epsilon(\omega_2)\omega_2^2\epsilon(\omega_3)\omega_3^2}\delta\left(\omega-\sum_i\omega_i\right),
	\ee
    which coincides with the one derived in Appendix D of Ref.\ \cite{gomezsalvador_Phys.Rev.B24SMREF} once we recall that $\chi_{\psi}(\omega)^{-1}=\omega^2\epsilon(\omega)$ and $E(\omega)\sim i\omega\psi(\omega)$. A comparison of the results obtained using the solution of the boundary-value problem or the response to a homogeneous perturbation is shown in {Fig.\ 13 of} the main text, both for the instantaneous (or mean-field) and the two-plasmon (or squeezing) contributions. {To assess the robustness of our conclusions, we also report here additional simulations obtained by varying key parameters used to generate that figure. We again consider a $T \sim T_c$ regime, with the plasma frequency softened to $\omega_J/\omega_J(T=0)=0.25$ (as in Fig.\ 13 of the main text), but we vary the damping parameter $\gamma$ from $\gamma/\omega_J(T=0)=1$ to $\gamma/\omega_J(T=0)=0.75$ in Fig.\ \ref{fig:c075}, and to $\gamma/\omega_J(T=0)=1.25$ in Fig.\ \ref{fig:c125}, values that represent reasonable variations in the proximity of the critical temperature. We assume a Lorentzian pump pulse with width $\sigma/\omega_J(T=0)=0.15$ (again as in Fig.\ 13 of the main text), and we vary its central frequency from $\Omega/\omega_J(T=0)=1$ to the three values $\Omega/\omega_J(T=0)=1.25,1.5,1.75$, shown columnwise from left to right in Figs.\ \ref{fig:c075} and \ref{fig:c125}. These additional checks confirm that the main effects of linear-response propagation on the 2D THz signal remain qualitatively the same as those discussed in Fig.\ 13 of the manuscript. In particular, both the two-plasmon and instantaneous contributions exhibit more pronounced variations upon changing the pump central frequency when the screening procedure of Ref.\ \cite{gomezsalvador_Phys.Rev.B24SMREF} is used, while these variations are smoothened when employing our propagation treatment, Eq.\ \ref{eq:eref}.}

\clearpage
\onecolumngrid

\end{document}